\documentclass[aps,prr,showpacs,footinbib,superscriptaddress,twocolumn,longbibliography]{revtex4-1}

\usepackage{amsfonts}
\usepackage{amsmath}
\usepackage{txfonts}
\usepackage{amssymb}
\usepackage{amsbsy} 
\usepackage{graphicx}
\usepackage{color}
\usepackage{mathdots}
\usepackage{braket}
\usepackage{utfsym}
\usepackage{hyperref}
\hypersetup{hypertex=true, colorlinks=true, linkcolor=blue, anchorcolor=blue, citecolor=blue,urlcolor=blue}
\begin{document}
\title{Non-Bloch band theory for non-Hermitian continuum systems}

\author{Yu-Min Hu}
\altaffiliation{These authors contributed equally to this work.}
\affiliation{Institute for Advanced Study, Tsinghua University, Beijing, 100084, China}

\author{Yin-Quan Huang}
\altaffiliation{These authors contributed equally to this work.}
\affiliation{Institute for Advanced Study, Tsinghua University, Beijing, 100084, China}

\author{Wen-Tan Xue}
\affiliation{Institute for Advanced Study, Tsinghua University, Beijing, 100084, China}
\affiliation{ Department of Physics, National University of Singapore, Singapore 117542, Singapore }

\author{Zhong Wang}
 \altaffiliation{ wangzhongemail@tsinghua.edu.cn }
\affiliation{Institute for Advanced Study, Tsinghua University, Beijing, 100084, China}
\begin{abstract}
One of the most pronounced non-Hermitian phenomena is the non-Hermitian skin effect,  which refers to the exponential localization of bulk eigenstates near the boundaries of non-Hermitian systems. Whereas non-Bloch band theory has been developed to describe the non-Hermitian skin effect in lattice systems, its counterpart in continuum systems still lacks a quantitative characterization. Here, we generalize the non-Bloch band theory to non-Hermitian continuum systems. In contrast to lattice systems for which the bulk Hamiltonian alone determines the non-Hermitian skin effect and energy spectrum, we find for continuum systems that the number of boundary conditions, i.e., the number of independent differential equations satisfied by wavefunctions at two boundaries, must also be included as essential information. We show that the appropriate discretization of continuum systems into lattice models requires matching the hopping range of the latter with the number of boundary conditions in the former. Furthermore, in periodic non-Hermitian continuum systems, we highlight the application of the transfer matrix in determining the generalized Brillouin zone. Our theory serves as a useful toolbox for investigating the rich non-Bloch physics in non-Hermitian continuum systems, such as photonic crystals, elastic media, and certain cold-atom systems. 
\end{abstract}

\maketitle
\section{Introduction}
Non-Hermitian systems arise from the intricate interplay between the degrees of freedom of systems and their surrounding environment. Over the past few years, a variety of novel phenomena have been uncovered within non-Hermitian systems, without analogs in Hermitian cases. This intriguing advancement has dramatically expanded the frontiers of non-Hermitian physics\cite{ashida2020non, bergholtz2021exceptional}.

One of the remarkable phenomena that emerges in non-Hermitian systems is the non-Hermitian skin effect (NHSE) \cite{yao2018edge,yao2018chern,kunst2018biorthogonal,lee2019anatomy,martinez2018non,zhang2022review,okuma2023non, lin2023topological,ding2022non}. In these systems, most eigenstates exhibit exponential accumulation towards the boundary in a generic open-boundary configuration. These skin modes under open boundary conditions (OBC) invalidate the Bloch-wave picture in lattice systems, leading to a non-Bloch band theory which is based on the generalized Brillouin zone (GBZ) \cite{yao2018edge,yao2018chern,yokomizo2019nonbloch,zhang2020correspondence}. Notably, the non-Bloch band theory is crucial for understanding non-Hermitian topology. The novel non-Bloch bulk-boundary correspondence has been experimentally observed in various domains, including quantum walk dynamics \cite{xiao2020non}, topological mechanics \cite{ghatak2020observation}, topolectrical circuits \cite{helbig2020generalized}, and photonic systems \cite{weidemann2020topological}.

The non-Bloch band theory exhibits extensive applicability across various domains of non-Hermitian systems. Aside from understanding the aforementioned non-Hermitian topology, non-Bloch band theory provides an efficient toolkit to obtain basic spectral information: the OBC spectrums and the localization length of skin modes. Moreover, NHSE also plays a crucial role in investigating non-Hermitian Green's functions \cite{xue2021simple,wanjura2020topological, Zirnstein2021bulk,okuma2021non,hu2023greens} and analyzing long-term relaxation dynamics \cite{song2019non,liu2020helical,longhi2019probing,longhi2019nonbloch, xiao2021observation,xue2022non,xiao2023observation}, thereby enriching our understanding of non-Hermitian dynamics. Furthermore, the context of non-Bloch band theory is further enriched by symmetry \cite{liu2019topological,okuma2020topological,yi2020non-hermitian,kawabata2020nonbloch,yang2020nonperturbative,yokomizo2021non} and dimensionality \cite{yao2018chern,kawabata2020higher-order,zhang2022universal,wang2022amoeba}. Notably, the non-Bloch band theory enables direct computation of various physical quantities (e.g., the energy spectrum) in the thermodynanmic limit (i.e. large-size limit), without the finite-size errors generated in the real-space calculations. 

In addition to non-Hermitian lattice models, non-Hermitian continuum systems are equally fascinating because of their ubiquitous presence in the real world. These systems offer valuable insights into non-Hermitian physics. For example, photonic crystals with dielectric loss serve as a natural platform for studying non-Hermitian phenomena \cite{Ding2015coalescence,hahn2016observation, Cerjan2016exceptional, feng2017non, el2018non, zhou2018observation, bandres2018topological, harari2018topological, Silveirinha2019topological, Lu2014topological, Ozawa2019topological}. They have been utilized to capture exceptional points \cite{Ding2015coalescence,hahn2016observation, Cerjan2016exceptional,feng2017non}, observe bulk Fermi arcs \cite{zhou2018observation}, and design topological insulator lasers \cite {bandres2018topological,harari2018topological}. As another example, non-Hermitian operators arise in elastic metamaterials which are subjected to non-conservative forces \cite{Shmuel2020linking,scheibner2020odd,scheibner2020non,chen2021realization, Braverman2021topological, fruchart2023odd}. In a broader sense, any continuum system that undergoes energetic exchange with its environment can exhibit non-Hermitian features. 

With the wide availability of non-Hermitian continuum systems in experimental setups, numerous studies have been conducted to propose the existence of NHSE in such systems. Examples include non-Hermitian elastic media \cite{scheibner2020non,chen2021realization}, photonic systems \cite{zhong2021nontrivial,Ochiai2022non,yan2021nonhermitian,Yokomizo2022non-hermitian_wave,FangHu2022Geometry,liu2023localization,zhu2023photonic,kokhanchik2023nonhermitian}, and cold atom platforms \cite{guo2022theoretical,li2020topological,liang2022dynamic,kokhanchik2023nonhermitian}. 

However, compared to the non-Bloch band theory in non-Hermitian lattice systems, a coherent framework is currently lacking to characterize NHSE in continuum systems. While interesting theoretical attempts have been made in this direction \cite{guo2022theoretical, Yokomizo2022non-hermitian_wave,longhi2021non,yuce2022non}, they are mainly tailored to describe specific classes of non-Hermitian continuum systems. Moreover, the connection between non-Hermitian lattice systems and their continuum counterparts remains unclear. Consequently, there is an urgent need for a general non-Bloch band theory applicable to non-Hermitian continuum systems, which would serve as a useful tool for investigating non-Bloch physics in these systems.

This work aims to develop the non-Bloch band theory in non-Hermitian continuum systems with continuous or discrete translation symmetry. The continuous translation symmetry in the bulk captures the behavior of uniform continuum systems, while the discrete translation symmetry is associated with periodic continuum systems. In particular, both types of systems can exhibit NHSE under open boundary conditions. We show that these continuum systems can be efficiently described using a non-Bloch band theory based on the concept of the generalized momentum space or the generalized Brillouin zone.

We start with investigating the non-Hermitian skin effect in homogeneous non-Hermitian continuum systems. In contrast to non-Hermitian lattice models, where the tight-binding Hamiltonian solely determines the GBZ, we discover that, in OBC continuum systems, the bulk Hamiltonian alone is insufficient to determine the generalized momentum space [see Eq. \eqref{eq:homogeneous_GBZ}], which is the analog of GBZ studied in lattice systems. Therefore we highlight the significant role of the number of boundary conditions in defining the generalized momentum space for continuum systems. We find that the number of independent differential equations satisfied by wavefunctions at two boundaries dramatically shapes the eigenstate wavefunction profiles in the bulk. To illustrate this point, we begin with two intuitive examples and then derive a concise formulation for general homogeneous non-Hermitian continuum systems.  Our findings indicate that the emergence of the NHSE in continuum systems arises from the intricate interplay between the non-Hermitian Hamiltonians and the boundary conditions.

Following the development of the non-Bloch band theory for homogeneous non-Hermitian continuum systems, our subsequent result establishes a connection between the GBZ in lattice models and the analogous generalized momentum space in continuum systems. This continuum-lattice correspondence is founded upon the identification of the number of boundary conditions at two edges of continuum systems with the hopping ranges in the corresponding lattice models. On one hand, this correspondence provides a practical approach to discretizing a non-Hermitian continuum Hamiltonian, which is beneficial for numerical simulations. On the other hand, it facilitates a comprehensive understanding of the low-energy (i.e., long wavelength)  physics in non-Hermitian lattice systems.

Moreover, we unveil that spatial periodicity further enriches the non-Bloch band theory in continuum systems. The periodic structure in continuum systems highlights the crucial role played by the transfer matrix along the spatial dimension. Notably, the form of the transfer matrix is very similar to the Floquet operator of a periodically driven non-Hermitian system. By analyzing the eigenvalues of the transfer matrix, we can efficiently define the GBZ and describe the NHSE in periodic continuum systems. Thus, in addition to the continuum-lattice correspondence, the transfer matrix offers an alternative avenue for exploring non-Bloch physics in periodic non-Hermitian continuum systems. 

The non-Bloch band theory can be readily extended to contain non-Hermitian continuum systems with local degrees of freedom. These degrees of freedom can be interpreted as spins of particles or as components of vector fields. This extension goes beyond mere mathematical generalization; it opens the door for practical applications in studying the  NHSE within non-Hermitian elastic media \cite{scheibner2020non,chen2021realization} and non-Hermitian photonic crystals \cite{zhong2021nontrivial, Ochiai2022non, yan2021nonhermitian, Yokomizo2022non-hermitian_wave, FangHu2022Geometry, liu2023localization, zhu2023photonic}. 

Before delving into the details, we summarize the main results of our work. Sec. \ref{sec:homogeneous_GBZ} presents the framework for describing the non-Hermitian skin effect in homogeneous non-Hermitian continuum systems. In Sec. \ref{sec:correspondence_discretization}, we outline the general approach for continuum-lattice correspondence. Sec. \ref{sec:periodic_GBZ} elaborates on the non-Bloch band theory for periodic non-Hermitian continuum systems. Finally, Sec. \ref{sec:matrix} extends our theory to non-Hermitian continuum systems with multiple local degrees of freedom. Collectively, these parts establish a theoretical framework for the non-Bloch band theory in general non-Hermitian continuum systems.

The remaining sections of this paper are organized as follows. In Sec. \ref{sec:homogeneous}, we introduce the phenomenon of NHSE in homogeneous non-Hermitian continuum systems. This is achieved through the analytical study of two instructive models. We then delve into the significance of boundary conditions and develop a general formulation for the theory of generalized momentum space in homogeneous continuum systems. In Sec. \ref{sec:correspondence}, we establish the connection between continuum models and lattice systems by providing a boundary-dependent discretization method. In Sec. \ref{sec:periodic}, we establish the non-Bloch band theory in periodic non-Hermitian continuum systems, where we emphasize the application of the transfer matrix. Finally, in Sec. \ref{sec:matrix}, we explore the non-Bloch band theory in non-Hermitian continuum systems with local degrees of freedom. We conclude our work in Sec. \ref{sec:discussion} and discuss the implications and potential future directions in non-Hermitian continuum systems.

\section{Homogeneous continuum systems}\label{sec:homogeneous}
In this section, we develop the non-Bloch band theory in homogeneous non-Hermitian continuum systems. Homogeneous systems establish a continuous translation symmetry in the bulk, although such symmetry can be violated at the boundaries. Starting with two analytically solvable models in Sec. \ref{sec:homogenerous_hatano} and Sec. \ref{sec:homogeneous_p4}, we find that the number of boundary conditions plays an indispensable role in determining the skin modes of non-Hermitian continuum systems. Following these intuitions, we establish the general non-Bloch band theory in Sec. \ref{sec:homogeneous_GBZ} and discuss its implications and applications subsequently. 
\subsection{Hatano-Nelson model}\label{sec:homogenerous_hatano}

As a warm-up, we begin with the  Hatano-Nelson model in the continuum  system. The Hamiltonian is given by $\hat H_{\text{HN}}=(\hat p+ig)^2/
2m=\hat p^2/2m+ig\hat p/m-g^2/2m$ \cite{hatano1996localization,hatano1997vortex}, where the momentum operator is $\hat p=-i\partial_x$ in the real-space representation. This Hamiltonian describes a massive particle moving in a one-dimensional system with a background imaginary gauge field. The corresponding stationary Schrodinger equation is
\begin{eqnarray}
 -\frac{1}{2m}\frac{\partial^2}{\partial x^2}\psi(x)+\frac{g}{m}\frac{\partial}{\partial x}\psi(x)-\frac{g^2}{2m}\psi(x)=E\psi(x).\label{eq:homogeneous_Hatano_Nelson}
\end{eqnarray}

We are interested in the eigenstates of this system under different boundary conditions. Starting with the periodic boundary conditions (PBC) $\psi(x)=\psi(x+L)$, the eigenstates are characterized by extended plane waves $\psi_{\text{PBC}}(x)=\frac{1}{\sqrt{L}}e^{ikx}$ with real-valued momentum $k\in2\pi\mathbb{Z}/L$, where $L$ is the system size. Consequently, the eigenenergies are given by $E_{\text{PBC}}=(k+ig)^2/2m$, which is complex if $g\ne0$.

However, when we take the OBC $\psi(0)=\psi(L)=0$, a representative eigenstate becomes $\psi_{\text{OBC}}(x)=Ae^{gx}\sin{(\kappa_nx)}$, with a real-valued eigenvalue $E_{\text{OBC}}=\kappa_n^2/2m$. Here, $A$ is an irrelevant normalization factor, and the discrete parameter $\kappa_n$ is $\kappa_n=n\pi/L$ with $n=1,2,\cdots$. Surprisingly, the eigenstates of the Hatano-Nelson model on an open chain are no longer extended. Instead, these eigenstates become exponentially localized to the boundary, where the localization length (i.e., skin depth) is determined by the strength of the imaginary gauge field. This is an explicit example of NHSE in continuum systems.

Originally proposed to discuss the localization-delocalization transition of the magnetic flux lines around the defects in superconductors, the Hatano-Nelson model has recently been studied as a simplified model of NHSE. Here, to demonstrate the role of boundary conditions in shaping the bulk wavefunctions in non-Hermitian continuum systems, we explore a variation of the Hatano-Nelson model, which is the one-dimensional Fokker-Planck equation in classical probability theory. In the latter case, two different types of boundary conditions naturally arise when describing different random walking processes.  As we will reveal through this work, boundary conditions play a more profound role in non-Hermitian continuum systems.

In particular, we consider a non-negative probability distribution function $P(x,t)$ that depends on the one-dimension spatial variable $x$ and the time $t$. The evolution of $P(x,t)$ is generated by a Fokker-Planck equation \cite{risken1996fokker}
\begin{equation}
\frac{\partial}{\partial t} P(x,t)=\frac{\partial}{\partial x}\left[-\mu(x)P(x,t)+\frac{\partial}{\partial x}D(x)P(x,t)\right],
\end{equation}
where the spatial-dependent $\mu(x) $ and $D(x)$ are the drift coefficient and the positive diffusion coefficient, respectively. Assuming that the eigenmodes of this differential equation are given by $P(x,t)=\psi(x)e^{-Et}$, we derive its corresponding stationary eigenequation:
\begin{equation}
	\begin{split}
	E\psi(x)=&-D(x)\frac{\partial^2}{\partial x^2}\psi(x)+\left[\mu(x)-2\left(\frac{\partial}{\partial x}D(x)\right)\right]\frac{\partial}{\partial x}\psi(x)\\
	&+\left[\frac{\partial}{\partial x}\mu(x)-\frac{\partial^2}{\partial x^2}D(x)\right]\psi(x).
	\end{split}\label{eq:general_stationary_fokker_planck}
\end{equation}

When the probability function is distributed within a finite domain $x\in[0,L]$, we need to impose suitable boundary conditions. In the language of a Fokker-Planck equation, there are two types of open boundary conditions: (a) the absorbing conditions $P(0,t)=P(L,t)=0$ and (b) the reflecting conditions $S(0,t)=S(L,t)=0$, where
$S(x,t)=\mu(x)P(x,t)-\partial_x[D(x)P(x,t)]$ is the probability current.

The absorbing boundary condition does not preserve the total probability, which can be interpreted as absorbing the Brownian particle at the boundary. Therefore this boundary condition leads to a decrease in the overall probability within the system. In contrast, the reflecting boundary condition preserves the total probability due to the zero probability current at the two edges. This scenario can be realized when a Brownian particle moves in an infinitely deep square potential well. 

In this section, we consider a simple setup where a Brownian particle with strong friction moves in an open chain $x\in[0,L]$. Additionally, we add a constant drift force along the $x$ direction. Under this circumstance, we can get a homogeneous stationary Fokker-Planck equation
\begin{equation}
-D\frac{\partial^2}{\partial x^2}\psi(x) +	\mu\frac{\partial}{\partial x}\psi(x)=E\psi(x),	\label{eq:homogeneous_stationary_Fokker_Planck}
\end{equation}
where the drift term $\mu$ is related to the external force while the diffusion term $D$ is determined by the random force and friction. This is the same as the Hanato-Nelson model in Eq. \eqref{eq:homogeneous_Hatano_Nelson}. Therefore it is expected that this classical system has skin modes as its eigenstates.

To capture the feature of NHSE, we consider the time evolution of $P(x,t)$ starting from an initial state $P(x,0)=\delta(x-x_0)$.  We analytically construct all the eigenstates under different types of open boundary conditions and discuss the classical stochastic dynamics. 

First, the absorbing boundary conditions $P(0,t)=P(L,t)=0$ ensure that the eigenstates satisfy $\psi(0)=\psi(L)=0$. After some calculations, we find that the eigenmodes and eigenenergies of Eq. \eqref{eq:homogeneous_stationary_Fokker_Planck} are
\begin{equation}\label{eq:homogeneous_stationary_Fokker_Planck_bc1_right_mode}
	 \psi_{n}(x)=\sqrt{\frac{2}{L}}e^{\frac{\mu}{2D} x}\sin(\kappa_nx),\quad E_n=D\kappa_n^2+\frac{\mu^2}{4D},
\end{equation} 
where $\kappa_n=\frac{n\pi}{L}$ and $n=1,2,\cdots$. These eigenmodes indeed exhibit NHSE, as their envelope function $e^{\frac{\mu}{2D}x}$ is localized to the right edge $x=L$ with a localization length $\lambda=\frac{2D}{\mu}$.

\begin{figure}[t]
	\centering
	\includegraphics[width=8.5cm]{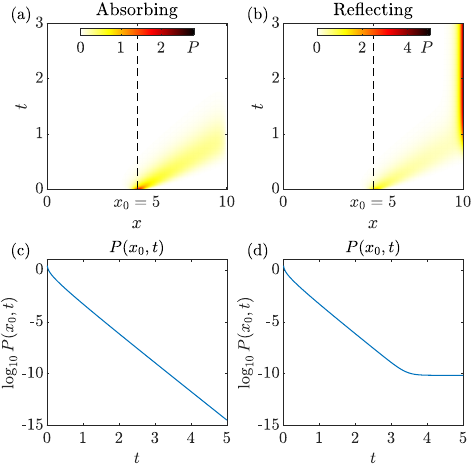}
	\caption{(a) $P(x,t)$ under the absorbing boundary conditions. (b) $P(x,t)$ under the reflecting boundary conditions. (c) $P(x_0,t)$ for the dashed line in (a). (d) $P(x_0,t)$ for the dashed line in (b). In all planes, the initial state is $P(x,0)=\delta(x-x_0)$. We choose a large cutoff $n_c=10^4$ for the summations in Eq. \eqref{eq:fokker_planck_absorbing_evolution} and Eq. \eqref{eq:fokker_planck_reflecting_evolution}. Other parameters: $\mu=5,D=1,L=10,x_0=5$. }\label{fig:fokker_planck}
\end{figure}

An eigenstate of a non-Hermitian Hamiltonian $\hat{H}$ is also called a ``right eigenstate'', to be distinguished from the ``left eigenstate'' which is an eigenstate of $\hat{H}^\dagger$. If we formally write the differential equation Eq. \eqref{eq:homogeneous_stationary_Fokker_Planck} as $\hat H\psi(x)=E\psi(x)$ with $\hat H=-D\partial_x^2+\mu\partial_x$, the eigenmodes $\psi_n(x)$ in Eq.\eqref{eq:homogeneous_stationary_Fokker_Planck_bc1_right_mode} are the right eigenstates of $\hat H$ with the boundary conditions $\psi(0)=\psi(L)=0$. These right eigenstates are denoted as $ \psi_{r,n}(x)$. (We explicitly use the subscript $r$ only for this section, whereas it is omitted in other sections for simplicity.) Correspondingly, the left eigenstates of $\hat H$, namely the eigenstates of $\hat H^\dagger=-D\partial_x^2-\mu\partial_x$, with proper boundary conditions  (see the Appendix \ref{appendix:left_eigenmode}), are given by $\psi_{l,n}(x)=\sqrt{\frac{2}{L}}e^{-\frac{\mu }{2D}x}\sin(\kappa_nx)$. Here we use the subscript $l$ to label the left eigenstates. Compared with the right eigenmodes $ \psi_{r,n}(x)$ in Eq. \eqref{eq:homogeneous_stationary_Fokker_Planck_bc1_right_mode}, these left eigenmodes $ \psi_{l,n}(x)$ are exponentially localized to the opposite direction. The detailed derivations of the right and left eigenmodes of $\hat H$ with the boundary conditions  $\psi(0)=\psi(L)=0$ are shown in Appendix \ref{appendix:left_eigenmode}, where we also show that the right and left eigenmodes satisfy the biorthogonal relation $\int_0^L\psi_{l,m}^*(x)\psi_{r,n}(x)dx=\delta_{m,n}$ and the completeness relation $	\sum_{n=1}^{+\infty}\psi_{r,n}(x)\psi_{l,n}^*(x^\prime)=\delta(x-x^\prime)$. As a result, the evolution of $P(x,t)$ starting from an initial state $P(x,0)= \delta(x-x_0)$ can be expressed as
\begin{equation}
	\begin{split}
	P(x,t)&=\sum_{n=1}^{\infty}e^{-E_nt}\psi_{r,n}(x)\psi^*_{l,n}(x_0)\\
	&=\frac{2}{L}e^{-\frac{\mu^2}{4D}t+\frac{\mu(x-x_0)}{2D}}\sum_{n=1}^{\infty}e^{-D\kappa_n^2t}\sin(\kappa_nx)\sin(\kappa_nx_0),
	\end{split}\label{eq:fokker_planck_absorbing_evolution}
\end{equation}
which is shown in Fig. \ref{fig:fokker_planck}(a). We observe that the Brownian particle is absorbed when it arrives at the boundary. More importantly, the NHSE of this system generates a chiral motion of the Brownian particle. While the probability distribution becomes broader due to the diffusion effect, the distribution center moves in the right direction due to NHSE. 

Next, we consider the reflecting boundary conditions $S(0,t)=S(L,t)=0$ where $S(x,t)=(\mu-D\partial_{x})P(x,t)$. In this case, the right eigenstates satisfy $ (\mu-D\partial_{x})\psi(x)|_{x=0}=(\mu-D\partial_{x})\psi(x)|_{x=L}=0$. As discussed in the Appendix \ref{appendix:left_eigenmode}, we can obtain the right and left eigenmodes of this model. The reflecting boundary conditions promise a steady state with a zero eigenvalue $E_0=0$. The right and left wavefunctions of the zero mode are given by
\begin{equation}
\psi_{r,0}(x)=\left(\frac{\mu/D}{e^{\mu L/D}-1}\right)^{\frac{1}{2}}e^{\frac{\mu}{D}x},\quad\psi_{l,0}(x) =\left(\frac{\mu/D}{e^{\mu L/D}-1}\right)^{\frac{1}{2}};
\end{equation}
Interestingly, the left eigenmode is independent of $x$, which is proved in Appendix \ref{appendix:left_eigenmode}. Other right and left eigenmodes are given by
\begin{equation}
	\begin{split}
	\psi_{r,n}(x)&=\left(\frac{2/L}{\mu^2+4\kappa_n^2D^2}\right)^{\frac{1}{2}}e^{\frac{\mu }{2D}x}(\mu\sin(\kappa_nx)+2\kappa_nD\cos(\kappa_n x)),\\	\psi_{l,n}(x)&=\left(\frac{2/L}{\mu^2+4\kappa_n^2D^2}\right)^{\frac{1}{2}}e^{-\frac{\mu }{2D}x}(\mu\sin(\kappa_nx)+2\kappa_nD\cos(\kappa_n x)).
	\end{split}
\end{equation}
Correspondingly, the eigenenergies are $E_n=D\kappa_n^2+{\mu^2}/{(4D)}$ where $\kappa_n=\frac{n\pi}{L}$ and $n=1,2,\cdots$ (see Appendix \ref{appendix:left_eigenmode} for detailed analysis). These right eigenstates $\psi_{r,n}(x)$ are enveloped by an envelope function $e^{\frac{\mu}{2D}x}$, exhibiting a similar skin profile to the eigenstates under the absorbing boundary conditions. 

Given an initial probability distribution $P(x,0)=\delta(x-x_0)$, we can use the spectral representation to find the evolution of $P(x,t)$, which is

\begin{equation}
	\begin{split}
		P(x,t)&=\psi_{r,0}(x)\psi^*_{l,0}(x_0)+\sum_{n=1}^{\infty}e^{-E_nt}\psi_{r,n}(x)\psi^*_{l,n}(x_0)\\
		&=\frac{\mu}{D}\frac{ e^{\frac{\mu x}{D}}}{e^{\frac{\mu L}{D}}-1}+\frac{2}{L}e^{-\frac{\mu^2}{4D}t+\frac{\mu(x-x_0)}{2D}}\sum_{n=1}^{\infty}\frac{e^{-D\kappa_n^2t}F_n(x)F_n(x_0)}{\mu^2+4\kappa_n^2D^2}.
	\end{split}\label{eq:fokker_planck_reflecting_evolution}
\end{equation}
In the above,  the auxiliary function is defined as $F_n(x)=\mu\sin(\kappa_nx)+2\kappa_nD\cos(\kappa_n x)$. This result is shown in Fig. \ref{fig:fokker_planck}(b). We observe that the NHSE of eigenmodes also generates the chiral motion of the Brownian particle. Different from the absorbing boundary conditions, the Brownian particle is confined within the system and evolves to a localized  steady state. 

In both cases, the Brownian particle initially in the bulk will move to the right boundary. This NHSE-induced chiral motion originates from the nonzero drift term and is independent of the details of boundary conditions. However, the long-time steady state relies on boundary details. Under the absorbing boundary conditions, the particle will be absorbed by the edge and the probability function will decay to zero after a long-time evolution [Fig. \ref{fig:fokker_planck}(a)]. In contrast, under the reflecting boundary conditions, the Brownian particle will remain in the system with a large probability of being found near the edge [Fig. \ref{fig:fokker_planck}(b)].

In addition to the chiral motion towards the boundary, the probability evolution $P(x_0,t)$ at the initial point $x=x_0$ also exhibits interesting features. As shown in Figs. \ref{fig:fokker_planck}(c) and (d), $P(x_0,t)\sim e^{-E_st}$ decays exponentially before reaching the steady-state value. Under both boundary conditions, $E_s\approx 6.57\approx{\mu^2}/{(4D)}$ where ${\mu^2}/{(4D)}$ serves as a saddle point at the bottom of the bulk OBC spectrum \cite{longhi2019probing}. The short-term evolution of $P(x_0,t)$ is not affected by the details of the boundary conditions, but it still holds the data of the OBC spectrum. This pure bulk feature can be seen clearly in Eq. \eqref{eq:fokker_planck_absorbing_evolution} and Eq. \eqref{eq:fokker_planck_reflecting_evolution}, where only the first time-dependent term $\exp(-\frac{\mu^2}{4D}t)$ survives when $L\to+\infty$ and then $t\to+\infty$. This feature has been understood in lattice models \cite{longhi2019probing}. The OBC-spectrum-based dynamical property of $P(x_0,t)$ can be applied to detect the existence of NHSE in continuum systems.

In summary, the analysis of the Hatano-Nelson model and the one-variable Fokker-Planck equation yields crucial insights into NHSE in continuum systems. Although the presence of a zero mode may depend on the specific choice of boundary conditions, the profiles of bulk eigenstates remain robust under both types of open boundary conditions. As we will explore in the subsequent section, it is the number of boundary conditions, rather than their specific details, that controls the behaviors of skin modes. This is a significant feature of NHSE in non-Hermitian continuum systems.

\subsection{Analytically solvable model $\hat H=\hat p^4$}\label{sec:homogeneous_p4}

Before introducing the general theory for the NHSE in continuum systems, we examine another analytically solvable model on a length-$L$ chain. The Hamiltonian is $\hat H=\hat{p}^4$. Despite its simplicity, this model characterizes basic aspects of non-Bloch physics in continuum systems.

The stationary Schrodinger equation of this Hamiltonian is
\begin{equation}
	\left(-i\partial_x\right)^4\psi(x)=E\psi(x),\quad x\in[0,L].
	\label{eq:homogeneous_analytical_p4}
\end{equation} 
We are interested in the eigenvalues and wavefunctions of the Schrodinger equation. To obtain these eigenstates, we have to impose suitable boundary conditions at $x=0$ and $x=L$. 

When we take PBC $\psi(x)=\psi(x+L)$, the Hamiltonian is Hermitian. The PBC spectrum is given by $E_{\text{PBC}}=k^4$ with a real-valued momentum $k\in2\pi\mathbb{Z}/L$, and the eigenstates are extended plane waves. This is similar to the aforementioned PBC Hatano-Nelson model.

What will happen if we put this model on a finite open chain? A naive expectation is that it is the same Hermitian Hamiltonian as in the PBC case. However, this intuition is incorrect. It is well-known that the Hermiticity of a differential operator closely relies on the boundary conditions for the wavefunctions on an open chain \cite{bonneau2001self,blank2008hilbert}. To show this, we take two arbitrary continuous wavefunctions $\psi_1(x)$ and $\psi_2(x)$ from a given Hilbert space in the region $x\in[0, L]$. The inner product of two wavefunctions is defined as $\braket{\psi_1|\psi_2}=\int_0^L\psi_1^*\psi_2\mathrm{d}x$. We omit $x$ of $\psi_{1,2}(x)$ for simplicity. The Hermiticity of an operator $\hat O$ requires that
\begin{equation}
	\braket{\psi_1|\hat O\psi_2}=\int_0^L\psi_1^*\left(\hat O\psi_2\right)\mathrm{d}x=\int_0^L\left(\hat O\psi_1\right)^*\psi_2\mathrm{d}x=\braket{\hat O\psi_1|\psi_2}.
	\label{eq:homogeneous_hermiticity}
\end{equation}
In other words, the operator $\hat O$ is Hermitian (self-adjoint) in the Hilbert space if it fulfills the above condition. As for $\hat H=\hat p^4$ on a length-$L$ chain, we can get
\begin{equation}
	\int_0^L\psi_1^* (-i\partial_x)^4\psi_2\mathrm{d}x=\left.\int_0^L\left((-i\partial_x)^4\psi_1\right)^*\psi_2\mathrm{d}x+f(x)\right|_0^L,
\end{equation}
where the boundary term $f(L)-f(0)$ comes from the integration by parts. The residual function $f(x)$ is \begin{equation}
 f(x)=\psi_1^*\psi^{\prime\prime\prime}_2-(\psi^\prime_1)^*\psi^{\prime\prime}_2+(\psi^{\prime\prime}_1)^*\psi^\prime_2-(\psi^{\prime\prime\prime}_1)^*\psi_2
\end{equation}
where the prime is a shorthand of spatial derivatives. Therefore, when two arbitrary wavefunctions $\psi_{1,2}(x)$ in the Hilbert space satisfy the Hermitian (self-adjoint) condition $f(0)-f(L)=0$, the Hamiltonian $\hat H=\hat p^4$ can be viewed as a Hermitian operator in the given Hilbert space. This result indicates that the Hermiticity of a continuum Hamiltonian closely depends on the boundary conditions.

Since we arbitrarily choose $\psi_{1,2}(x)$ in the Hilbert space, we should carefully impose the boundary conditions at $x=0, L$ for all the wavefunctions in the Hilbert space. While the Hermitian condition $f(0)-f(L)=0$ is automatically fulfilled by the PBC $\psi(x)=\psi(x+L)$, the OBC situation becomes more interesting since the boundary conditions at two ends can be taken independently. In a general case, with an arbitrary energy $E$, there are four complex wave vectors that satisfy the characteristic equation $E=k^4$. The corresponding wavefunction is $\psi_E(x)=\sum_{m=1}^4c_me^{ik_mx}$ where $k_m$'s are four roots of $E=k^4$. These four undetermined coefficients $ c_m$ should be fixed by four independent boundary conditions at two ends.

As a first example, if we take the boundary condition as $\psi(0)=\psi^\prime(0)=\psi(L)=\psi^\prime(L)=0$ for all the wavefunctions in the Hilbert space, the arbitrary two of them provide $f(0)=f(L)=0$. Thus, $\hat H=\hat p^4$ is Hermitian under this type of boundary condition. Alternatively, we can choose another type of boundary condition as $\psi(0)=\psi^\prime(0)=\psi^{\prime\prime}(0)=\psi(L)=0$. In this case, there are $3$ $(1)$ boundary equations at the left (right) edge. Two arbitrary wavefunctions $\psi_{1,2}(x)$ in this Hilbert space give rise to $f(L)-f(0)=-(\psi^\prime_1(L))^*\psi^{\prime\prime}_2(L)+(\psi^{\prime\prime}_1(L))^*\psi^{\prime}_2(L)\ne0$. Therefore the Hamiltonian is non-Hermitian under this type of boundary condition.

In conclusion, by studying a simple example, we have demonstrated that the Hermiticity of a continuum Hamiltonian is closely related to boundary conditions. This result implies that the eigenstates of a continuum Hamiltonian may exhibit different behaviors under different boundary conditions, showing NHSE in one case while being extended in another case. 

To see this connection more clearly, we explicitly solve the wavefunctions and spectrum of $\hat H=\hat p^4$ on an open chain. We take three different types of open boundary conditions, which are shown in Table \ref{table:three_OBC}. Here, we define an integer $n_l>0$ ($n_r>0$) to represent the number of boundary equations at the left (right) edge. These two numbers are constrained by $n_r+n_l=4$. The three kinds of boundary conditions, labeled (i) to (iii), are represented by $(n_l, n_r)=(1,3)$, $(2,2)$, and $(3,1)$, respectively. We will discuss other types of boundary conditions later.

\begin{table*}[]

\caption{Three types of open boundary conditions and the corresponding bulk profiles for the Hamiltonian $\hat H=\hat p^4$.}\label{table:three_OBC}
\begin{tabular}{|c|c|c|c|c|c|}
\hline
Index & $(n_l,n_r)$ & Left boundary conditions& Right boundary conditions& Bulk wave functions $(k\in\mathbb{R}_+)$ & Non-Hermitian skin effect\\ \hline
(i) &$(1,3)$     &        $\psi(0)=0$                  &         $\psi(L)=\psi^{\prime}(L)=\psi^{\prime\prime}(L)=0$                  &       $\psi(x)\sim e^{k (L-x)}(ie^{ik(L-x)}+e^{-ik(L-x)})$              &   $\usym{2714}$   \\ \hline
(ii) &$(2,2)$     &        $\psi(0)=\psi^{\prime}(0)=0$                  &       $\psi(L)=\psi^{\prime}(L)=0$                      &                     $\psi(x)\sim \sin(k x)-\cos(k x)$&    $\usym{2718}$  \\ \hline
(iii) & $(3,1)$     &              $\psi(0)=\psi^{\prime}(0)=\psi^{\prime\prime}(0)=0$               &                $\psi(L)=0$             &                     $\psi(x)\sim e^{k x}(ie^{ikx}+e^{-ikx})$&    $\usym{2714}$ \\ \hline
\end{tabular}
\end{table*}

\subsubsection{Case (i): $(n_l, n_r)=(1,3)$}\label{sec:homogeneous_p4_case1}

We first investigate the case (i), where the four independent boundary equations are given by $\psi(0)=0$ and $\psi(L)=\psi^{\prime}(L)=\psi^{\prime\prime}(L)=0$. We know that the Hamiltonian in this case is a non-Hermitian operator. Keeping this in mind, we solve the eigenstates for Eq. \eqref{eq:homogeneous_analytical_p4} under this kind of boundary conditions. 

When considering an arbitrary complex energy $E$, the characteristic equation $E=k^4$  has four complex roots $k_{1},\ k_2,\ k_3,\ k_4$. These complex wave vectors are related by $k_1=-ik_2=-k_3=ik_4=k_0=E^{1/4}$, where we define $k_0$ as a specific root. The four complex roots are related to $k_0$ by some phase factors. The general wavefunction is the superposition of these four modes:
\begin{equation}
	\psi(x)=\sum_{m=1}^4c_me^{ik_mx}.
	\label{eq:homogeneoous_p4_wave}
\end{equation}

Substituting this wavefunction into the four boundary equations, we get a set of constraints on $c_m$:
\begin{equation}
		\begin{cases}
			\sum_{m=1}^4c_m=0,\\
			\sum_{m=1}^4e^{ik_mL}c_m=0,\\
			\sum_{m=1}^4ik_me^{ik_mL}c_m=0,\\
			\sum_{m=1}^4(ik_m)^2e^{ik_mL}c_m=0.
		\end{cases}
\end{equation}
The existence of a nonzero solution requires that the determinant of the coefficient matrix of $c_m$ is zero. Namely, we have the following equation
\begin{equation}
	\left|\begin{matrix}
		1 & 1& 1 &1\\
		e^{ik_0L}&e^{-k_0L}&e^{-ik_0L}&e^{k_0L}\\
		ik_0e^{ik_0L}&-k_0e^{-k_0L}&-ik_0e^{-ik_0L}&k_0e^{k_0L}\\
		(ik_0)^2e^{ik_0L}&(-k_0)^2e^{-k_0L}&(-ik_0)^2e^{-ik_0L}&(k_0)^2e^{k_0L}
	\end{matrix}\right|=0.
\end{equation}
This leads to a constraint on $k_0$:
\begin{equation}
	\sin(k_0L)=\sinh(k_0L).
	\label{eq:homogeneous_p4_first_condition}
\end{equation}

It turns out that only the complex $k_0$ that satisfies this constraint can describe the eigenstates under the given open boundary conditions. The solution $k_0$ of this equation provides the OBC energy spectrum by the identity $E=k_0^4$.

It is easy to check that a nonzero real (imaginary) number $k_0$ does not solve Eq. \eqref{eq:homogeneous_p4_first_condition}. This is because the absolute value of one side of Eq. \eqref{eq:homogeneous_p4_first_condition} is always smaller than one, while that of the other side is exponentially large in the limit $L\to+\infty$. As a result, we have to solve this equation with a complex momentum $k_0=a+ib$ where  the real numbers $a,b\ne 0$. 

Eq. \eqref{eq:homogeneous_p4_first_condition} can be written as $e^{iaL-bL}-e^{-iaL+bL}=ie^{aL+ibL}-ie^{-aL-ibL}$. When $L\to+\infty$, we have to impose $a=\pm b$ to approximately fulfill this equation. Both cases give rise to $e^{aL}(e^{iaL}-ie^{-iaL})=e^{-aL}(e^{-iaL}-ie^{iaL})$. When $a>0$, $e^{2iaL}=i$ leads to $aL=(4\nu+1)\pi/4$; when $a<0$, $e^{-2iaL}=i$ leads to $aL=-(4\nu+1)\pi/4$. In both cases, $\nu=0,1,2,\cdots$. Together with the condition $a=\pm b$, we find $k_0=(\pm1\pm i)(4\nu+1)\pi/4L$ in the thermodynamic limit $L\to+\infty$. 

\begin{figure}[t]
	\centering
	\includegraphics[width=8.5cm]{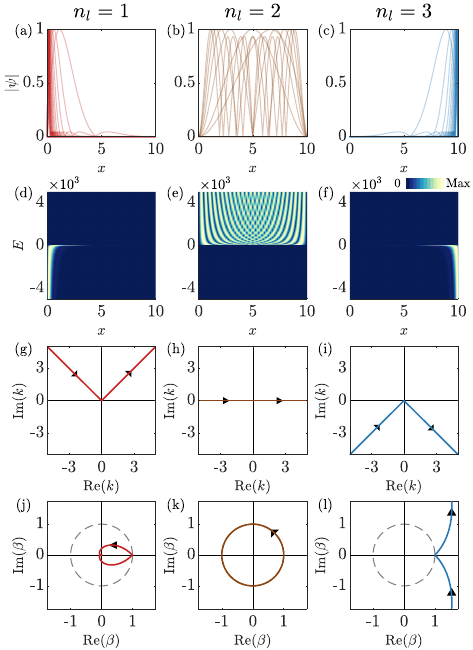}
	\caption{NHSE in the continuum model $\hat H=\hat p^4$ under different boundary conditions. Each column represents a case labeled by $n_l$, the number of boundary equations at the left edge. (a)$\sim$(c): the profiles of wavefunctions normalized by their maximum amplitudes. (d)$\sim$(f): the eigenstate profiles for different energies. (g)$\sim$(i): the generalized momentum space in the complex $k$ plane. (j)$\sim$(l): the generalized momentum space in the complex $\beta$-plane, which is obtained from (g)$\sim$(i) by the map $\beta=e^{ik}$. Dashed lines in (j)$\sim$(l) represent the unit circle.}
	\label{fig:analytical_p4}
\end{figure}

The above result immediately provides all four solutions of $E=k^4$, which are $k_1=-ik_2=-k_3=ik_4=k_0$. Correspondingly, the energy of the eigenstate is given by a real negative number $E_\nu=-(4\nu+1)^4\pi^4/64L^4$. When $L$ is sufficiently large, $k_0$ can effectively be viewed as a continuous variable $(\pm1\pm i)\tilde{k}$ with $\tilde{k}\in\mathbb{R}_+$. The energy is $E=-4\tilde{k}^4<0$ [Fig. \ref{fig:analytical_p4}(d)]. Notably, the unbounded negative spectrum reveals the non-Hermitian nature of our problem.

We note that,  when $L$ is finite, the quantitative results of the discrete energy spectrum require a small correction, as Eq. \eqref{eq:homogeneous_p4_first_condition} is a transcendental equation whose exact solutions are not accessible through the above analysis. Nevertheless, qualitative properties such as negative energies remain valid. 

With $k_0$ satisfying Eq. \eqref{eq:homogeneous_p4_first_condition}, we are ready to express the corresponding wavefunction as
\begin{equation}
	\psi(x)\sim e^{ik_0(x-L)}+ie^{-k_0(x-L)}-e^{-ik_0(x-L)}-ie^{k_0(x-L)}.
\end{equation}
In the large-$L$ limit where $k_0=(1+i)\tilde k$ with $\tilde{k}\in\mathbb{R}_+$, the norm of the wavefunction in the bulk region $(0\ll x\ll L)$ is dominated by $|\psi(x)|\sim e^{\tilde k(L-x)}$. This result reveals the existence of skin modes localized to the left boundary $x=0$ [Fig. \ref{fig:analytical_p4}(a)].

\subsubsection{Case (ii): $(n_l, n_r)=(2,2)$}

Now we turn to case (ii) whose boundary conditions are $\psi(0)=\psi^{\prime}(0)=0$ and $\psi(L)=\psi^{\prime}(L)=0$. It is easy to see that this is a Hermitian case.

Using the same wavefunction ansatz as in Eq. \eqref{eq:homogeneoous_p4_wave} and following the same procedure, the four boundary constraints, in this case, lead to
\begin{equation}
	\cosh(k_0L)\cos(k_0L)=1,
	\label{eq:homogeneous_p4_second_condition}
\end{equation}

We still assume that $k_0=a+ib$ with $a,b\in\mathbb{R}$. If $a\ne0,b\ne0$, we will have $\cosh((a+ib)L)\cosh((b-ia)L)=1$, which is impossible since the left-hand side is divergent with $L\to+\infty$. If $b=0$, $\cos(aL)\cosh(aL)=1$ leads to $\cos(aL)\approx0$ when $L\to+\infty$. Therefore, in the large-$L$ limit, we can find that $aL=(2\nu+1)\pi/2$ with $\nu=0,1,2,\cdots$. A similar result can be obtained for the assumption $a=0$. To sum up, we find $k_0=(2\nu+1)\pi/2L$, and the four wave vectors satisfy $k_1=-ik_2=-k_3=ik_4=k_0$. Consequently, the discrete spectrum under this kind of boundary condition is given by a set of positive numbers $E_\nu=(2\nu+1)^4\pi^4/16L^4$. In the large-$L$ limit, the energy spectrum effectively becomes $E(k_0)=k_0^4$ with $k_0\in\mathbb{R}_+$ [Fig. \ref{fig:analytical_p4}(e)].

When $k_0$ satisfies Eq. \eqref{eq:homogeneous_p4_second_condition}, the wavefunctions can be written as
	\begin{equation}
		\begin{split}
			\psi(x)\sim&\left(1 - 2e^{-k_0L} \cos(k_0 L)+e^{-2k_0L}\right) \sin(k_0 x)\\
			& - \left(1 - 2e^{-k_0L}\sin(k_0 L)-e^{-2k_0L}\right)\cos(k_0 x) \\
			& +\left(\cos(k_0L)-\sin(k_0L)-e^{-k_0L}\right)e^{-k_0(L-x)}\\
			&+\left(1-e^{-k_0L}\cos(k_0L)-e^{-k_0L}\sin(k_0L)\right)e^{-k_0x}.
		\end{split}\label{eq:homogeneous_p4_second_wave}
	\end{equation}
It is easy to check that this wavefunction indeed satisfies the boundary conditions $\psi(0)=\psi(L)=\psi^\prime(0)=0$ and $\psi^\prime(L)\sim4k_0e^{-k_0L}\left(\cos(k_0L)\cosh(k_0L)-1\right)=0$, as long as $k_0$ follows Eq. \eqref{eq:homogeneous_p4_second_condition}. In the large-$L$ limit with $k_0\in\mathbb{R}_+$, the dominant term in the bulk region $(0\ll x\ll L)$ is given by $\psi(x)\sim \sin(k_0x)-\cos(k_0x)$, which is the superposition of extended plane waves. As a result, the wavefunctions do not exhibit exponential decay or growth behavior, indicating the nonexistence of NHSE in the Hermitian system [Fig. \ref{fig:analytical_p4}(b)].

\subsubsection{Case (iii): $(n_l, n_r)=(3,1)$}

Finally, we investigate case (iii) under the boundary conditions $\psi(0)=\psi^{\prime}(0)=\psi^{\prime\prime}(0)=0$ and $\psi(L)=0$. This case can be viewed as the spatial inversion of case (i) by replacing $x$ with $L-x$. Therefore this is also a non-Hermitian case.

Following the same procedure in Sec. \ref{sec:homogeneous_p4_case1}, with $k_1=-ik_2=-k_3=ik_4=k_0=E^{1/4}$, the four boundary constraints lead to the equation:
\begin{equation}
	\sin(k_0L)=\sinh(k_0L).	\label{eq:homogeneous_p4_third_condition}
\end{equation}

We find that Eq. \eqref{eq:homogeneous_p4_third_condition} coincides with Eq. \eqref{eq:homogeneous_p4_first_condition}. This is because this case is the spatial inversion of case (i). As a result, we get the energy spectrum $E=-4\tilde{k}^4$ with $\tilde{k}\in\mathbb{R}_+$ [Fig. \ref{fig:analytical_p4}(f)], which is the same as in case (i). The unbounded negative spectrum reveals its non-Hermitian nature. However, the inversion of boundary conditions modifies the form of wavefunctions, which are given by
\begin{equation}
		\psi(x)\sim -e^{ik_0x}-ie^{-k_0x}+e^{-ik_0x}+ie^{k_0x},
\end{equation}
with $k_0$ satisfying Eq. \eqref{eq:homogeneous_p4_third_condition}. In the large-$L$ limit where $k_0=(1+i)\tilde k$ with $\tilde{k}\in\mathbb{R}_+$, the wavefunction in the bulk region $(0\ll x\ll L)$ is dominated by $|\psi(x)|\sim e^{\tilde kx}$. Therefore these results reveal the existence of NHSE localized to the right boundary $x=L$ [Fig. \ref{fig:analytical_p4}(c)].

In conclusion, we have shown that boundary conditions dramatically shape the energy spectrum and wavefunction profiles of a non-Hermitian continuum system, even though we focus on the same bulk Hamiltonian. The results are summarized in Table \ref{table:three_OBC}. Compared to the results in Sec. \ref{sec:homogenerous_hatano}, we find that the number of boundary conditions, rather than their specific details, significantly constrains the form of bulk eigenstates in non-Hermitian continuum systems. This simple model provides the essential intuition for the general theory of NHSE in the continuum, which is developed in the following section. 

\subsection{Genernalized momentum space}\label{sec:homogeneous_GBZ}

Building upon the understanding that boundary conditions play an important role in characterizing the eigenstates of a continuum system, we can now develop the non-Bloch band theory in general non-Hermitian continuum systems.  Analogous to lattice models, the non-Bloch theory aims to describe the NHSE in continuum systems when subjected to various boundary conditions.

We start from a one-dimensional stationary Schrodinger's equation $\hat H\psi(x)=E\psi(x)$ on an open chain. A generic Hamiltonian $\hat H$ in homogeneous non-Hermitian continuum systems is
\begin{equation}
	\hat H=\sum_{m=0}^{n}a_m\hat p^m,\quad x\in[0,L],
	\label{eq:homogeneous_general_continuum_H}
\end{equation}
where $a_m$ are spatially independent complex constants. If we first ignore the boundary conditions, with an arbitrary complex energy $E$, there are $n$ independent wavefunctions $\phi_m(x)=e^{ik_mx}$ with $m=1,\cdots,n$. The complex wave vectors $k_m$ of these modes correspond to the roots of the characteristic equation
\begin{eqnarray}\label{eq:homogeneous_general_continuum_charact}
    E=h(k)=\sum_{m=0}^{n}a_m k^m.
\end{eqnarray}
Given a specific complex energy $E$, these $k_m$ roots are ordered by their imaginary parts $\text{Im}[k_1(E)]\ge\text{Im}[k_2(E)]\ge\cdots\ge\text{Im}[k_n(E)]$.  Here, we use an order-$n$ polynomial $h(k)=\sum_{m=0}^{n}a_m k^m$ to represent the Hamiltonian in the complex momentum space.

A generic wavefunction with energy $E$ is the superposition of these $n$ modes: 
\begin{equation}
	\psi(x)=\sum_{m=1}^{n}c_m\phi_m(x).
	\label{eq:homogeneous_GBZ_wave}
\end{equation}
The coefficients $c_m$ should be uniquely determined by $n$ independent homogeneous boundary equations at $x=0,L$. For simplicity, we assume that there are $n_l$ ($0<n_l<n$) homogeneous boundary equations $\psi(0)=\psi'(0)=\cdots=\psi^{(n_l-1)}(0)=0$ at $x=0$, and correspondingly, $n_r=n-n_l$ equations $\psi(L)=\psi'(L)=\cdots=\psi^{(n_r-1)}(L)=0$ at $x=L$. Then we get $n$ linear independent equations for the coefficients $c_m$:
\begin{equation}
	\begin{cases}
		\sum_{m=1}^n(ik_m)^{p}c_m=0,\quad &p=0,1,\cdots,n_l-1;\\		\sum_{m=1}^n(ik_m)^{p}e^{ik_mL}c_m=0,\quad& p=0,1,\cdots,n_r-1.
	\end{cases}
\label{eq:homogeneous_OBC}
\end{equation}

The condition for the existence of nonzero solutions is given by the zero determinant of the coefficient matrix:
 \begin{equation}
 	\left|\begin{matrix}
 		1 & 1& \cdots &1\\
 		ik_1&ik_2&\cdots&ik_n\\
 		\vdots&\vdots&\ddots&\vdots\\
 		(ik_1)^{n_l-1}&(ik_2)^{n_l-1}&\cdots&(ik_n)^{n_l-1}\\
 		e^{ik_1L}&e^{ik_2L}&\cdots&e^{ik_nL}\\
 		ik_1e^{ik_1L}&ik_2e^{ik_2L}&\cdots&ik_ne^{ik_nL}\\
 		\vdots&\vdots&\ddots&\vdots\\
 		(ik_1)^{n_r-1} e^{ik_1L}&(ik_2)^{n_r-1} e^{ik_2L}&\cdots&(ik_n)^{n_r-1} e^{ik_nL}		
 	\end{matrix}\right|=0.
 \label{eq:homogeneous_bigdet}
 \end{equation}

The order $\text{Im}{(k_1)}\ge\text{Im}(k_2)\ge\cdots\ge\text{Im} (k_n)$ leads to $\left|e^{ik_1L}\right|\le\left|e^{ik_2L}\right|\le\cdots\le\left|e^{ik_nL}\right|$. Therefore the two dominant terms on the left-hand side (LHS) of Eq. \eqref{eq:homogeneous_bigdet} are
\begin{equation}
	\begin{split}
 		\text{LHS}=&g_1e^{i(k_n+k_{n-1}+\cdots+k_{n-n_r+2})L}e^{ik_{n-n_r+1}L}+\\
 		&g_2e^{i(k_n+k_{n-1}+\cdots+k_{n-n_r+2})L}e^{ik_{n-n_r}L}+\cdots,
 	\end{split}\label{eq:homogeneous_LHS}
\end{equation}
where $g_{1,2}$ are irrelevant constants. When $L\to+\infty$, to satisfy Eq. \eqref{eq:homogeneous_bigdet}, we have to impose the condition $\text{Im}(k_{n-n_r})=\text{Im}(k_{n-n_r+1})$, namely $\text{Im}(k_{n_l})=\text{Im}(k_{n_l+1})$. Noting that this condition depends on $E$ by the characteristic equation $E=h(k)$, we can introduce the following equation:
 \begin{equation}
 	\text{Im}[k_{n_l}(E)]=\text{Im}[k_{n_l+1}(E)].
 	\label{eq:homogeneous_GBZ}
 \end{equation}
Since the complex vectors $k_m$ are functions of a complex energy $E$, encoded in the characteristic equation Eq. \
 \eqref{eq:homogeneous_general_continuum_charact}, the above equation becomes an equation of $E$. Therefore the collection of energies satisfying Eq. \eqref{eq:homogeneous_GBZ} offers the OBC spectrum $\{E_{\text{OBC}}\}$ in the thermodynamic limit under the given boundary conditions. In addition, the collection of $\{k_{n_l}(E_{\text{OBC}}),\, k_{n_l+1}(E_{\text{OBC}})\}$ forms a one-dimensional curve in the complex $k$ plane, dubbed \emph{generalized momentum space}. 

The OBC wavefunction with energy $E\in\{E_{\text{OBC}}\}$ exhibits the bulk profile $|\psi_E(x)|\sim\exp(-\text{Im}[k_{n_l}(E)]x)$. Compared to $c_{n_l}$ and $c_{n_l+1}$, the contribution of other $c_{m}e^{ik_mx}$ terms with $m\ne n_l,n_l+1$ is exponentially small in the bulk, which is already seen in the discussion of the solvable model in Sec. \ref{sec:homogeneous_p4} . Therefore the nonzero imaginary parts of the generalized momentum space indicate the existence of NHSE on an open chain. With the wavefunction profile being $|\psi_E(x)|\sim\exp(-\text{Im}[k_{n_l}(E)]x)$,  $\text{Im}[k_{n_l}(E)]=\text{Im}[k_{n_l+1}(E)]>0$ implies that the skin modes are localized to the left boundary, while $\text{Im}[k_{n_l}(E)]=\text{Im}[k_{n_l+1}(E)]<0$ indicates the wavefunctions are squeezed to the right boundary.

The equation Eq. \eqref{eq:homogeneous_GBZ} is a central result of this section. We emphasize that $n_l$ and $n_r$, the number of boundary conditions at two edges, determine the form of the equation for the generalized momentum space. As a result, we establish the non-Bloch theory for homogeneous non-Hermitian continuum systems. This theory is based on the introduction of the generalized momentum space, which plays the same role as the GBZ in non-Hermitian lattice models \cite{yao2018edge,yokomizo2019nonbloch}.

\subsection{Role of boundary conditions}\label{sec:number_of_boundary_condition}

After developing the theory of the generalized momentum space, in this part, we elucidate the important role of the boundary conditions. It is the number of boundary conditions at two edges, not the details of them, that determines the form of the equation Eq. \eqref{eq:homogeneous_GBZ}. We already encounter examples of this conclusion in Sec. \ref{sec:homogenerous_hatano} and Sec. \ref{sec:homogeneous_p4}. Here, we make a general argument about the role of boundary conditions.

In the derivations in Sec. \ref{sec:homogeneous_GBZ}, we impose the boundary conditions of an open chain as $\psi(0)=\psi'(0)=\cdots=\psi^{(n_l-1)}(0)=0$ at $x=0$ and $\psi(L)=\psi'(L)=\cdots=\psi^{(n_r-1)}(L)=0$ at $x=L$. These boundary constraints can be extended to more general cases. For example, we can impose $n_l$ boundary conditions at $x=0$ as $\psi^{(n)}(0)=\psi^{(n-1)}(0)=\cdots=\psi^{(n-n_l+1)}(0)=0$ and $n_r$ boundary conditions at $x=L$ as $\psi^{(n)}(L)=\psi^{(n-1)}(L)=\cdots=\psi^{(n-n_r+1)}(L)=0$, with the condition $n=n_l+n_r$. Following the same procedure as in Eq. \eqref{eq:homogeneous_GBZ_wave}$\sim$\eqref{eq:homogeneous_LHS}, we get the same equation as Eq. \eqref{eq:homogeneous_GBZ}.

Moreover, we can even set the homogeneous boundary conditions as a linear combination of derivatives at each edge. To be specific, we set $n_l$ independent boundary constraints at $x=0$ as $\sum_{j=0}^{n}b_{p,j}\psi^{(j)}(0)=0$ with $p=1,2,\cdots,n_l$ and $n_r=n-n_l$ independent boundary constraints at $x=L$ as $\sum_{j=0}^{n}b_{p,j}\psi^{(j)}(L)=0$ with $p=n_l+1,n_l+2,\cdots,n$. If the coefficients $b_{p,j}$ in these equations do not depend on $L$, the above procedure in Eq. \eqref{eq:homogeneous_GBZ_wave}$\sim$\eqref{eq:homogeneous_LHS} will provide the same equation of the generalized momentum space.

The different choices of boundary equations with the same $n_l$ and $n_r$ only change the values of the coefficients $g_1,\, g_2,\,\cdots$ in Eq. \eqref{eq:homogeneous_LHS}. However, the details of the boundary conditions do not change the scaling behavior of the dominant terms in Eq. \eqref{eq:homogeneous_LHS}. As a result, the equation to determine the generalized momentum space survives for different choices of $b_{p,j}$'s, leading to the same wavefunction profiles in the bulk region. Although the wavefunctions near two boundaries do rely on the explicit form of boundary conditions, this dependence becomes exponentially small in the bulk.

We provide a detailed example illustrating the impact of boundary details on the eigenstates of the model $\hat{H} = \hat{p}^4$. By setting the number of boundary conditions to $n_l = n_r = 2$, we will show that the details of the boundary equations primarily affect the wavefunction amplitudes near the boundaries. However, the bulk profiles of the eigenstates remain unchanged, irrelevant to the  boundary details at two edges.

We rewrite the stationary differential equation Eq. \eqref{eq:homogeneous_analytical_p4} as
\begin{equation}
	\partial_x^4u(x)=\omega^2u(x),\quad x\in[0,L].
\end{equation}

By interpreting $u(x)$ as the deflection of a one-dimensional length-$L$ elastic beam and $\omega$ as the natural frequency of its free vibration mode, this differential equation describes the eigenmodes of a homogeneous dynamic Euler-Bernoulli beam \cite{Euler-Bernoulli}. For simplicity, we set the elastic modulus, mass density, and the second moment of area of the cross-section to be constants. Therefore we can omit them by rescaling the coordinates.

To find the eigenmodes of the Euler-Bernoulli beam, we need to impose suitable boundary conditions at two ends. In case (ii) of Sec. \ref{sec:homogeneous_p4}, we set the boundary conditions as $u(0)=u'(0)=0$ and $u(L)=u'(L)=0$.  In the language of elastic beam theory, they are fixed boundary conditions at two edges. Furthermore, for a cantilevered beam that is fixed at $x=0$ and free at $x=L$, the boundary conditions are given by $u(0)=u'(0)=0$ and $u''(L)=u'''(L)=0$. Following the same procedure as case (ii) of Sec. \ref{sec:homogeneous_p4}, the eigenmodes with frequency $\omega$ are given by
\begin{eqnarray}
\begin{split}
	u(x)\sim&\left(1 + 2e^{-k_0L} \cos(k_0 L)+e^{-2k_0L}\right) \sin(k_0 x)\\
	& - \left(1 + 2e^{-k_0L}\sin(k_0 L)-e^{-2k_0L}\right)\cos(k_0 x) \\
	& +\left(\sin(k_0L)-\cos(k_0L)-e^{-k_0L}\right)e^{-k_0(L-x)}\\
	&+\left(1+e^{-k_0L}\cos(k_0L)+e^{-k_0L}\sin(k_0L)\right)e^{-k_0x}.
\end{split}\label{eq:homogeneous_beam_1}
\end{eqnarray}
In the above, $k_0^4=\omega^2$ satisfies the constraint $\cosh(k_0L)\cos(k_0L)+1=0$, which is similar to Eq. \eqref{eq:homogeneous_p4_second_condition}. This constraint comes from the requirement for the existence of nonzero solutions under the boundary conditions $u(0)=u^{\prime}(0)=0$ and $u^{\prime\prime}(L)=u^{\prime\prime\prime}(L)=0$. The resulting $k_0$ is real-valued in the thermodynamic limit.

Another example of $n_l=n_r=2$ is given by the boundary conditions $u^{\prime\prime}(0)=u^{\prime\prime\prime}(0)=0$ and $u^{\prime\prime}(L)=u^{\prime\prime\prime}(L)=0$, which describe a free-free beam without supports at two ends. Together with $k_0^4=\omega^2$, the requirement for nonzero solutions leads to $\cosh(k_0L)\cos(k_0L)-1=0$, the same as in Eq. \eqref{eq:homogeneous_p4_second_condition}. Correspondingly, the eigenmodes are given by
\begin{equation}
	\begin{split}	
		u(x)\sim&\left(1 - 2e^{-k_0L} \cos(k_0 L)+e^{-2k_0L}\right) \sin(k_0 x)\\
		& - \left(1 - 2e^{-k_0L}\sin(k_0 L)-e^{-2k_0L}\right)\cos(k_0 x) \\
		& -\left(\cos(k_0L)-\sin(k_0L)-e^{-k_0L}\right)e^{-k_0(L-x)}\\
		&-\left(1-e^{-k_0L}\cos(k_0L)-e^{-k_0L}\sin(k_0L)\right)e^{-k_0x}.
	\end{split}\label{eq:homogeneous_beam_2}
\end{equation}

In summary, we have studied three types of boundary conditions with the same number of boundary equations $n_l=n_r=2$: fixed-fixed boundary conditions $u(0)=u^{\prime}(0)=0$ and $u(L)=u^{\prime}(L)=0$; fixed-free boundary conditions $u(0)=u^{\prime}(0)=0$ and $u^{\prime\prime}(L)=u^{\prime\prime\prime}(L)=0$; free-free boundary conditions $u^{\prime\prime}(0)=u^{\prime\prime\prime}(0)=0$ and $u^{\prime\prime}(L)=u^{\prime\prime\prime}(L)=0$. In particular, in the thermodynamic limit with $k_0\in\mathbb{R}$, the profiles of the corresponding eigenmodes in Eq. \eqref{eq:homogeneous_p4_second_wave}, Eq. \eqref{eq:homogeneous_beam_1} and Eq. \eqref{eq:homogeneous_beam_2} exhibit the same bulk behavior [i.e., $u(x)\sim \sin(k_0x)-\cos(k_0x)$], although their explicit coefficients differ from each other. These results show the nonexistence of NHSE in all three cases. We conclude that the details of boundary conditions are irrelevant to the bulk profiles of the eigenstates.

Putting the results in this section and Sec. \ref{sec:homogeneous_p4}  together, our findings of this analytical example explicitly show that it is the number of homogeneous boundary conditions, rather than their detailed forms, that determines the properties of generalized momentum space and the bulk profiles of the eigenstates. While the given example is a Hermitian system without NHSE, this conclusion is also applicable to general non-Hermitian continuum systems with NHSE.

In conclusion, it is the number of independent homogeneous boundary conditions at two ends of an open chain that determines the eigenstate profiles and OBC spectrum of non-Hermitian continuum systems. Namely, the values of $n_l$ and $n_r$ determine the equation Eq. \eqref{eq:homogeneous_GBZ}.

\subsection{Applications of the theory of generalized momentum space}\label{sec:applications_of_GBZ}

After establishing the theory of generalized momentum space for homogeneous non-Hermitian continuum systems, we discuss its applications to previous analytically solvable models.

To facilitate applications, we follow the procedure of the auxiliary GBZ method developed in lattice models \cite{yang2020non}. We extend this method to non-Hermitian continuum systems. Because the points on the generalized momentum space form pairs with the same imaginary parts, these pairs differ from each other by a real number. Specifically, we can solve an auxiliary equation $h(k)=h(k+k_a)$ where $k_a\in\mathbb{R}$ is an auxiliary real number. The complex roots $k(k_a)$ of the auxiliary equation are located on the curve called auxiliary generalized momentum space, which consists of the solutions of $ 	\text{Im}[k_{i}(E)]=\text{Im}[k_{i+1}(E)]$ with $i=1,2,\cdots,n-1$. $k_i(E)$ are the $i$-th roots of $E=h(k)$. From these solutions, we can select the relevant $k$ points that satisfy Eq. \eqref{eq:homogeneous_GBZ}. Scanning $k_a$ over $\mathbb{R}$, we will get the full generalized momentum space and, consequently, the bulk OBC spectrum.

\begin{figure*}[t]
	\centering
	\includegraphics[width=\textwidth]{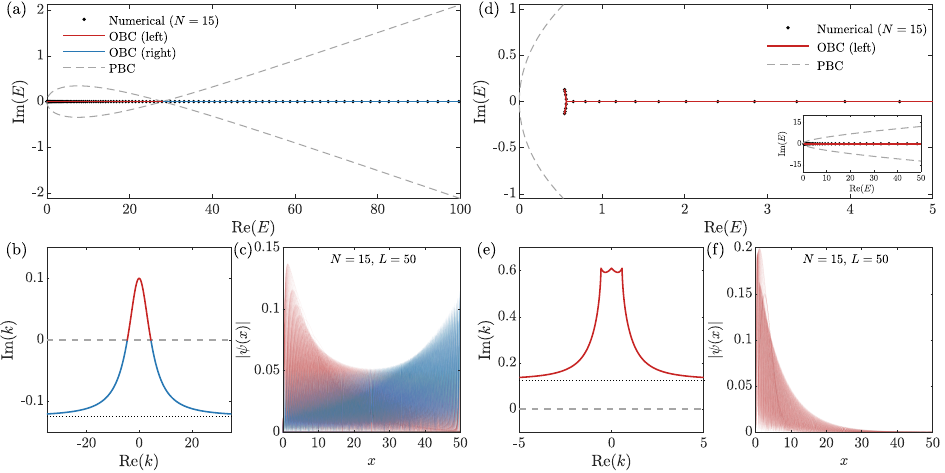}
	\caption{The energy spectrum (a), the generalized momentum space in the complex $k$ plane (b), and the low-energy eigenstates (c) for the model in Eq. \eqref{eq:homogeneous_general_p4}. We take the boundary conditions as $n_l=n_r=2$. The parameters are given by $(a_4,\, a_3,\, a_2,\, a_1,\, a_0)=(0.02,\, 0.01i,\, 1,\, -0.2i,\, 0)$. The solid red (blue) lines in (a,b) represent the eigenstates localized at the left (right) edge of the open chain. These results are obtained by the theory of generalized momentum space in continuum systems. The dashed gray lines demonstrate the Bloch states under PBC. The dotted line in (b) is the asymptotic limit of the generalized momentum space. The numerical OBC eigenstates (black points) in (a) and the typical OBC wavefunctions in (c) are obtained from the corresponding lattice model. The lattice model has $N=15$ sites per unit length, and the total length is given by $L=50$ ($x\in[0,L]$). That is, the total number of sites is $NL=750$. (d)$\sim$(f) have similar meanings to (a)$\sim$(c) except that $(a_4,\, a_3,\, a_2,\, a_1,\, a_0)=(1,\, -0.5i,\, 1,\, -1.5i,\, 0)$.}
	\label{fig:generalp4}
\end{figure*}

It is an easy task to check that the theory of generalized momentum space for homogeneous non-Hermitian continuum systems predicts the NHSE of the Hatano-Nelson model in Sec. \ref{sec:homogenerous_hatano}. With the characteristic equation $E=h(k)=(k+ig)^2/2m$, the auxiliary equation $h(k)=h(k+k_0)$ gives rise to $k=-k_0/2-ig$ as the points on the auxiliary generalized momentum space, where $k_0\in\mathbb{R}$ is a real number. Then the condition $\text{Im}[k_{1}(E)]=\text{Im}[k_{2}(E)]$ leads to $k_1=-k_0/2-ig$ and $k_2=k_0/2-ig$ as paired points on the generalized momentum space. This result reveals that the bulk profile of the wavefunctions is proportional to $e^{gx}$, which is the same as the exact solution in Sec. \ref{sec:homogenerous_hatano}. Meanwhile, $E=k_0^2/2m$ is the corresponding real-valued OBC energy. 

Furthermore, we can also apply the theory of generalized momentum space to the Hamiltonian $\hat H=\hat{p}^4$ discussed in Sec. \ref{sec:homogeneous_p4}. The different boundary conditions are labeled as $(n_l,n_r)$. These three cases share the same characteristic equation $E=h(k)=k^4$. Therefore the auxiliary equation $h(k)=h(k+k_a)$ with $k_a\in\mathbb{R}$ gives rise to $4(k/k_a)^3+6(k/k_a)^2+4k/k_a+1=0$, which provides the solution of the auxiliary generalized momentum space: $k/k_a\in\{-1/2,-(1+i)/2,-(1-i)/2\}$. Since $k_a\in\mathbb{R}$, the auxiliary generalized momentum space consists of the real axis and the diagonal lines of each quadrant of the complex $k$ plane. These three lines are the solutions of $\text{Im}[k_{i}(E)]=\text{Im}[k_{i+1}(E)]$ with $i=1,2,3$. Here, $k_{i}(E)$ are the roots of the characteristic equation $E=k^4$, ordered by their imaginary parts. The auxiliary generalized momentum space is coincident in all three cases since they share the same characteristic equation.

The difference in generalized momentum spaces comes from the number of boundary conditions at two ends. Firstly, in case (i) with $n_l=1$, the equation $\text{Im}[k_{1}(E)]=\text{Im}[k_{2}(E)]$ indicates that the generalized momentum space is the two diagonal lines of the upper-half complex $k$ plane [Fig. \ref{fig:analytical_p4}(g) and (j)]. In this sense, the wavefunctions are localized to the left edge, and the bulk spectrum does not have a lower bound [Fig. \ref{fig:analytical_p4}(a) and (d)]. Secondly, in case (ii) with $n_l=2$, the equation $\text{Im}[k_{2}(E)]=\text{Im}[k_{3}(E)]$ provides a generalized momentum space lying on the real axis of the complex $k$ plane [Fig. \ref{fig:analytical_p4}(h) and (k)], indicating the nonexistence of NHSE [Fig. \ref{fig:analytical_p4}(b) and (e)]. Lastly, in case (iii) with $n_l=3$, $\text{Im}[k_{3}(E)]=\text{Im}[k_{4}(E)]$ picks out the two diagonal lines of the lower half of the complex $k$ plane [Fig. \ref{fig:analytical_p4}(i) and (l)]. Consequently, the eigenstates are localized to the right end, and the spectrum is not lower-bounded [Fig. \ref{fig:analytical_p4}(c) and (f)].

In conclusion, the theory of generalized momentum space, which depends on the number of boundary conditions, successfully reproduces both the
bulk spectrum and the eigenstate profiles in Sec. \ref{sec:homogenerous_hatano} and Sec. \ref{sec:homogeneous_p4}. This method is very efficient since we obtain the spectrum information by simply solving a few algebraic equations.

\subsection{General $\hat p^4$ model}\label{sec:homogenerous_general_p4}

To demonstrate the power of the theory of generalized momentum space in homogeneous non-Hermitian continuum systems, we study a general non-Hermitian continuum model
\begin{equation}
	\hat H_4=a_4\hat p^4+a_3\hat p^3+a_2\hat p^2+a_1\hat p+a_0,\quad x\in[0,L].
	\label{eq:homogeneous_general_p4}
\end{equation}
This model can be viewed as a modified Hatano-Nelson model with higher orders of momentum operators. Although it is difficult to get an analytical expression of the OBC eigenstates, the OBC spectrum and the localization length of skin modes can be efficiently extracted from the generalized momentum space theory.

In this section, we set the boundary conditions as $n_l=n_r=2$. Namely, there are two boundary constraints at each edge. We have shown that the details of boundary conditions are not relevant because we focus on the bulk spectrum and NHSE. These properties are determined solely by the number of boundary conditions at two edges.

Under this type of boundary conditions, the non-Hermiticity of this model comes from complex parameters. Specifically, we take $a_3,\, a_1$ to be imaginary numbers and $a_4,\, a_2,\, a_0$ to be real numbers. Under this circumstance, the system has a (generalized) parity-time (PT) symmetry as $\hat H_4^*=\hat H_4$ since $\hat p^*=-\hat p$.  This symmetry forces the eigenvalues in the bulk spectrum either to be  real-valued or to form complex-conjugate pairs \cite{bender1998real,bender2002generalized,Mostafazadeh2002i,Mostafazadeh2002ii,bender2007making}.

In this model, the equation to determine generalized momentum space with $n_l=n_r=2$ is given by $\text{Im}[k_2(E)]=\text{Im}[k_3(E)]$. $k_i(E)$ are the four roots of the characteristic equation $E=h(k)=\sum_{m=0}^{4}a_mk^m$, ordered by $\text{Im}[k_1(E)]\ge\text{Im}[k_2(E)]\ge\text{Im}[k_3(E)]\ge\text{Im}[k_4(E)]$. Solving the equation  $\text{Im}[k_2(E)]=\text{Im}[k_3(E)]$ provides the generalized momentum space as well as the OBC spectrum of $\hat H_4$ under the given boundary conditions.

We present two sets of parameters in Fig. \ref{fig:generalp4}. For the parameters in Fig. \ref{fig:generalp4}(a)$\sim$(c), while the PBC spectrum together with the infinite point forms closed loops, the arc-like bulk spectrum lies inside the PBC spectrum. The mismatch between the PBC and OBC spectrums indicates the existence of NHSE on an open chain. This property is consistent with the lattice GBZ theory, where the nontrivial point-gap topology of the PBC spectrum implies the existence of NHSE in the OBC system \cite{zhang2020correspondence, okuma2020topological}. In addition, the generalized momentum space passes through the real axis of the complex $k$ plane, showing evidence of bipolar NHSE \cite{song2019real}. Namely, some eigenstates are localized at the left edge, while others are localized at the right edge. For the parameters in Fig. \ref{fig:generalp4}(d)$\sim$(f), the bulk spectrum on the open chain has complex-conjugate energy pairs. 

The PT symmetry of this model manifests itself in the symmetric distribution of the OBC spectrum with respect to the real axis. Interestingly, while the complex PBC spectrum always forms a loop, the OBC spectrum can be purely real-valued [Fig. \ref{fig:generalp4}(a)] or have complex-conjugate energy pairs [Fig. \ref{fig:generalp4}(d)]. This is called non-Bloch PT symmetry, which is caused by the interplay between PT symmetry and NHSE \cite{longhi2019probing,longhi2019nonbloch, xiao2021observation}. The non-Bloch PT symmetry plays an indispensable role in non-Hermitian dynamics, as it affects the long-time relaxation of non-Hermitian OBC systems. In non-Hermitian lattice models, the non-Bloch PT symmetry breaking has a geometric origin. Some cusps appear on the lattice GBZ after the PT symmetry transition \cite{hu2024geometric}. With the theory of generalized momentum space, we can directly extend the concept of geometric cusps to non-Hermitian continuum systems. As shown in Fig. \ref{fig:generalp4}, the real-valued OBC spectrum [Fig. \ref{fig:generalp4}(a)] is linked to the smooth generalized momentum space [Fig. \ref{fig:generalp4}(b)], while the complex-valued OBC spectrum [Fig. \ref{fig:generalp4}(d)] is linked to the generalized momentum space with cusps [Fig. \ref{fig:generalp4}(e)]. This feature indicates that the geometric perspective on the non-Bloch PT symmetry breaking remains valid in non-Hermitian continuum systems.

Moreover, we observe that the generalized momentum space of this concrete model asymptotically approaches the line $\operatorname{Im}(k)=-\text{Im}(a_3/(4a_4)$ of the complex plane  [the dotted lines in Figs. \ref{fig:generalp4}(b) and (e)]. The reason is that, in the high-energy region, the dominant term in the Hamiltonian becomes $\hat h_{\text{eff}}=a_4 (\hat p+a_3/(4a_4)^4$, giving rise to the generalized momentum space as $\text{Im}(k)=-\text{Im}(a_3/(4a_4)$ under the boundary conditions $n_r=n_l=2$.

The example shown in this section reveals the applicability of the theory of generalized momentum space in homogeneous non-Hermitian continuum systems. With the knowledge of the number of boundary conditions, we can easily determine the bulk OBC spectrum and the decay length of the skin modes. This is the central idea of generalized momentum space theory.

\section{Correspondence between continuum and lattice}\label{sec:correspondence}

\subsection{General theory}\label{sec:correspondence_discretization}

We have developed the theory of generalized momentum space for homogeneous non-Hermitian continuum systems. Naturally, the following question arises: What is the relation between the generalized momentum space in continuum systems and the lattice GBZ in lattice models? This section aims to address this question. We will eventually bridge the gap between non-Hermitian lattice models and continuum systems. Specifically, we find that the boundary conditions play a key role in discretizing the bulk Hamiltonian of a non-Hermitian continuum model.

To begin with, we consider a single-band lattice system with the smallest lattice distance $\delta=1/N$, where $N$ is the number of sites per unit length. We expect that in the limit $N\to+\infty$, the lattice model naturally returns to a continuum model. Inversely, by substituting differential operators with finite differences, a discrete lattice can be used to imitate the physics of a continuum Hamiltonian.

The conventional method to derive a lattice model is to replace the momentum operator $\hat p=-i\partial_x$ with a symmetric form of the finite difference:
\begin{equation}
	\hat p\psi(x)\to\frac{\psi(x+\delta)-\psi(x-\delta)}{2i\delta}.
	\label{eq:correspondence_conventional_discretization}
\end{equation}
In this sense, the higher powers of the momentum operator can be efficiently derived by the matrix product of this form. Combined with the coefficients in Eq. \eqref{eq:homogeneous_general_continuum_H}, this finite difference procedure provides a unique lattice model. Immediately, the non-Bloch band theory in the lattice gives rise to the lattice GBZ, which is the starting point to describe the OBC spectrum and wavefunctions. It should be noted that this procedure depends only on the continuum Hamiltonian in the bulk, regardless of the boundary conditions.

However, the theory of generalized momentum space in continuum systems demonstrates that a continuum Hamiltonian, subjected to different boundary conditions, can exhibit significantly distinct bulk properties. In particular, different choices of $n_l$ and $n_r$ at the two edges give rise to different generalized momentum spaces. It seems impossible for the unique lattice GBZ obtained by the procedure in Eq. \eqref{eq:correspondence_conventional_discretization} to generate many different generalized momentum spaces when taking the continuum limit. This contradiction indicates the breakdown of the conventional discretization protocol given by Eq. \eqref{eq:correspondence_conventional_discretization}. Therefore we shall develop a new self-consistent discretization procedure that relies on the number of boundary conditions at each edge.

To proceed, we first review the GBZ theory in non-Hermitian lattice models. Let us consider a generic single-band lattice model on an open chain whose tight-binding Hamiltonian is given by
\begin{equation}
	\hat H=\sum_i\sum_{m=-l}^{r}t_m\hat a^\dagger_{i}\hat a_{i+m}.
	\label{eq:correspondence_lattice_H}
\end{equation}
Here, $l$ ($r$) is the hopping range from the left (right) direction. $\hat a_i$ and $\hat a_i^\dagger$ are bosonic (or fermionic) annihilation and creation operators, respectively. With the definition of the non-Bloch Hamiltonian $h(\beta)=\sum_{m=-l}^{r}t_m\beta^m$, the lattice GBZ is given by 
\begin{equation}
 |\beta_{l}(E)|=|\beta_{l+1}(E)|,
\end{equation}
where $|\beta_1(E)|\le\cdots\le|\beta_{l+r}(E)|$ are $l+r$ complex roots of the characteristic equation $E=h(\beta)$ \cite{yao2018edge,yokomizo2019nonbloch, zhang2020correspondence}.

The stationary Schrodinger equation of the OBC lattice model implies that the number of boundary equations is $l$ ($r$) near the left (right) edge. These $l+r$ boundary equations are distinguished from the bulk equation, determining the GBZ equation $|\beta_{l}(E)|=|\beta_{l+1}(E)|$ in lattice systems. Notably, this property is the same as the continuum theory developed in Sec. \ref{sec:homogeneous_GBZ}. Therefore, to establish a correct correspondence between the lattice GBZ and the generalized momentum space, the number of boundary equations at the two edges of continuum systems and of lattice models should be identified, namely, $n_l=l$ and $n_r=r$. As a result, we can expect $\text{Im}[k_{n_l}(E)]=\text{Im}[k_{n_l+1}(E)]$ from $|\beta_{l}(E)|=|\beta_{l+1}(E)|$ by replacing $\beta$ with $e^{ik\delta}$ and taking the limit $\delta\to0$. In this sense, we need to expand $h(\beta)\equiv h(e^{ik\delta})$ to the order of $k^{l+r}$ to obtain a self-consistent continuum Hamiltonian $h(k)$.

Inversely, starting from a continuum model with $n_l$ ($n_r$) boundary equations at the left (right) end, we can get a lattice model with a proper approximation of differential operators. To satisfy the correspondence between the generalized momentum space and the lattice GBZ, we require that the number of boundary equations at two edges remains unchanged during the discretization procedure. Therefore we should identify the number of boundary conditions in continuum systems as the hopping ranges in lattice systems, which provides $l=n_l$ and $r=n_r$.

With this continuum-lattice correspondence, we can express the $p$-th order derivative $\psi^{(p)}(x)=\partial_x^p\psi(x)$ as the superposition of the wavefunctions $\psi(x+j\delta)$ on discrete lattice sites, where $j=-n_l,-n_l+1,\cdots,0,\cdots,n_r-1,n_r$. This leads to the identification $\psi^{(p)}(x)\to\sum_{q=0}^{n} D_{pq}\psi(x+(q-n_l)\delta)$ for $p=0,\cdots,n$ with $n=n_l+n_r$. In this sense, the $(n+1)$-by-$(n+1)$ matrix
$D_{pq}$ characterizes the coefficients of the lattice approximation for these differential operators. From the opposite perspective, we can expand $\psi(x+(q-n_l)\delta)$ with $q=0,\cdots,n$ as the Taylor series $\psi(x+(q-n_l)\delta)=\sum_{p=0}^{n}\frac{(q-n_l)^p\delta^p}{p!}\psi^{(p)}(x)+o(\delta^{n})$. To the order $o(\delta^{n})$, the inverse of the $D$ matrix is given by $(D^{-1})_{qp}={(q-n_l)^p\delta^p}/{p!}$ with $q,p=0,\cdots,n$.

Therefore the continuum Hamiltonian Eq. \eqref{eq:homogeneous_general_continuum_H} with $n_l$ ($n_r$) boundary conditions at the left (right) edge can be approximated as $\hat H\psi(x)=\sum_{p=0}^{n}a_p\left(-i{\partial_x}\right)^p\psi(x)\to\sum_{p=0}^{n}\sum_{q=0}^{n}a_p(-i)^pD_{pq}\psi(x+(q-n_l)\delta)$. This identification shows that the hopping elements of the lattice model are given by
\begin{equation}
	t_m=\sum_{p=0}^{n}a_p(-i)^pD_{p,(m+n_l)},
\end{equation}
where $m=-n_l,\cdots,0,\cdots,n_r$. This equation stands as a central result of this section, as it establishes a crucial connection between the coefficients of homogeneous continuum Hamiltonians and the hopping elements of tight-binding lattice models. It is the number of boundary conditions that determines this correspondence, which provides an appropriate way to get effective lattice models for non-Hermitian continuum systems.

To illustrate this continuum-lattice correspondence, we investigate the lattice versions of the homogeneous continuum models in Sec. \ref{sec:homogeneous} and compare the lattice GBZ with the generalized momentum space.

\subsection{Homogeneous Hatano-Nelson model}\label{sec:correspondence_Hatano}

We start with the continuum Hatano-Nelson model in Sec. \ref{sec:homogenerous_hatano}. With $n_l=n_r=1$, the decay length of skin modes is determined by the imaginary gauge field $g$. Here we elucidate the relationship between the generalized momentum space and the lattice GBZ for this simple model.

The procedure developed in Sec. \ref{sec:correspondence_discretization} leads to a $D$ matrix being
\begin{equation}
	D=\left(\begin{matrix}
	0&1&0\\
	-\frac{1}{2\delta}& 0&\frac{1}{2\delta}\\
	\frac{1}{\delta^2}&-\frac{2}{\delta^2}&\frac{1}{\delta^2}
	\end{matrix}\right).
\end{equation}
$\delta$ is the distance between the two nearest-neighbor sites. The column index of $D$ is
$\{-1+n_l,0+n_l,1+n_l\}$ with $n_l=n_r=1$, and the row
index is $\{0,1,2\}$ which comes from the highest order of the momentum operator $n=2$. This $D$ matrix is consistent with the traditional discretization procedure Eq. \eqref{eq:correspondence_conventional_discretization} for the operators $\hat p,\hat p^2$.  This discretization procedure provides a lattice model:
\begin{equation}
	\hat H=\frac{1}{2m\delta^2}\sum_i (-1-g\delta)\hat a^\dagger_{i+1}\hat a_{i}+(-1+g\delta)\hat a^\dagger_{i}\hat a_{i+1}+(2-g^2\delta^2)\hat a^\dagger_{i}\hat a_{i}.
\end{equation}

This lattice model can be solved analytically by similarity transformation under open boundary conditions. Following the procedure of the lattice GBZ theory, we find that the discrete Hatano-Nelson model has a circular GBZ on the complex $\beta$-plane. The radius of this circular lattice GBZ is
\begin{equation}
	|\beta|=\sqrt{\frac{1+g\delta}{1-g\delta}}.
\end{equation}

Noting that the lattice GBZ and the generalized momentum space are related to each other by $\beta=e^{ik\delta}$, we can reverse this relation to get $e^{ik}=\beta^{1/\delta}$. Then we obtain
\begin{equation}
	|e^{ik}|=\lim_{\delta\to0}|\beta|^{{1}/{\delta}}	=\lim_{\delta\to0}\left(\frac{1+g\delta}{1-g\delta}\right)^{\frac{1}{2\delta}}=e^g.
\end{equation}

Therefore we find $\text{Im}(k)=-g$ in the limit $\delta\to0$. The nonzero imaginary part of the wave vector $k$ indicates the exponential localization of the wavefunctions. This result is consistent with the previous results of generalized momentum space in the continuum Hatano-Nelson model, revealing the validity of the continuum-lattice correspondence.

\subsection{The analytically solvable model $\hat H=\hat p^4$}

\begin{figure*}[t]
	\centering
	\includegraphics[width=\textwidth]{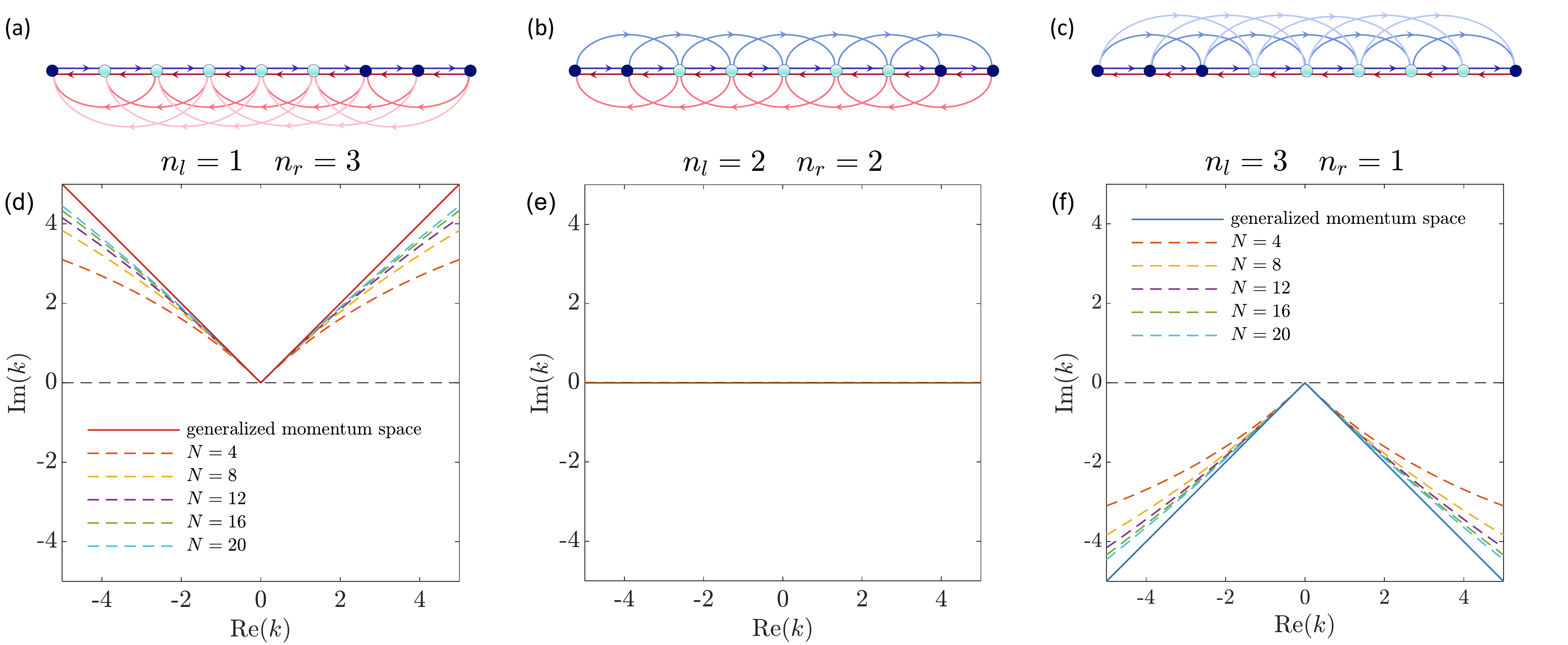}
	\caption{(a)$\sim$(c): the effective lattices for the  $\hat H=\hat p^4$ model. The hopping ranges of these lattices are determined by the number of boundary conditions in the continuum system. In these plots, the light blue sites are located in the bulk, while the dark blue sites stay at the boundaries. (d)$\sim$(f): the comparison between the generalized momentum space in Fig. \ref{fig:analytical_p4}(g)$\sim$(i) (solid lines in each plane) and the lattice GBZ in the corresponding lattice models. The discretization parameter $N$ is the number of sites per unit length.}
	\label{fig:hopping_generalp4}
\end{figure*}

To further show the correspondence between the hopping range in lattice models and the number of boundary conditions in continuum systems, we apply the procedure developed in Sec. \ref{sec:correspondence_discretization} to the analytically solvable model $\hat H=\hat p^4$ under three different kinds of boundary conditions. Consequently, we obtain three different lattice models with different hopping ranges [Fig. \ref{fig:hopping_generalp4}(a)$\sim$(c)]. For example, the effective lattice model with boundary conditions $n_l=1$ and $n_r=3$ has one hopping element from the left direction and three hopping elements from the right direction, as shown in Fig. \ref{fig:hopping_generalp4}(a).

Choosing the lattice constant as $\delta=1/N$ and calculating the $D$ matrix explicitly in each case, we get three non-Bloch Hamiltonians for different lattice models:
\begin{subequations}
	\begin{align}
	h_N^{1,3}(\beta)&=N^4(\beta-1)^4\beta^{-1};\\	
	h_N^{2,2}(\beta)&=N^4(\beta-1)^4\beta^{-2};\label{eq:correspondence_P4_discretization_second_case}\\
	h_N^{3,1}(\beta)&=N^4(\beta-1)^4\beta^{-3}.
	\end{align}
\end{subequations}
In the above, $h_N^{n_l,n_r}(\beta)$ is the non-Bloch Hamiltonian of the corresponding lattice model, where $N$ is the number of sites per unit length and $n_l$ ($n_r$) is the hopping range from the left (right) direction.

From the non-Bloch band theory in non-Hermitian lattice models, the lattice GBZ is given by $|\beta_{n_l}(E)|=|\beta_{n_l+1}(E)|$, where $|\beta_{1}(E)|\le|\beta_{2}(E)|\le|\beta_{3}(E)|\le|\beta_{4}(E)|$ are four roots of the lattice characteristic equation $E=h_N^{n_l,n_r}(\beta)$. For each specific $N$, the lattice GBZ equation gives rise to the GBZ set $\{\beta_N^{n_l,n_r}\}$. Since the lattice GBZ and the generalized momentum space are related by $\beta=e^{ik\delta}$ with $\delta=1/N$, we get $k_N^{n_l,n_r}=-iN\ln(\beta_N^{n_l,n_r})$. Therefore we can compare the results of $\{k_N^{n_l,n_r}\}$ with the generalized momentum space shown in Figs. \ref{fig:analytical_p4}(g)$\sim$(i).

Figs. \ref{fig:hopping_generalp4}(d) and (f) reveal that the lattice approximation is in good agreement with the exact generalized momentum space. These numerical results indicate that the number of boundary conditions is crucial for the correspondence between lattice models and continuum systems.

We do not show the numerical results for the Hermitian lattice model Eq. \eqref{eq:correspondence_P4_discretization_second_case} where $n_l=n_r=2$. Because this case does not have skin effect [Fig, \ref{fig:analytical_p4}(b)], the lattice GBZ is the conventional Brillouin zone $|\beta_N^{2,2}|=1$. Therefore the real-valued $k_N^{2,2}\in\{-N\pi,N\pi\}$ is coincident with the generalized momentum space [Fig. \ref{fig:hopping_generalp4}(e)].

\subsection{General $\hat p^4$ model}\label{sec:correspondence_general_p4}

To elucidate the continuum-lattice correspondence in general non-Hermitian systems, we study the generic continuum model in Eq. \eqref{eq:homogeneous_general_p4} under the boundary conditions $n_l=n_r=2$. Under this circumstance, the boundary-dependent discretization procedure leads to a $5\times 5$ matrix $D$, which is determined by the discretization parameter $\delta=1/N$. We then get a single-band lattice model with the next-nearest-neighbor hopping terms from both directions [Fig. \ref{fig:hopping_generalp4}(b)]. The effective non-Bloch Hamiltonian of this lattice model is
\begin{equation}
	h_4(\beta)=t_{-2}\beta^{-2}+t_{-1}\beta^{-1}+t_{0}+t_{1}\beta+t_{2}\beta^2,
\label{eq:correspondence_tight_binding_general_p4}
\end{equation}
where the hopping elements are given by
\begin{subequations}
	\begin{align}
		t_{-2}&=+{a_4}{N ^4}-\frac{i}{2} a_3 N ^3+\frac{1}{12}a_2N ^2-\frac{i}{12 } a_1 N,\\
		t_{-1}&=-{4 a_4}{N^4}+{i a_3}{{N} ^3}-\frac{4 }{3}a_2 N^2+\frac{2 i }{3}a_1 N,\\
		t_0&=+{6 a_4}{N^4}+\frac{5}{2 } a_2N ^2+a_0,\\
		t_1&=-{4 a_4}{N^4}-{i a_3}{{N} ^3}-\frac{4 }{3}a_2 N^2-\frac{2 i }{3}a_1 N,\\
		t_2&=+{a_4}{N ^4}+\frac{i}{2} a_3 N ^3+\frac{1}{12}a_2N ^2+\frac{i}{12 } a_1 N.
\end{align}\label{eq:correspondence_hopping_general_p4}
\end{subequations}

Specifically, we choose $N=15$ and $L=50$ such that there are $NL=750$ sites for the OBC lattice. Notably, the low-energy parts of the OBC spectrum from the lattice model [black diamond points in Fig. \ref{fig:generalp4}(a) and (d)] are coincident with the bulk energies obtained by the theory of generalized momentum space in continuum systems. Moreover, the localization of wavefunctions also reveals the existence of NHSE [Fig. \ref{fig:generalp4}(c) and (f)].

\begin{figure}[h]
	\centering
	\includegraphics[width=8.5cm]{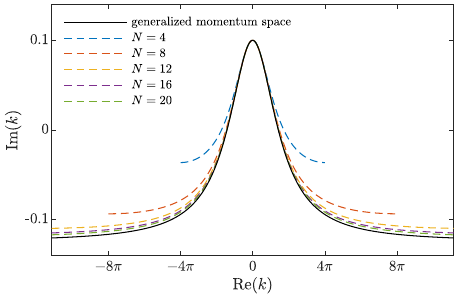}
	\caption{The comparison between the generalized momentum space (black solid line) and the lattice GBZ (colored dashed lines). This plot is obtained from the model in Eq. \eqref{eq:homogeneous_general_p4} under the boundary conditions $n_l=n_r=2$. The discretization parameter $N$ is the number of sites per unit length. Parameters: $(a_4,\, a_3,\, a_2,\, a_1,\, a_0)=(0.02,\, 0.01i,\, 1,\, -0.2i,\, 0)$.}
	\label{fig:discrete_general_p4}
\end{figure}

For each specific $N$, we employ the lattice GBZ equation $|\beta_2(E)|=|\beta_3(E)|$ to calculate the lattice GBZ $\{\beta_N\}$. Comparing the results $k_N=-iN\ln(\beta_N)$ with the generalized momentum space [Fig. \ref{fig:discrete_general_p4}(b)], we observe that the lattice results $k_N$ are approaching the exact generalized momentum space as $N$ increases. This result also shows the validity of the discretization procedure.

\section{Periodic continuum systems}\label{sec:periodic}

After a thorough exploration of homogeneous non-Hermitian continuum systems and their connections to lattice models, we move on to periodic systems in this section. The periodicity naturally appears in state-of-the-art experiments, such as photonic crystals and cold atom platforms. Therefore, to facilitate the study of the NHSE in periodic non-Hermitian continuum systems, it is highly desirable to develop a general non-Bloch band theory that takes periodicity into account.  

The periodic Hamiltonian in the bulk has a discrete translation symmetry $\hat H(x)=\hat H(x+a)$, where the lattice constant $a$ (i.e., the length of a unit cell) is much smaller than the system size $L$.  With the structure of periodically repeated unit cells, a periodic continuum system contains an infinite number of orbitals in each unit cell, and therefore, has infinite energy bands. Due to the discrete translation symmetry, we expect to obtain a compact generalized Brillouin zone in the complex plane. In this sense, the GBZ in periodic continuum systems should be similar to the GBZ in multiband lattice models where each unit cell contains more than one orbital, although the latter case usually has a finite number of energy bands. Then the non-Bloch band theory based on GBZ can be utilized to predict the bulk spectrum and the localization length of skin modes in periodic non-Hermitian continuum systems.

We first study the periodic Hatano-Nelson model, where both the imaginary gauge potential and the energy potential are periodic functions. Notably, while it is hard to obtain all the information about the OBC eigenstates in a periodic continuum system, we can explicitly find the decay length of OBC skin modes and exactly construct the GBZ of the periodic Hatano-Nelson model. 

Following the insights obtained from the periodic Hatano-Nelson model, we then properly define the GBZ in arbitrary non-Hermitian periodic continuum systems. Our definition highlights the significance of the transfer matrix in ordinary differential equations, and therefore, provides a general formulation to study NHSE in periodic continuum systems.

Finally, we apply our theory to a concrete example, where we add a periodic potential to the general model in Sec. \ref{sec:homogenerous_general_p4}. We also compare this transfer-matrix method to the discretization approach in Sec. \ref{sec:correspondence_discretization}. The numerical results reveal that both methods produce the same low-energy part of the OBC spectrum. With these two methods in hand, our non-Bloch band theory can be directly applied to arbitrary periodic non-Hermitian continuum systems.

\subsection{Periodic Hatano-Nelson model}\label{sec:periodic_hatano_nelson}
To begin with, we study the periodic Hatano-Nelson model on a length-$L$ chain, whose Hamiltonian is \cite{Yokomizo2022non-hermitian_wave,longhi2021non}
\begin{eqnarray}
	\hat H_{\text{HN}}=\frac{(\hat p+ig(x))^2}{
	2m}+V(x),\quad x\in[0,L].
\label{eq:periodic_Hatano_Nelson}
\end{eqnarray}
In the above, the gauge field $g(x)=g(x+a)$ and the potential $V(x)=V(x+a)$ are two complex-valued periodic functions. The parameter $a$ is the lattice constant, and we set the integer $L/a\gg1$. In other words, there are $L/a$ unit cells. We also take the open boundary conditions as $\psi(0)=\psi(L)=0$, i.e. $n_l=n_r=1$. These boundary conditions are the same as the homogeneous Hatano-Nelson model in Sec. \ref{sec:homogenerous_hatano}.

As shown below, the periodic Hatano-Nelson model has an interesting property: All OBC eigenstates of this continuum system share the same localization length determined by the complex gauge field $g(x)$. This is similar to the homogeneous case in Sec. \ref{sec:homogenerous_hatano}.  To show this property, we need to prove that the model in Eq. \eqref{eq:periodic_Hatano_Nelson}, under the OBC $n_l=n_r=1$, has a circular GBZ presented in Eq. \eqref{eq:periodic_Hatano_Nelson_GBZ}. This circular GBZ is a circle around the origin of the complex plane, and its radius provides the localization length of the eigenstates. 

We define $m=1/2$, $a_1(x)=2ig(x)$ and $a_0(x)=-g(x)^2+\partial_xg(x)+V(x)$ for simplicity. These periodic functions enable us to obtain the stationary Schrodinger equation for the eigenstate $\psi(x)$:
\begin{equation}
		\hat p^2\psi(x)+a_1(x)\hat p\psi(x)+(a_0(x)-E)\psi(x)=0,\ x\in[0,L].
		\label{eq:periodic_general_p2}
\end{equation}
where $E$ is the eigenvalue of $\psi(x)$. Ignoring the boundary conditions first, we take a general non-Bloch ansatz for the wavefunction in the bulk: 
\begin{equation}
 \psi_{\beta}(x)=\beta^{x/a}f_\beta(x)
 \label{eq:periodic_Hatano_Nelson_non_bloch_ansatz}
\end{equation}
where $\beta$ labels an independent eigenstate for the fixed energy $E$. The scaling factor $\beta$ provides the localization length of the wavefunction ansatz. Besides, $f_\beta(x)=f_\beta(x+a)$ is the periodic part of the wavefunction. In this sense, we observe that \begin{equation}
    \psi_\beta(x+a)=\beta\psi_\beta(x).\label{eq:periodic_Hatano_Nelson_non_bloch_ansatz_trans}
\end{equation} 
This non-Bloch ansatz is similar to the Bloch theorem in periodic Hermitian systems, where we take the Bloch ansatz as $e^{ikx}f_{k}(x)$ with a real quasimomentum $k$. Unlike the conventional Bloch ansatz, the complex $\beta$ in the non-Bloch ansatz is crucial in non-Hermitian systems because $|\beta|$ in Eq. \eqref{eq:periodic_Hatano_Nelson_non_bloch_ansatz} can describe localized skin modes. We anticipate that the non-Bloch band theory will generate a GBZ from which we can extract $\beta$ to characterize the OBC eigenstates.

It should be noted that the periodic Hatano-Nelson model in Eq. \eqref{eq:periodic_general_p2} is a second-order ordinary differential equation. Therefore, with a fixed energy $E$, there exist two independent solutions of non-Bloch factors: $\beta_1(E)$ and $\beta_2(E)$. To simplify the notation, we use $\psi_1(x)$ and $\psi_2(x)$,  rather than $\psi_{\beta_{1,2}}(x)$, to represent these two wavefunctions.

To proceed, we define a Wronskian determinant as
\begin{equation}
	W(x)=\left|\begin{matrix}
		\psi_1(x)&\psi_2(x)\\
		\hat p\psi_1(x)&\hat p\psi_2(x)
	\end{matrix}\right|.
\end{equation}
Then $\hat pW(x)$ can be expressed as
\begin{equation}
	\begin{split}
		\hat pW(x)&=\left|\begin{matrix}
		\hat p\psi_1(x)&\hat p\psi_2(x)\\
		\hat p\psi_1(x)&\hat p\psi_2(x)\end{matrix}\right|+\left|\begin{matrix}
		\psi_1(x)&\psi_2(x)\\
		\hat p^2\psi_1(x)&\hat p^2\psi_2(x)\end{matrix}\right|\\
		&=-a_1(x)\left|\begin{matrix}
		\psi_1(x)&\psi_2(x)\\
		\hat p\psi_1(x)&\hat p\psi_2(x)
		\end{matrix}\right|=-a_1(x)W(x).
		\end{split}\label{eq:periodic_Hatano_Nelson_Wronskian}
\end{equation}

In the above, we use the relation $\hat p^2\psi_{}(x)=-a_1(x)\hat p\psi(x)+(E-a_0(x))\psi(x)$ for two independent eigenstates $\psi_{1,2}(x)$. It is worth noting that the differential equation Eq. \eqref{eq:periodic_Hatano_Nelson_Wronskian}
describes how the Wronskian $W(x)$ is transported in real space. This property is crucial for determining the GBZ in the periodic Hatano-Nelson model. On the one hand, Eq. \eqref{eq:periodic_Hatano_Nelson_Wronskian} gives rise to $W(x+a)=\exp\left({-i\int_x^{x+a}a_1(x') dx'}\right)W(x)$. On the other hand, 
the definition of $W(x)$ leads to the relation $W(x+a)=\beta_1(E)\beta_2(E)W(x)$. Therefore the two solutions $\beta_1(E)$ and $\beta_2(E)$ in the order-2 periodic Hatano-Nelson model satisfy the following identity
\begin{equation}\label{eq:periodic_Hatano_Nelson_rootproduct}
	\beta_1(E)\beta_2(E)=\exp\left({-i\int_x^{x+a}a_1(x') dx'}\right)=\exp\left({-i\int_0^{a}a_1(x') dx'}\right),
\end{equation}
where we use the fact that $a_1(x+a)=a_1(x)$ is a periodic function. 

Now we take the boundary conditions into account. The general OBC wavefunction ansatz under the boundary conditions $n_r=n_l=1$ should be
\begin{equation}
 \psi(x)=c_1\psi_1(x)+c_2\psi_2(x),\quad x\in[0,L].
\end{equation}
The coefficients $c_1$ and $c_2$ are determined by the boundary conditions. For example, in the case of $\psi(0)=\psi(L)=0$, we have
\begin{equation}
 \left\{\begin{split}
 &c_1\psi_1(0)+c_2\psi_2(0)=0\\
 &c_1\beta_1(E)^{L/a}\psi_1(0)+c_2\beta_2(E)^{L/a}\psi_2(0)=0
 \end{split}\right.
\end{equation}
where we use the identity $\psi_{1,2}(L)=\beta_{1,2}(E)^{L/a}\psi_{1,2}(0)$ since $L/a$ is assumed to be an integer. The existence of nontrivial solutions for $c_{1,2}$ requires that
\begin{eqnarray}
 \left|\begin{matrix}
 \psi_1(0) & \psi_2(0)\\
 \beta_1(E)^{L/a}\psi_1(0)&\beta_2(E)^{L/a}\psi_2(0)
 \end{matrix}\right|=0
\end{eqnarray}
Because $\psi_{1,2}(0)$ are generally nonzero, in the thermodynamic limit $L\to\infty$, we obtain the GBZ equation for the periodic Hatano-Nelson model: \begin{equation}
 |\beta_1(E)|=|\beta_2(E)|.
\end{equation}
Finally, the compact GBZ should be a circle with the radius 
\begin{equation}\label{eq:periodic_Hatano_Nelson_GBZ}
	|\beta_1(E)|=|\beta_2(E)|=\exp\left({\frac{1}{2}\text{Im}\left(\int_0^{a}a_1(x^\prime) dx^\prime\right)}\right).
\end{equation}

Notably, the average value of $\operatorname{Im}[a_1(x)/2]$, which can be interpreted as the average strength of the imaginary gauge field, determines the GBZ in the periodic Hatano-Nelson model. This circular GBZ indicates that all OBC skin modes have the same localization length.

Following the idea in Sec. \ref{sec:homogenerous_hatano}, the periodic Hatano-Nelson model can be experimentally realized in a classical stochastic process, which is described by a periodic Fokker-Planck equation. In this case, the Fokker-Planck equation is generated by a periodic drift term $\mu(x)=\mu(x+a)$ and a positive periodic diffusion term $D(x)=D(x+a)$. Consequently, a periodic structure in Eq. \eqref{eq:general_stationary_fokker_planck} leads to a circular GBZ whose radius is given by
\begin{equation}
|\beta_1(E)|=|\beta_2(E)|=\exp\left({\frac{1}{2}\int_0^{a}D^{-1}(x^\prime)\mu(x^\prime) dx^\prime}\right). 
\end{equation}
If we interpret the Fokker-Planck equation as a Brownian particle moving in a periodic system, the sign of the local integral $\int_0^{a}D^{-1}(x^\prime)\mu(x^\prime) dx^\prime$ generates a preferable direction of the particle's motion.

Non-Hermitian photonic crystals offer another experimental platform for studying non-Bloch physics. With a multilayer structure of two alternately stacked dielectric media, non-Hermitian photonic crystals can manifest NHSE under open boundary conditions, which is effectively described by a periodic Hatano-Nelson model \cite{Yokomizo2022non-hermitian_wave}.  

\subsection{Generalized Brillouin zone in periodic systems }\label{sec:periodic_GBZ}

The picture behind Eq. \eqref{eq:periodic_Hatano_Nelson_Wronskian} suggests that the translation generated by $\hat p$ should play an important role in studying the GBZ in periodic continuum systems. To make this clear, we study a general periodic non-Hermitian continuum model:
\begin{equation}
	\hat H=\sum_{m=0}^{n}a_m(x)\hat p^m,\quad x\in[0,L].
\end{equation}
Here, $a_m(x)=a_m(x+a)$ are complex periodic functions with the lattice constant $a$. The number of unit cells is $L/a$. The stationary Schrodinger equation in real space is given by
\begin{equation}
	\sum_{m=0}^{n}a_m(x)\left(-i\partial_x\right)^m\psi(x)=E\psi(x),\quad x\in[0,L].\label{eq:periodic_general_eigen}
\end{equation}
$\psi(x)$ and $E$ are the eigenstate's wavefunction and energy, respectively.

Without loss of generality, we focus on the case $a_n(x)=1$. In general setup, as long as $a_n(x)$ is always nonzero for $x\in[0, L]$, this situation can be transformed into the case $a_n(x)=1$ by rescaling the coefficients $a_m(x)\to {a_m(x)}/{a_n(x)}$ for $m\ne0$ and $a_0(x)-E\to {(a_0(x)-E)}/{a_n(x)}$. 

Following the idea of the transfer matrix in non-Hermitian lattice models \cite{kunst2019non}, we introduce an $n$-component vector $\ket{\phi(x)}=(\psi(x),\hat p\psi(x),\cdots,\hat p^{n-1}\psi(x))^T$. The Eq. \eqref{eq:periodic_general_eigen} becomes
\begin{equation}
	\hat p^n\psi(x)=E\psi(x)-\sum_{m=0}^{n-1}a_m(x)\hat p^m\psi(x).
	\label{eq:perodic_general_pneigen}
\end{equation}
Then we get a relation between two vectors $\ket{\phi(x)}$ and $\hat p\ket{\phi(x)}=-i\frac{\partial}{\partial x}\ket{\phi(x)}$:
\begin{equation}
	-i\frac{\partial}{\partial x}\ket{\phi(x)}=A(x)\ket{\phi(x)},\label{eq:periodic_general_floquet}
\end{equation}
where $A(x)$ is an $n\times n$ matrix
\begin{equation}
	A(x)=\left(\begin{matrix}
		0 &1&0&\cdots&0&0\\
		0 &0&1&\cdots&0&0\\
		\vdots&\vdots&\vdots&\ddots&\vdots&\vdots\\
		0 &0&0&\cdots&0&1\\
		E-a_0(x)&-a_1(x)&-a_2(x)&\cdots&-a_{n-2}(x)&-a_{n-1}(x)
	\end{matrix}\right).\label{eq:periodic_general_A_matrix}
\end{equation}

Eq. \eqref{eq:periodic_general_floquet} shows that the translation property of $\ket{\phi(x)}$ is generated by the momentum operator $\hat p$. Because the periodic matrix $A(x)$ satisfies $A(x)=A(x+a)$, Eq. \eqref{eq:periodic_general_floquet} is a non-Hermitian Floquet problem of $\ket{\phi(x)}$, which looks like a time-dependent Schrodinger equation if we replace $x$ with $t$. We can formally express the solutions of Eq. \eqref{eq:periodic_general_floquet} as
\begin{equation}
	\ket{\phi(x)}=\mathcal{P}_xe^{i\int_0^xA(x')dx'}\ket{\phi(0)}=T(x,0)\ket{\phi(0)},\label{eq:periodic_general_floquet_wavefunction}
\end{equation}
where $\mathcal{P}_x$ is the path-order operator and $\ket{\phi(0)}$ is the initial value at a reference point $x=0$. The $n\times n$ matrix 
\begin{equation}
    T(x,0)=\mathcal{P}_xe^{i\int_0^xA(x')dx'}
\end{equation} is the transfer matrix from $0$ to $x$, which depends on both the coefficients $a_m(x)$ and the energy $E$ \footnote{It is arbitrary to choose the initial reference point $x_0$ in the definition of the transfer matrix. Here we fix $x_0=0$ for simplicity. }.

In a periodic system, the transfer matrix $T(x,0)$ satisfies the relation $T(x+a,0)=T(x,0)T_a$, where we define the transfer matrix inside a unit cell as 
\begin{equation}
T_a\equiv T(a,0)=\mathcal{P}_xe^{i\int_0^a A(x')dx'}.\label{eq:periodic_general_transfer_matrix}
\end{equation}
Notably, the transfer matrix $T_a$, resembling the Floquet operator in periodically driven quantum systems, plays a central role in connecting the wavefunctions between different unit cells. Assuming that $T_a$ is diagonalizable, we have
\begin{equation}
	T_a=\sum_{i=1}^n\beta_i\ket{\xi_i^R}\bra{\xi_i^L},
\end{equation}
where $\ket{\xi_i^{R,L}}$ are the right/left eigenvectors of $T_a$. We order the eigenvalues by their norms $|\beta_1|\le|\beta_2|\le\cdots\le|\beta_n|$. These eigenstates of $T_a$ generate a series of functions 
\begin{equation}
    \ket{\phi_i(x)}=T(x,0)\ket{\xi_i^R}.
\end{equation}
They follow a property that 
\begin{equation}
\ket{\phi_i(x+a)}=T(x+a,0)\ket{\xi_i^R}=T(x,0)T_a\ket{\xi_i^R}=\beta_i\ket{\phi_i(x)}.
\end{equation}
By choosing the first component of the $n$-component vector $\ket{\phi_i(x)}$, we can construct non-Bloch waves $\psi_i(x)=\braket{1|T(x,0)|\xi_i^R}$ which satisfy $\psi_i(x+a)=\beta_i\psi_i(x)$. Here, we use $\bra{m}$ to extract the $m$-th component of the vector $\ket{\phi(x)}$. The general wavefunction of the eigenstates in Eq. \eqref{eq:periodic_general_eigen} can be expressed as the superposition of $n$ independent $\psi_i(x)$:
\begin{equation}
	\psi(x)=\sum_{i=1}^nc_i\psi_i(x)=\sum_{i=1}^nc_i\braket{1|T(x,0)|\xi_i^R}.\label{eq:periodic_general_wave}
\end{equation}
Similarly, the derivatives of $\psi(x)$ can be expressed as 
\begin{equation}
	\hat p^m\psi(x)=\sum_{i=1}^nc_i\braket{m+1|T(x,0)|\xi_i^R}.
\end{equation}
Under this circumstance, we find that
\begin{equation}
\begin{split}
    \hat p^m\psi(0)=&\sum_{i=1}^nc_i\braket{m+1|\xi_i^R},\\
     \hat p^m\psi(L)=&\sum_{i=1}^nc_i\beta_i^{{L}/{a}}\braket{m+1|\xi_i^R}.
\end{split}
\end{equation}
To fully determine the coefficients $c_i$, we have to impose suitable boundary conditions. 

The PBC $\psi(x)=\psi(x+L)$ requires that there should be at least one eigenstate of $T_a$ whose eigenvalue $\beta_{i_0}$ satisfies $\beta_{i_0}^{L/a}=1$. Consequently, $|\beta_{i_0}|=1$ when $L\to+\infty$. In this case, the coefficients are given by $c_{i}=\delta_{i,i_0}$, and the wavefunction of the PBC system is $\psi(x)\sim\braket{1|T(x,0)|\xi_{i_0}^R}$. This is exactly the extended Bloch wave since  $\psi(x+a)=\beta_{i_0}\psi(x)$ with $|\beta_{i_0}|=1$. 

However, the situation is different when we consider open boundary conditions. To determine the coefficients $c_i$ in Eq. \eqref{eq:periodic_general_wave}, we need $n$ independent boundary constraints at two ends $x=0,L$. Following the procedure in Sec. \ref{sec:homogeneous_GBZ}, if there are $n_l$ ($n_r$) homogeneous independent boundary equations at the left (right) edge, with the constraint $n=n_r+n_l$, there will be $n$ linear independent equations of $c_i$. The condition for the nonzero solutions of $c_i$ is given by
 \begin{equation}
	\left|\begin{matrix}
	g_{1,1} & g_{1,2}& \cdots &g_{1,n}\\
	g_{2,1}&g_{2,2}&\cdots&g_{2,n}\\
	\vdots&\vdots&\ddots&\vdots\\
	g_{n_l,1}&g_{n_l,2}&\cdots&g_{n_l,n}\\
	g_{n_l+1,1}\beta_{1}^{L/a} & g_{n_l+1,2}\beta_{2}^{L/a}& \cdots &g_{n_l+1,n}\beta_{n}^{L/a}\\
	g_{n_l+2,1}\beta_{1}^{L/a}&g_{n_l+2,2}\beta_{2}^{L/a}&\cdots&g_{n_l+2,n}\beta_{n}^{L/a}\\
	\vdots&\vdots&\ddots&\vdots\\
	g_{n,1}\beta_{1}^{L/a}&g_{n,2}\beta_{2}^{L/a}&\cdots&g_{n,n}\beta_{n}^{L/a}
	\end{matrix}\right|=0.
	\label{eq:periodic_general_homogeneous_bigdet}
\end{equation}
Here the coefficients $g_{p,q}$ are constants determined by the Hamiltonian, the details of boundary conditions, and the energy. These constants do not depend on the system size $L$. Thus, when $L\to+\infty$, the leading terms of the left-hand side (LHS) of Eq. \eqref{eq:periodic_general_homogeneous_bigdet} are
\begin{equation}
\begin{split}
	\text{LHS}=&C_{1}\left[\beta_n\beta_{n-1}\cdots\beta_{n-(n_r-2)}\beta_{n-(n_r-1)}\right]^{L/a}\\
	&+C_{2}\left[\beta_n\beta_{n-1}\cdots\beta_{n-(n_r-2)}\beta_{n-n_r}\right]^{L/a}\\
	&+\cdots.
\end{split}
\end{equation}
In the thermodynamic limit, the condition $\text{LHS}=0$ leads to $|\beta_{n-(n_r-1)}|=|\beta_{n-n_r}|$. Together with $n=n_r+n_l$, we obtain
\begin{equation}
	|\beta_{n_l}(E)|=|\beta_{n_l+1}(E)|\label{eq:periodic_general_GBZ}
\end{equation}
where we explicitly show the dependence on the energy $E$. This is the GBZ equation in general periodic non-Hermitian continuum systems, which is a central result of this section. It should be noted that the transfer matrix $T_a$ is indispensable in defining the GBZ in periodic continuum systems.

The energies $E$ satisfying Eq. \eqref{eq:periodic_general_GBZ} give rise to the OBC spectrum where there are $n_l$ $(n_r)$ boundary constraints at the left (right) end of an open chain. Simultaneously, $\beta_{n_l}, \beta_{n_l+1}$ of this equation provide the corresponding GBZ on the complex $\beta$-plane. The condition $|\beta_{n_l}(E)|=|\beta_{n_l+1}(E)|\ne1$ indicates the existence of OBC skin modes with energy $E$.

In conclusion, based on the transfer matrix, we have established the non-Bloch band theory for periodic non-Hermitian continuum systems. The transfer matrix inside a unit cell is given by Eq. \eqref{eq:periodic_general_transfer_matrix}, which depends on the Hamiltonian $\hat H$ and the energy $E$. With these ingredients, the characteristic equation is formally defined as
\begin{equation}
	f(\beta,E)\equiv\det(\beta\mathcal{I}_{n}-T_a)=0,\label{eq:periodic_general_characteristic_equation}
\end{equation}
where $\mathcal{I}_{n}$ is an $n\times n$ identity. The $n$ roots of this equation can be ordered by their norms $|\beta_1(E)|\le|\beta_2(E)|\le\cdots\le|\beta_n(E)|$. Finally, if there are $n_r/n_l$ independent homogeneous boundary constraints at the right/left edges, with the condition $n=n_r+n_l$, we get the GBZ equation Eq. \eqref{eq:periodic_general_GBZ}.

Here are some remarks. First, in general systems, we cannot explicitly find $T_a$ due to the noncommutative nature of $A(x)$ at different positions. $T_a$ is determined by $E$ in a complicated way. As a result, there are an infinite number of energy bands in periodic continuum systems.

\begin{figure}
	\centering
	\includegraphics[width=8.5cm]{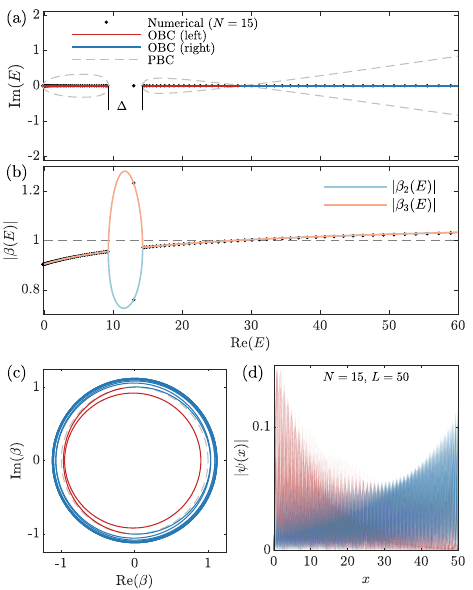}
	\caption{The eigenstates of the periodic continuum model Eq. \eqref{eq:periodic_general_p4} under the open boundary conditions $n_l=n_r=2$. (a) Energy spectrum: the red (blue) lines represent OBC eigenstates localized at the left (right) edge, which are obtained by the lattice GBZ theory. $\Delta$ denotes the energy gap between the first and second non-Bloch bands of the OBC spectrum. The PBC spectrum is shown as gray dashed lines, obtained by diagonalizing $h_{N=15}(\beta\to e^{ik})$ in Eq. \eqref{eq:periodic_general_p4_non_Bloch_H} with $k\in[0,2\pi]$. The black points are the numerical OBC energies, which are obtained from the brute-force calculation of an OBC lattice Hamiltonian generated by Eq. \eqref{eq:periodic_general_p4_non_Bloch_H}. (b) $|\beta_{2,3}(E)|$ changes with respect to a real $E$. The solid lines come from the transfer-matrix method; the black points are obtained from solving the characteristic equation Eq. \eqref{eq:periodic_genernal_p4_char_equation}  with the energies coming from the brute-force calculation of an OBC lattice Hamiltonian. (c) GBZ for the lattice model in Eq. \eqref{eq:periodic_general_p4_non_Bloch_H} with $N=15$. The red (blue) lines represent that these eigenstates are inside (outside) the unit circle $|\beta|=1$. (d) Typical low-energy eigenstate profiles. These eigenstates are obtained by brute-force diagonalization of an OBC lattice Hamiltonian with $N=15$ and $L=50$ ($x\in[0,L]$). Therefore this chain contains 50 unit cells with $15$ sites in each cell. Parameters: $(a_4,\, a_3,\, a_2,\, a_1,\, a_0)=(0.02,\, 0.01i,\, 1,\, -0.2i,\, 0)$, $V(x)=V_0\sin(2\pi x)$ with $V_0=5$.}
	\label{fig:periodic_p4_energy}
\end{figure}

Second, if all $a_m(x)$ are constants that are independent of $x$, $T_a$ can be evaluated exactly. With the map $\beta\to e^{ika}$, we reproduce the equation Eq. \eqref{eq:homogeneous_GBZ} of generalized momentum space in homogeneous non-Hermitian continuum systems.

Third, this theory can be easily generalized to periodic continuum systems with local degrees of freedom where all $a_m(x)$ are $q\times q$ matrices. We will go back to this generalization in Sec. \ref{sec:matrix_periodic}.

After establishing the non-Bloch band theory in periodic non-Hermitian continuum systems, in the subsequent sections, we will apply this theory to two instructive models and discuss the NHSE and spectrum under open boundary conditions.
\subsection{Revisit the periodic Hatano-Nelson model}
An immediate application of the non-Bloch band theory in Sec. \ref{sec:periodic_GBZ} is to utilize the transfer-matrix method to derive the circular GBZ [Eq. \eqref{eq:periodic_Hatano_Nelson_GBZ}] of the periodic Hatano-Nelson model introduced in Sec. \ref{sec:periodic_hatano_nelson}. 

As a second-order differential equation,  Eq. \eqref{eq:periodic_general_p2} gives rise to the $A(x)$ matrix of dimension 2:
   \begin{equation}
	A(x)=\left(\begin{matrix}
		0 &1\\
		E-a_0(x)&-a_1(x)
	\end{matrix}\right).\label{eq:periodic_hatano_nelson_A_matrix}
\end{equation}
Then the transfer matrix inside a unit cell is given by $    T_a=\mathcal{P}_xe^{i\int_0^a A(x')dx'} $, whose eigenvalues provide two non-Bloch wave vectors $\beta_1(E)$ and $\beta_2(E)$. We can show that \footnote{Eq. \eqref{eq:periodic_hatano_nelson_det} stems from the fact that $\det [\mathcal{P}_xe^{i\int_0^a A(x')dx'}]=\det[\lim\limits_{N\to+\infty}\prod_{n=1}^Ne^{iA(\frac{na}{N})\frac{a}{N}}]=\lim\limits_{N\to+\infty}\prod_{n=1}^N\det[e^{iA(\frac{na}{N})\frac{a}{N}}]=\lim\limits_{N\to+\infty}e^{i\sum_{n=1}^N\operatorname{Tr}[A(\frac{na}{N})]\frac{a}{N}}=e^{i\int_0^a\operatorname{Tr}[A(x^\prime)]\mathrm{d}x^\prime}$.} 
\begin{equation}\label{eq:periodic_hatano_nelson_det}
    \begin{split}
        \beta_1(E)\beta_2(E)=\det [\mathcal{P}_xe^{i\int_0^a A(x')dx'}]=e^{i\int_0^a\operatorname{Tr}[A(x^\prime)]\mathrm{d}x^\prime}.
    \end{split}
\end{equation}
This equation is the same as Eq. \eqref{eq:periodic_Hatano_Nelson_rootproduct},  leading to the circular GBZ in Eq. \eqref{eq:periodic_Hatano_Nelson_GBZ}. 

\subsection{General $\hat p^4$ model with a periodic potential}

\begin{figure*}[t]
	\centering
	\includegraphics[width=\textwidth]{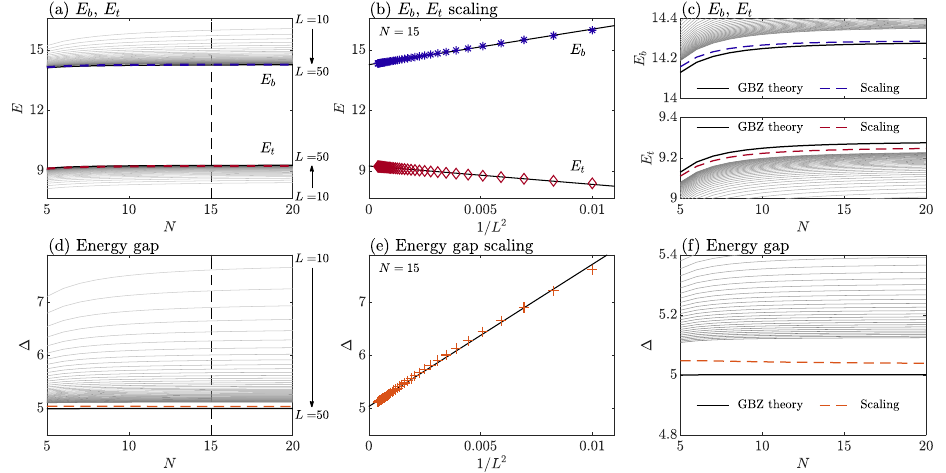}
	\caption{$E_b$ and $E_t$. Gray lines in (a) and (c) present the highest energy $E_t$ of the first non-Bloch band and the lowest energy $E_b$ of the second non-Bloch band. The discretization parameter $N$ varies from 5 to 20 and the lattice length $L$ changes from 10 to 50. These results are calculated by brute-force real-space diagonalization, where the OBC Hamiltonian is generated by Eq. \eqref{eq:periodic_general_p4_non_Bloch_H}. The red/blue dashed lines denote the vertical intercepts shown in (b). We are not interested in the possible edge states. The solid black lines are saddle point energies obtained by the lattice GBZ theory. (b) The scaling of $E_b$ and $E_t$ with respect to $1/L^2$ for $N=15$. (d) and (f) show the spectrum gap $\Delta=E_b-E_t$ obtained from (a). The orange line represents the vertical intercept shown in (e). The black line is the gap obtained from the lattice GBZ theory. (e) The scaling of $\Delta$ with respect to $1/L^2$ for $N=15$. The model and other parameters are the same as in Fig. \ref{fig:periodic_p4_energy}.}
	\label{fig:periodic_p4_gap}
\end{figure*}

To demonstrate the applications of the non-Bloch band theory, we consider a general periodic non-Hermitian continuum model:
\begin{equation}
	\hat H=a_4\hat p^4+a_3\hat p^3+a_2\hat p^2+a_1\hat p+a_0+V(x),\quad x\in[0,L].
	\label{eq:periodic_general_p4}
\end{equation}
The potential $V(x)$ is a periodic function $V(x)=V(x+1)$, where we set the lattice constant $a=1$ for simplicity. If $V(x)=0$, this model goes back to the homogeneous $\hat p^4$ model in Eq. \eqref{eq:homogeneous_general_p4}. In this section, we set the same parameters $a_0,a_1,\cdots,a_4$ as Fig. \ref{fig:generalp4}(a)$\sim$(c). Additionally, we add a periodic potential $V(x)=V_0\sin(2\pi x)$ with $V_0=5$. The boundary conditions are the same as Fig. \ref{fig:generalp4}(a)$\sim$(c), where $n_l=n_r=2$. 

Interestingly, like $\hat H_4$ in Eq. \eqref{eq:homogeneous_general_p4}, if we impose real parameters $a_4,a_2,a_0$ and imaginary parameters $a_3,a_1$, introducing a real-valued potential $V(x)$ still preserves the PT symmetry $\hat H^* =\hat H$. This symmetry can also lead to a possible real-valued OBC spectrum.

Now we apply the non-Bloch band theory in Sec. \ref{sec:periodic_GBZ} to the Hamiltonian in Eq. \eqref{eq:periodic_general_p4}. The corresponding $A(x)$ matrix is given by
\begin{eqnarray}
 	A(x)=\left(\begin{matrix}
		0 &1&0&0\\
		0 &0&1&0\\
 0 &0&0&1\\
		\frac{E-a_0-V(x)}{a_4}&-\frac{a_1}{a_4}&-\frac{a_2}{a_4}&-\frac{a_3}{a_4}
	\end{matrix}\right).\label{eq:periodic_general_p4_A_matrix}
\end{eqnarray}
Since $A(x)$ and $A(x')$ do not commute with each other, it is difficult to express the transfer matrix $T_a$ explicitly. Practically, we can use a discretized approximation to efficiently evaluate $T_a$:
\begin{eqnarray}
 T_a=\mathcal{P}_xe^{i\int_0^aA(x)dx}\approx e^{i\frac{1}{N}A(\frac{N-1}{N})}e^{i\frac{1}{N}A(\frac{N-2}{N})}\cdots e^{i\frac{1}{N}A(\frac{1}{N})}e^{i\frac{1}{N}A(0)}
\end{eqnarray}
where we already set the lattice constant $a=1$ for simplicity. The discretization parameter $N$ represents the number of lattice sites in a unit cell. The distance between two nearest sites is given by $\delta=1/N$. Therefore, given a complex energy $E$, we can obtain an effective transfer matrix $T_a$ and its eigenvalues $|\beta_1(E)|\le|\beta_2(E)|\le|\beta_3(E)|\le|\beta_4(E)|$. With the boundary conditions $n_l=n_r=2$, the GBZ equation $|\beta_2(E)|=|\beta_3(E)|$ picks up the OBC energies.

Due to the PT symmetry, many OBC eigenstates of the Hamiltonian can have real-valued energies. Choosing $N=15$ and sweeping a real-valued $E$ along the axis, we observe that the resulting  $|\beta_2(E)|$ and $|\beta_3(E)|$ coincide with each other within some intervals of $E$ [Fig. \ref{fig:periodic_p4_energy}(b)]. Based on the GBZ theory, these energy intervals provide the OBC energy bands. The corresponding $|\beta_{2}(E)|$ and $|\beta_{3}(E)|$ offer the decay length of the skin modes. 

To compare the transfer matrix method in Sec. \ref{sec:periodic_GBZ} to the lattice discretization in Sec. \ref{sec:correspondence_general_p4}, we apply the discretization procedure in Sec. \ref{sec:correspondence_general_p4} to the periodic model Eq. \eqref{eq:periodic_general_p4}. 

The lattice version of the homogeneous case, specifically with $V_0=0$, is presented in Sec. \ref{sec:correspondence_general_p4}. In this section, the periodic potential $V(x)$ merely affects the onsite potential of the lattice, leaving the hopping elements unchanged. The single-band tight-binding Hamiltonian in Eq. \eqref{eq:correspondence_tight_binding_general_p4} is simply modified by adding the discrete periodic potential $V(i/N)\delta_{ij}$. This modification transforms the single-band lattice model into a multiband model with $N$ energy bands. Eventually, the non-Bloch Hamiltonian of this effective lattice model can be expressed as an $N\times N$ matrix:
\begin{widetext}
\begin{equation}
	\setcounter{MaxMatrixCols}{20}
	h_{N}(\beta)=\left(\begin{matrix}
		t_0+V(0)&t_1&t_2&0&0 &\cdots&0&0&0&t_{-2}\beta^{-1}&t_{-1}\beta^{-1}\\
		t_{-1}&t_0+V(\frac{1}{N})&t_1&t_2 &0&\cdots&0&0&0&0&t_{-2}\beta^{-1}\\
		t_{-2}&	t_{-1}&	t_0+V(\frac{2}{N})&t_1&t_2&\cdots&0&0&0&0&0\\
		\vdots&\vdots&\vdots&\vdots&\vdots&\ddots&\vdots&\vdots&\vdots&\vdots&\vdots\\			0&0&0&0&0&\cdots&t_{-2}&t_{-1}&	t_0+V(\frac{N-3}{N})&t_1&t_2\\
		t_2\beta&0&0&0&0&\cdots&0&t_{-2}&	t_{-1}&t_0+V(\frac{N-2}{N})&t_1\\
	t_1\beta&t_2\beta&0&0&0&\cdots&0&0&t_{-2}&t_{-1}&t_0+V(\frac{N-1}{N})
	\end{matrix}\right),
\label{eq:periodic_general_p4_non_Bloch_H}
\end{equation}
\end{widetext}
where $V(x)=V(x+1)$ as defined. The hopping elements $\{t_{-2},\, t_{-1},\, t_0,\, t_1,\, t_2\}$ are the same as Eq. \eqref{eq:correspondence_hopping_general_p4}.

Choosing $N=15$, we investigate Eq. \eqref{eq:periodic_general_p4_non_Bloch_H} as a non-Hermitian multiband lattice model. Since the characteristic equation
\begin{equation}
	f_N(\beta,E)=\det(E\mathcal{I}_{N}-h_N(\beta))=0\label{eq:periodic_genernal_p4_char_equation}
\end{equation}
has four roots $|\beta_{1}(E)|\le|\beta_{2}(E)|\le|\beta_{3}(E)|\le|\beta_{4}(E)|$ in this case, the lattice GBZ equation is given by $|\beta_{2}(E)|=|\beta_{3}(E)|$. Then the non-Bloch band theory for multiband lattice systems directly gives the lattice GBZ and OBC spectrum [Fig. \ref{fig:periodic_p4_energy}].

Notably, the numerical results obtained from the lattice model, as depicted in Fig. \ref{fig:periodic_p4_energy}(a), exhibit remarkable consistency with the results derived from the transfer-matrix method, as illustrated in Fig. \ref{fig:periodic_p4_energy}(b). The agreement between these two methods indicates that they can serve as reliable benchmarks for each other. Consequently, they efficiently provide accurate wavefunctions and OBC spectrums of a periodic non-Hermitian continuum system, particularly in {the low-energy part of the OBC spectrum.

Upon comparing the OBC spectrum of the periodic model [Fig. \ref{fig:periodic_p4_energy}(a)] to that of the homogeneous model [Fig. \ref{fig:generalp4}(a)], we observe that the additional periodic potential $V(x)$ has a significant impact primarily on the low-energy segment of the spectrum, regardless of whether PBC or OBC are considered. 

In the OBC case, the periodic potential induces a spectrum gap in the low-energy region, and this gap is proportional to the magnitude of $V_0$. The appearance of this gap can be attributed to the periodicity-induced coupling between two degenerate non-Bloch modes whose complex wave vectors differ by $2\pi$. Within the low-energy portion of the spectrum, two degenerate modes satisfying this condition can be gapped through the perturbation theory. Therefore an energy gap $\Delta$ is anticipated in the OBC spectrum. In the specific case illustrated in Fig. \ref{fig:periodic_p4_energy}(a), it is expected that $\Delta\sim V_0=5$.

The determination of the gap value can be challenging due to finite-size effects in the effective lattice models. These finite-size effects arise from two factors: the discretization parameter $N$ and the length of the lattice $L$. Fortunately, given a specific $h_N(\beta)$, the finite-size effects induced by the lattice length $L$ can be avoided.  This is because the OBC spectrum gap can be efficiently calculated by the GBZ theory for multiband lattice models, where we do not need to introduce the lattice length $L$. 

The energy gap $\Delta$ is defined as the energy difference between the lowest eigenstate of the second band and the highest eigenstate of the first band. These two states correspond to the endpoints of each OBC energy band. Their energies satisfy $\beta_{2}(E)=\beta_{3}(E)$, which are \emph{saddle points} of the characteristic equation Eq. \eqref{eq:periodic_genernal_p4_char_equation}. The  saddle points are given by the following equations:
\begin{equation}
	f_N(\beta,E)=\frac{\partial}{\partial\beta}f_N(\beta,E)=0.
	\label{eq:periodic_general_p4_GBZ_Gap}
\end{equation}

By solving these equations and identifying the two relevant energy values, we can determine the energy gap $\Delta$ without finite-size effects arising from the lattice length $L$. Notably, calculating the energy gap $\Delta$ by the GBZ method only requires solving two algebraic equations, which is computationally more efficient compared to the real-space diagonalization method.

In Fig. \ref{fig:periodic_p4_gap}, we present the lowest energy ($E_b$) of the second band, the highest energy ($E_t$) of the first band, and the corresponding energy gap ($\Delta=E_b-E_t$). These values are obtained through a brute-force real-space diagonalization method, considering various discretization parameters $N$ and lattice lengths $L$ [see Fig. \ref{fig:periodic_p4_gap}(a),(d)]. Interestingly, at the end point of an energy band, the energy exhibits a scaling behavior that is approximately proportional to $1/L^2$. This scaling behavior can be understood by expanding the energy dispersion around a saddle point $\beta_s$, where $\beta_s$ represents the solution of Eq. \eqref{eq:periodic_general_p4_GBZ_Gap} in the complex $\beta$-plane. The dispersion near a saddle point can be approximated as $E(\beta)-E(\beta_s)\sim (\beta-\beta_s)^2$, and a finite system induces a scaling behavior $|\beta-\beta_s|\sim 1/L$. As a result, this scaling behavior enables us to estimate saddle point energies as $L$ approaches the infinity [Fig. \ref{fig:periodic_p4_gap}(b),(e)]. Nevertheless, it is important to note that this scaling behavior is merely an approximation to mitigate the finite-size effect induced by $L$. Consequently, it is challenging to obtain accurate gap values in the thermodynamic limit by using the brute-force diagonalization method [see Fig. \ref{fig:periodic_p4_gap}].

The saddle point equation, Eq. \eqref{eq:periodic_general_p4_GBZ_Gap}, offers an alternative approach to determining the energy gap $\Delta$ without requiring the large-$L$ limit. The results obtained by this algebraic method are shown as solid black lines in Fig. \ref{fig:periodic_p4_gap}. When comparing these results with those obtained by the brute-force diagonalization method, we observe that the algebraic approach does not show finite-size effects induced by the lattice length $L$. The calculation of energy gaps highlights the efficiency and reliability of the non-Bloch band theory in non-Hermitian continuum systems.

\section{Continuum systems with local degrees of freedom}\label{sec:matrix}
 \subsection{Homogeneous continuum systems}\label{sec:matrix_homogeneous}

To further generalize the non-Bloch band theory in non-Hermitian continuum systems, we introduce a new type of homogeneous non-Hermitian continuum systems, which have local degrees of freedom at each position in the real space. The generic Hamiltonian for such systems can be expressed as:
\begin{equation}
	\hat H=\sum_{m=0}^na_m\hat p^m, \quad x\in[0,L],
	\label{eq:matrix_H}
\end{equation}
where the coefficients $a_m$ become $q\times q$ constant matrices. The index $q$ labels the local degrees of freedom in this system. The $q$-band matrix-valued homogeneous system can be understood as a collection of $q$ 1D homogeneous subsystems, similar to the form in Eq. \eqref{eq:homogeneous_general_continuum_H}, with additional couplings between these subsystems. From a mathematical perspective, the evolution of the entire system can be described by a set of linear differential equations, e.g. the Maxwell equations. Here we focus on the eigenstates and energy spectrum of this system under appropriate boundary conditions.

The stationary Schrodinger equation of the Hamiltonian Eq. \eqref{eq:matrix_H} is defined as
\begin{equation}
	\sum_{m=0}^na_m\left(-i\partial_x\right)^m\psi(x)=E\psi(x), \quad x\in[0,L],
\end{equation}
where the vector $\psi(x)=(\psi_1(x),\psi_2(x),\cdots,\psi_q(x))^T $ represents a q-component wavefunction. We can solve the problem under the PBC $\psi(x)=\psi(x+L)$ by defining the Bloch Hamiltonian $H(k)=\sum_{i=0}^n a_ik^i$. The PBC spectrum is given by the eigenvalues of this $q\times q$ Bloch Hamiltonian $H(k)$ with $k\in\mathbb{R}$.

However, when considering an open chain, it is necessary to specify the boundary conditions at $x=0$ and $x=L$. In this context, we impose the following boundary conditions:
\begin{subequations}
	\begin{align}
		&\psi(0)=\psi'(0)=\cdots=\psi^{(n_l-1)}(0)=0,\\
		&\psi(L)=\psi'(L)=\cdots=\psi^{(n_r-1)}(L)=0.
	\end{align}
	\label{eq:matrix_OBC}
\end{subequations}

Similar to Eq. \eqref{eq:homogeneous_OBC}, we also require $n=n_r+n_l$. It is important to note that $\psi(x)$ is a $q$-component vector and there are $qn$ boundary conditions to be satisfied. To determine the OBC spectrum, we should focus on the number of boundary equations on each edge rather than the specific details of these equations. We formally extend the Bloch Hamiltonian $H(k)$ to the complex $k$ plane. Then the characteristic equation with energy $E$ is given by:
\begin{equation}
	f(k,E)=\det[E-H(k)].
\end{equation}
 In general cases, the characteristic equation $f(k,E)=0$ has $qn$ roots, which can be ordered by their imaginary parts $\text{Im}[{k_1}(E)]\ge\text{Im}[k_2(E)]\ge\cdots\ge\text{Im} [k_{qn}(E)]$. The general wavefunction is described by $\psi(x)=\sum_{j=1}^{nq}c_je^{ik_jx}\varphi_j$, where $\varphi_j$ is the right eigenstate of $H(k_j)$ with the eigenvalue being $E$. Following the same procedure described in Sec. \ref{sec:homogeneous_GBZ}, we find that Eq. \eqref{eq:matrix_OBC} provides $nq$ independent constraints on the coefficients $c_j$. The requirement for the existence of nontrivial solutions gives rise to the equation of the generalized momentum space under the boundary conditions in Eq. \eqref{eq:matrix_OBC}, which can be expressed as:
\begin{equation}\label{eq:matrix_homogeneous_GBZ}
	\text{Im}[k_{qn_l}(E)]=\text{Im}[k_{qn_l+1}(E)].
\end{equation}
This equation of the generalized momentum space, similar to Eq. \eqref{eq:homogeneous_GBZ}, determines the OBC spectrum and the localization length of the skin modes simultaneously. 

The generalized momentum space can be obtained by the method of auxiliary generalized momentum space developed in Sec. \ref{sec:applications_of_GBZ}. Specifically, we first evaluate the common roots $k$ of two algebraic equations $f(k,E)=0$ and $f(k+k_0,E)=0$, where $k_0\in\mathbb{R}$ is a real auxiliary parameter. The common roots can be found by employing the resultant to eliminate the variable $E$ in these two equations (see Ref. \cite{yang2020non} for details). Then the resultant provides an algebraic equation $    \operatorname{Res}_E[f(k,E),f(k+k_0,E)]=0$ of the variable $k$. The solutions $k(k_0)$ of this equation lead to the auxiliary generalized momentum space. The points on the auxiliary generalized momentum space satisfy $\operatorname{Im}[k_i(E)]=\operatorname{Im}[k_{i+1}(E)]$ with $i=1,2,\cdots,nq-1$, where  $\text{Im}[{k_1}(E)]\ge\text{Im}[k_2(E)]\ge\cdots\ge\text{Im} [k_{qn}(E)]$ are roots of the characteristic equation $f(k,E)=0$. Finally, the generalized momentum space of this model can be picked out by checking Eq. \eqref{eq:matrix_homogeneous_GBZ}.

As an example, we consider two Hatano-Nelson models with local couplings, whose Hamiltonian reads
\begin{eqnarray}\label{eq:coupled_Hatano_Nelson}
    \hat H=\begin{pmatrix}
        \hat p^2+a_1\hat p+a_0 & V_{AB} \\
        V_{BA} & \hat p^2+b_1\hat p+b_0
    \end{pmatrix}.
\end{eqnarray}
$\hat H_A=\hat p^2 + a_1\hat p+a_0$ and $\hat H_B=\hat p^2 + b_1\hat p+b_0$ represent two Hatano-Nelson models like Eq. \eqref{eq:homogeneous_Hatano_Nelson}; $V_{AB}$ and $V_{BA}$ are the couplings between these two models. $A$ and $B$ can be viewed as spin indexes or other types of internal degrees of freedom. The Bloch Hamiltonian of this model is given by \begin{eqnarray}
    H(k)=\begin{pmatrix}
        k^2+a_1k+a_0 & V_{AB} \\
        V_{BA} &k^2+b_1k+b_0
    \end{pmatrix}.
\end{eqnarray}
The real-space wavefunction of this model is a 2-component vector $\psi(x)=(\psi_A(x),\psi_B(x))^T$. To analyze the eigenstates and energies of this system on an open chain with $0\le x\le L$, we consider the open boundary conditions $\psi(x)=\psi(L)=0$, which leads to
\begin{eqnarray}
    \psi_A(0)= \psi_B(0)=\psi_A(L)= \psi_B(L)=0.
\end{eqnarray}
Namely, there are $qn_l=2$ $(qn_r=2)$ boundary conditions at the left (right) boundary. The characteristic equation $\det[E-h(k)]=0$ contains four roots $k_1(E),\ k_2(E),\ k_3(E),\ k_4(E)$, ordered by their imaginary parts. The boundary conditions then give rise to the equation of generalized momentum space:
\begin{eqnarray}\label{eq:coupled_Hatano_Nelson_GBZ}
    \operatorname{Im}[k_2(E)]=\operatorname{Im}[k_3(E)].
\end{eqnarray}
Solving this equation by the above method of auxiliary generalized momentum space, we can obtain the OBC spectrum and generalized momentum space of this model [Fig. \ref{fig:coupled_hatano_nelson}].
\begin{figure}
    \centering
    \includegraphics[width=8cm]{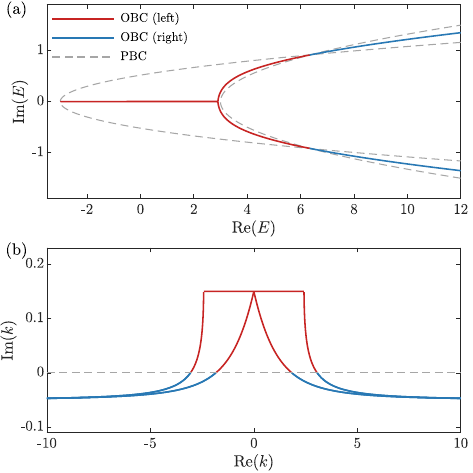}
    \caption{The OBC spectrum (a) and the generalized momentum space (b) of the model in Eq. \eqref{eq:coupled_Hatano_Nelson}. The red (blue) lines represent the eigenstates localized at the left (right) edge; gray dashed lines show the PBC eigenstates. The parameters are given by $a_1=0.5i,\ b_1=-0.3i,\ a_0=-b_0=3,\ V_{AB}=V_{BA}=0.1$. }`
    \label{fig:coupled_hatano_nelson}
\end{figure}

The non-Bloch band theory for homogeneous continuum systems with local degrees of freedom not only represents a direct theoretical generalization but also carries broad implications for state-of-the-art experiments. Local degrees of freedom naturally appear when studying spinful particles in cold atom physics or vector fields in non-Hermitian photonic systems. In this context, the non-Bloch band theory offers valuable insights and tools for exploring the broader non-Bloch physics of such systems.
\subsection{Periodic continuum systems}\label{sec:matrix_periodic}
The combination of the ideas in Sec. \ref{sec:periodic_GBZ} and Sec. \ref{sec:matrix_homogeneous} immediately provides the non-Bloch band theory in periodic non-Hermitian continuum systems where there are local degrees of freedom. In this case, the Hamiltonian is given by

\begin{equation}
	\hat H=\sum_{m=0}^{n}a_m(x)\hat p^m,\quad x\in[0,L].
\end{equation}
We consider the matrix-valued periodic complex  functions $a_m(x)=a_m(x+a)$, where $a$ represents the lattice constant and $L/a\gg1$ is an integer. These functions are $q\times q$ matrices. Without loss of generality, we set $a_n(x)=\mathcal{I}_q$, where $\mathcal{I}_q$ is the $q\times q$ identity matrix. Under this assumption, we can define $A(x)$ in Eq. \eqref{eq:periodic_general_A_matrix} and $T_a$ in Eq. \eqref{eq:periodic_general_transfer_matrix} in a similar way. Now $A(x)$ and $T_a$ become two $qn\times qn$ matrices. As a result, the characteristic equation Eq. \eqref{eq:periodic_general_characteristic_equation} for $\beta$ becomes an algebraic equation of order $qn$. The roots of this equation can be ordered based on their norms, such that $|\beta_1(E)|\le|\beta_2(E)|\le\cdots\le|\beta_{qn}(E)|$. Considering the presence of $qn_l$ boundary constraints at the left edge and $qn_r$ boundary constraints at the right edge, the GBZ equation is derived as \begin{equation}
    |\beta_{qn_l}(E)|=|\beta_{qn_l+1}(E)|,
\end{equation}which is similar to the GBZ equation in Sec. \ref{sec:periodic_GBZ}. 

This GBZ equation is a building block of the non-Bloch band theory in periodic non-Hermitian continuum systems with local degrees of freedom. It directly applies to the study of the non-Hermitian skin effect in photonic crystals \cite{zhong2021nontrivial, Ochiai2022non,yan2021nonhermitian, Yokomizo2022non-hermitian_wave, FangHu2022Geometry, liu2023localization,zhu2023photonic}. Hence, our general theory offers a promising approach to investigating the non-Bloch properties in periodic photonic systems. 

\section{Conclusions}\label{sec:discussion}

In this work, we generalize the non-Bloch band theory from non-Hermitian lattice models to continuum systems. 

First, we uncover the significant role of boundary conditions in studying non-Hermitian continuum systems. The number of boundary conditions at the edges of an open chain prominently controls the fundamental properties of the system, including the structure of generalized momentum space or generalized Brillouin zone, the localization of skin modes, and the OBC spectrum. Several representative models such as the Hatano-Nelson model with or without a periodic structure and the homogeneous/periodic general $\hat p^4$ model, have been extensively studied in this regard. The application of the non-Bloch band theory to these models allows for the characterization and understanding of the rich non-Bloch properties in a variety of non-Hermitian continuum systems. 

In addition, we have also established a self-consistent correspondence between continuum systems and lattice models. This connection is based on mapping the number of boundary conditions in the OBC continuum systems into the hopping ranges in the corresponding lattice models. The continuum-lattice correspondence allows us to study non-Hermitian continuum systems by discretizing them into lattice models and then employing the well-established lattice version of the non-Bloch band theory. Following this procedure, we efficiently computed energy gaps of a periodic non-Hermitian continuum model. Based on the saddle points on the GBZ, we proposed a computationally efficient method to determine the spectrum gaps without finite-size effects. The continuum-lattice correspondence provides a valuable perspective on non-Hermitian continuum systems.

Furthermore, we demonstrate the application of the transfer matrix in periodic non-Hermitian continuum systems. It can be used to identify the GBZ in such systems. From this perspective, the transfer matrix provides a valuable tool for analyzing the band structure and characterizing non-Bloch physics in periodic non-Hermitian continuum systems. 

Whereas our theory is readily applicable to a number of problems in non-Hermitian continuum systems, there are also many questions for further investigation. For example, topology in non-Hermitian continuum systems has been attracting great attention across various fields. The point-gap topology of the PBC spectrum gives rise to the NHSE in lattice systems \cite{zhang2020correspondence, okuma2020topological}. This topological connection between the PBC and OBC spectra is expected to also provide valuable insight into the emergence of NHSE in non-Hermitian continuum systems.

Moreover, the interplay between NHSE and spatial dimensionality is also a subject of great interest \cite{yao2018chern,kawabata2020higher-order,zhang2022universal,wang2022amoeba}. While the non-Bloch band theory in higher-dimensional lattice models has recently been developed based on the Amoeba formulation \cite{wang2022amoeba}, the extension to higher-dimensional continuum systems deserves further investigation.

Finally, symmetries in non-Hermitian systems contribute to rich non-Bloch physics \cite{liu2019topological,okuma2020topological,yi2020non-hermitian,kawabata2020nonbloch,yang2020nonperturbative,yokomizo2021non}. An intriguing example is the symmetry-protected $\mathbb{Z}_2$ NHSE \cite{okuma2020topological,yi2020non-hermitian,kawabata2020nonbloch}. Exploring the existence and properties of $\mathbb{Z}_2$ NHSE in continuum systems would be a stimulating avenue for future research. Additionally, it would be worthwhile to investigate other types of symmetry-enriched NHSE phenomena that would arise in continuum systems. These studies would contribute to a deeper understanding of the role of symmetries in shaping the non-Bloch physics of continuum systems.

\begin{acknowledgments}
This work was supported by the National Natural Science Foundation of China (Grants No. 12125405), and the National Key R\&D Program of China (No. 2023YFA1406702). 
\end{acknowledgments}
\appendix

\section{Eigenmodes of homogeneous Fokker-Planck equation}\label{appendix:left_eigenmode}
In this Appendix, we present a detailed discussion on the eigenstate properties of the homogeneous Fokker-Planck equations in Sec. \ref{sec:homogenerous_hatano}. For the sake of clarity, we rewrite the stationary equation in Eq. \eqref{eq:homogeneous_stationary_Fokker_Planck} for the $n$-th right eigenmode $P_n(x,t)=\psi_{r,n}(x)e^{-E_nt}$:
\begin{equation}
		-D\frac{\partial^2}{\partial x^2}\psi_{r,n}(x)+\mu\frac{\partial}{\partial x}\psi_{r,n}(x)=E_n\psi_{r,n}(x),\quad x\in[0,L].
		\label{eq:periodic_general_homogenenous_fokker_planck}
\end{equation}

By finding the right eigenmodes $\psi_{r,n}(x)$ and the corresponding left eigenmodes $\psi_{l,n}(x)$ under some suitable boundary conditions, we can use the completeness relation
\begin{equation}
	\sum_n\psi_{r,n}(x)\psi_{l,n}^*(x^\prime)=\delta(x-x^\prime)
\end{equation} to express any initial state $P(x,0)$ as
\begin{equation}
	P(x,0)=\sum_n\left(\int_0^L\psi_{l,n}^*(x^\prime)P(x^\prime,0)\mathrm{d}x\right)\psi_{r,n}(x).\label{eq:periodic_general_superposition}
\end{equation}
The corresponding time evolution is given by
\begin{equation}
	P(x,t)=\sum_n e^{-E_nt}\left(\int_0^L\psi_{l,n}^*(x^\prime)P(x^\prime,0)\mathrm{d}x\right)\psi_{r,n}(x).\label{eq:periodic_general_evolution_superposition}
\end{equation}

Therefore we have to investigate the orthogonality and completeness relations of the right and left eigenmodes. To do this, we consider two kinds of boundary conditions: absorbing and reflecting.
\subsection{Absorbing boundary conditions}\label{appedenix:absorbing}
The absorbing boundary conditions $P(0,t)=P(L,t)=0$ give rise to $\psi_{r,n}(0)=\psi_{r,n}(L)=0$ for the right eigenstates. To find the left and right eigenmodes of Eq. \eqref{eq:periodic_general_homogenenous_fokker_planck} simultaneously, we apply a similarity transformation to the right eigenstates: $\psi_{r,n}(x)=\exp(\frac{\mu}{2D}x)\tilde\psi_{r,n}(x)$. Correspondingly, the left eigenstates are transformed by $\psi_{l,n}(x)=\exp(-\frac{\mu}{2D}x)\tilde\psi_{l,n}(x)$. Under the similarity transformation, the boundary conditions for the right eigenstates $\psi_{r,n}(0)=\psi_{r,n}(L)=0$ become $\tilde\psi_{r,n}(0)=\tilde\psi_{r,n}(L)=0$. Now we get a new eigenequation
\begin{equation}
	-D\frac{\partial^2}{\partial x^2}\tilde\psi_{r,n}(x)+\frac{\mu^2}{4D}\tilde\psi_{r,n}(x)=E_n\tilde\psi_{r,n}(x),\quad x\in[0,L].
\end{equation}
Under the boundary conditions $\tilde\psi_{r,n}(0)=\tilde\psi_{r,n}(L)=0$, this eigenequation is nothing but the stationary equation for a massive particle moving in an infinitely deep square well. This is a standard Hermitian problem in an elementary quantum mechanics textbook. Then we can get its left and right eigenmodes immediately:
\begin{equation}
	\tilde\psi_{r,n}(x)=\tilde\psi_{l,n}(x)=\sqrt{\frac{2}{L}}\sin(\kappa_nx),\quad E_n=D\kappa_n^2+\frac{\mu^2}{4D},
\end{equation}
where $\kappa_n=\frac{n\pi}{L}$ with $n=1,2,3,\cdots$. Here, we use the fact that the left and right eigenmodes are coincident in Hermitian systems and satisfy the same boundary conditions. In this sense, after an inverse similarity transformation, we find that the boundary conditions of the left eigenstates $\psi_{l,n}(x)$ in the non-Hermitian problem are also $\psi_{l,n}(0)=\psi_{l,n}(L)=0$. Performing the inverse of the similarity transformation, we get the left and right eigenmodes for Eq. \eqref{eq:periodic_general_homogenenous_fokker_planck} under absorbing boundary conditions:
\begin{equation}
	\begin{split}
\psi_{r,n}(x)&=\sqrt{\frac{2}{L}}e^{\frac{\mu}{2D}x}\sin(\kappa_nx),\\
\psi_{l,n}(x)&=\sqrt{\frac{2}{L}}e^{-\frac{\mu}{2D}x}\sin(\kappa_nx).
\end{split}
\end{equation}
The eigenenergies are given by $E_n=D\kappa_n^2+{\mu^2}/(4D)$. It is easy to check that these eigenstates fulfill the biorthogonal relation:
\begin{equation}
	\int_0^L\psi_{l,m}^*(x)\psi_{r,n}(x)\mathrm{d}x=\frac{2}{L}\int_0^L\sin(\kappa_mx)\sin(\kappa_nx)\mathrm{d}x=\delta_{m,n}.
\end{equation}
Besides, the completeness relation can be proved to be
\begin{equation}
	\begin{split}
	\sum_{n=1}^{+\infty}\psi_{r,n}(x)\psi_{l,n}^*(x^\prime)&=e^{\frac{\mu}{2D}(x-x^\prime)}\frac{2}{L}\sum_{n=1}^{+\infty}\sin(\kappa_nx)\sin(\kappa_nx^\prime)\\
	&=e^{\frac{\mu}{2D}(x-x^\prime)}\frac{1}{L}\sum_{n=-\infty}^{+\infty}\sin(\kappa_nx)\sin(\kappa_nx^\prime).
	\end{split}
\end{equation}
In the above, we use the fact that $\sin(\kappa_nx)\sin(\kappa_nx^\prime)$ is an even function of $\kappa_n$ and $\sin(\kappa_nx)\sin(\kappa_nx^\prime)=0$ for $\kappa_n=0$. Noting that $x,x^\prime\in[0,L]$ and rewriting $\sin(\kappa_nx)\sin(\kappa_nx^\prime)$ as
\begin{equation}
\frac{1}{4}\left(e^{i\kappa_n(x-x^\prime)}+e^{-i\kappa_n(x-x^\prime)}-e^{i\kappa_n(x+x^\prime)}-e^{-i\kappa_n(x+x^\prime)}\right),
\end{equation}
we find that the summation of $\kappa_n=\frac{n\pi}{L}$ with $n\in\mathbb{Z}$ leads to
\begin{equation}
	\sum_{n=1}^{+\infty}\psi_{r,n}(x)\psi_{l,n}^*(x^\prime)=e^{\frac{\mu}{2D}(x-x^\prime)}\delta(x-x^\prime)=\delta(x-x^\prime).
\end{equation}
Here, we use the identity $\frac{1}{2L}\sum_{n=-\infty}^{+\infty}\exp(i\frac{\pi}{L} xn)=\frac{\pi}{L}\delta(\frac{\pi}{L}x)=\delta(x)$.

Therefore we have proved the biorthogonality and completeness relations of the left and right eigenmodes. Together with Eq. \eqref{eq:periodic_general_evolution_superposition}, the time evolution of an initial state $P(x,0)=\delta(x-x_0)$ under absorbing boundary conditions can be expressed as
\begin{equation}
	\begin{split}
		P(x,t)&=\sum_{n=1}^{\infty}e^{-E_nt}\psi_{r,n}(x)\psi^*_{l,n}(x_0)\\
		&=\frac{2}{L}e^{-\frac{\mu^2}{4D}t+\frac{\mu(x-x_0)}{2D}}\sum_{n=1}^{\infty}e^{-D\kappa_n^2t}\sin(\kappa_nx)\sin(\kappa_nx_0),
	\end{split}
\end{equation}
which is the same as Eq. \eqref{eq:fokker_planck_absorbing_evolution} in the main text.
\subsection{Reflecting boundary conditions}
The situation in the reflecting boundary conditions is similar, though the calculation is more tedious. The reflecting boundary conditions $S(0,t)=S(L,t)=0$ with $S(x,t)=(\mu-D\partial_{x})P(x,t)$ provide the following boundary constraints on the right eigenmodes in Eq. \eqref{eq:periodic_general_homogenenous_fokker_planck}:
\begin{equation}
	\left.(\mu-D\partial_{x})\psi_{r,n}(x)\right|_{x=0}=\left.(\mu-D\partial_{x})\psi_{r,n}(x)\right|_{x=L}=0 \label{eq:periodic_general_fokker_planck_reflecting}
\end{equation}
We apply the same similarity transformation to the right and left eigenmodes: $\psi_{r,n}(x)=\exp(\frac{\mu}{2D}x)\tilde\psi_{r,n}(x)$ and $\psi_{l,n}(x)=\exp(-\frac{\mu}{2D}x)\tilde\psi_{l,n}(x)$. Then we can get a new eigenequation
\begin{equation}
		-D\frac{\partial^2}{\partial x^2}\tilde\psi_{r,n}(x)+\frac{\mu^2}{4D}\tilde\psi_{r,n}(x)=E_n\tilde\psi_{r,n}(x),\quad x\in[0,L].
		\label{eq:periodic_general_fokker_planck_reflect_simimar_eigen}
\end{equation}
However, the similarity transformation changes the boundary conditions of these new right eigenmodes. They become
\begin{equation}
	\left.({\mu}/{2}-D\partial_{x})\tilde\psi_{r,n}(x)\right|_{x=0}=\left.({\mu}/{2}-D\partial_{x})\tilde\psi_{r,n}(x)\right|_{x=L}=0.
	\label{eq:periodic_general_fokker_planck_reflect_similar_boundary}
\end{equation}

Before we move on, we prove an important fact that the differential operator $-D\partial_x^2+\frac{\mu^2}{4D}$ in Eq. \eqref{eq:periodic_general_fokker_planck_reflect_simimar_eigen} is Hermitian under the boundary conditions Eq. \eqref{eq:periodic_general_fokker_planck_reflect_similar_boundary}. The real constant part $\frac{\mu^2}{4D}$ is obviously Hermitian. We then calculate the integral
\begin{equation}
	\begin{split}
	\int_0^L\phi_1^*(x)& (-\partial_x^2)\phi_2(x)\mathrm{d}x\\
	&=\left.\int_0^L\left((-\partial_x^2)\phi_1(x)\right)^*\phi_2(x)\mathrm{d}x+g(x)\right|_0^L,		
\end{split}
\end{equation}
where $\phi_1(x),\phi_2(x)$ are two arbitrary functions satisfying Eq. \eqref{eq:periodic_general_fokker_planck_reflect_similar_boundary}. The boundary term is given by $g(x)=\phi_2(x)\partial_x\phi_1(x)-\phi_1(x)\partial_x\phi_2(x)$. Thus, we have
\begin{equation}
	\begin{split}
		g(0)=\phi_2(0)\partial_x\phi_1(0)-\phi_1(0)\partial_x\phi_2(0)=0,\\		g(L)=\phi_2(L)\partial_x\phi_1(L)-\phi_1(L)\partial_x\phi_2(L)=0,\\
	\end{split}
\end{equation}
where we use Eq. \eqref{eq:periodic_general_fokker_planck_reflect_similar_boundary} to get $\partial_x\phi_\alpha(x)|_{x=0,L}=\frac{\mu}{2D}\phi_\alpha(x)|_{x=0,L}$ with $\alpha=1,2$. Therefore we have proved that
\begin{equation}
	\int_0^L\phi_1^*(x)(-\partial_x^2)\phi_2(x)\mathrm{d}x		=\int_0^L\left((-\partial_x^2)\phi_1(x)\right)^*\phi_2(x)\mathrm{d}x
\end{equation}
under the boundary conditions Eq. \eqref{eq:periodic_general_fokker_planck_reflect_similar_boundary}. Namely, the eigenproblem defined in Eq. \eqref{eq:periodic_general_fokker_planck_reflect_simimar_eigen} under the boundary conditions Eq. \eqref{eq:periodic_general_fokker_planck_reflect_similar_boundary} is a Hermitian problem. Thus, the left and right eigenmodes coincide with each other, and the left eigenmodes in this Hermitian problem share the same boundary conditions as the right eigenmodes in Eq.\eqref{eq:periodic_general_fokker_planck_reflect_similar_boundary}:
\begin{equation}
	\left.({\mu}/{2}-D\partial_{x})\tilde\psi_{l,n}(x)\right|_{x=0}=\left.({\mu}/{2}-D\partial_{x})\tilde\psi_{l,n}(x)\right|_{x=L}=0.
	\label{eq:periodic_general_fokker_planck_reflect_similar_boundary_left}
\end{equation}
Now we are in a position to find the eigenstates for Eq. \eqref{eq:periodic_general_fokker_planck_reflect_simimar_eigen}. There is an edge mode with zero eigenvalue
\begin{equation}
		\tilde\psi_{r,0}(x)=\tilde\psi_{l,0}(x)=\left(\frac{\mu/D}{e^{\mu L/D}-1}\right)^{\frac{1}{2}}e^{\frac{\mu}{2D}x},\quad E_0=0.
\end{equation}
Besides, an infinite number of bulk eigenmodes are given by
\begin{equation}
\begin{split}
	\tilde\psi_{r,n}(x)=\tilde\psi_{l,n}(x)=\frac{\sqrt{2}\left[\mu\sin(\kappa_nx)+2D\kappa_n\cos(\kappa_nx)\right]}{\left[L(\mu^2+4D^2\kappa_n^2)\right]^{\frac{1}{2}}}.
\end{split}
\end{equation}

The corresponding bulk eigenvalues are $E_n=D\kappa_n^2+\mu^2/(4D)$ where $\kappa_n=\frac{n\pi}{L}$ with $n=1,2,3,\cdots$. Performing the inverse of the similarity transformation provides the left and right eigenmodes for Eq. \eqref{eq:periodic_general_homogenenous_fokker_planck} under the reflecting boundary conditions Eq. \eqref{eq:periodic_general_fokker_planck_reflecting}. The final results are shown below. The wavefunctions of the zero mode are given by
\begin{equation}
	\begin{split}
		\psi_{r,0}(x)&=\left(\frac{\mu/D}{e^{\mu L/D}-1}\right)^{\frac{1}{2}}e^{\frac{\mu}{D}x},\quad\psi_{l,0}(x) =\left(\frac{\mu/D}{e^{\mu L/D}-1}\right)^{\frac{1}{2}}.\\
\end{split}\label{eq:periodic_general_reflect_zerostates}
\end{equation}
Interestingly, the left eigenmode with zero energy is a constant.  Other bulk eigenstates are 
\begin{equation}
	\begin{split}
		\psi_{r,n}(x)&=e^{\frac{\mu}{2D}x}\frac{\sqrt{2}\left[\mu\sin(\kappa_nx)+2D\kappa_n\cos(\kappa_nx)\right]}{\left[L(\mu^2+4D^2\kappa_n^2)\right]^{\frac{1}{2}}},	\\
        \psi_{l,n}(x)&=e^{-\frac{\mu}{2D}x}\frac{\sqrt{2}\left[\mu\sin(\kappa_nx)+2D\kappa_n\cos(\kappa_nx)\right]}{\left[L(\mu^2+4D^2\kappa_n^2)\right]^{\frac{1}{2}}}.
	\end{split}\label{eq:periodic_general_reflect_eigenstates}
\end{equation}
The right eigenstates are easy to check by substituting them back to the eigenequation Eq. \eqref{eq:periodic_general_homogenenous_fokker_planck} under the reflecting boundary conditions Eq. \eqref{eq:periodic_general_fokker_planck_reflecting}. Given that $\tilde \psi_{l,n}(x)$ in the Hermitian problem satisfy the boundary conditions in Eq. \eqref{eq:periodic_general_fokker_planck_reflect_similar_boundary_left}, the inverse similarity transformation shows that the left eigenstates $\psi_{l,n}(x)$ in the original non-Hermitian eigenproblem satisfy the boundary conditions 
\begin{equation}\label{eq:periodic_general_fokker_planck_reflect_boundary_left}
\partial_x\psi_{l,n}(x)|_{x=0}=\partial_x\psi_{l,n}(x)|_{x=L}=0.\end{equation} 
It is  straightforward to check that the left eigenstates in Eq.\eqref{eq:periodic_general_reflect_zerostates} and Eq. \eqref{eq:periodic_general_reflect_eigenstates} fulfill the boundary conditions in Eq.\eqref{eq:periodic_general_fokker_planck_reflect_boundary_left}.

To further confirm our results, we show that the eigenmodes in Eq. \eqref{eq:periodic_general_reflect_zerostates} and Eq. \eqref{eq:periodic_general_reflect_eigenstates} satisfy the biorthogonality and completeness relations. First, we check that \begin{equation}
\int_{0}^L\psi_{l,0}^*(x)\psi_{r,0}(x)\mathrm{d}x=\frac{\mu/D}{e^{\mu L/D}-1}\int_0^Le^{\frac{\mu}{D}x}\mathrm{d}x=1. \label{eq:periodic_general_reflect_biorth_1}
\end{equation}
Then, we calculate
\begin{widetext}
\begin{equation}
	\begin{split}
		\int_{0}^L\psi_{l,0}^*(x)\psi_{r,n}(x)\mathrm{d}x&=\left(\frac{2\mu/D}{L(e^{\mu L/D}-1)(\mu^2+4D^2\kappa_n^2)}\right)^{\frac{1}{2}}\int_0^Le^{\frac{\mu}{2D}x}(\mu\sin(\kappa_nx)+2D\kappa_n\cos(\kappa_nx))\mathrm{d}x\\
		&=\left(\frac{2\mu/D}{L(e^{\mu L/D}-1)(\mu^2+4D^2\kappa_n^2)}\right)^{\frac{1}{2}}\int_0^Le^{\frac{\mu}{2D}x}(\mu+2D\frac{\partial}{\partial x})\sin(\kappa_nx)\mathrm{d}x\\
		&=\left(\frac{2\mu/D}{L(e^{\mu L/D}-1)(\mu^2+4D^2\kappa_n^2)}\right)^{\frac{1}{2}}\int_0^L\sin(\kappa_nx)(\mu-2D\frac{\partial}{\partial x})e^{\frac{\mu}{2D}x}\mathrm{d}x\\
		&=0.
	\end{split}\label{eq:periodic_general_reflect_biorth_2}
\end{equation}
In the above, we use the integration by parts, and the boundary term $e^{\frac{\mu}{2D}x}\sin(\kappa_nx)|_0^L=0$ should vanish as $\kappa_n=\frac{n\pi}{L}$. Similarly, we get
\begin{equation}
		\int_{0}^L\psi_{r,0}^*(x)\psi_{l,n}(x)\mathrm{d}x=\left(\frac{2\mu/D}{L(e^{\mu L/D}-1)(\mu^2+4D^2\kappa_n^2)}\right)^{\frac{1}{2}}\int_0^Le^{\frac{\mu}{2D}x}(\mu\sin(\kappa_nx)+2D\kappa_n\cos(\kappa_nx))\mathrm{d}x=0.\label{eq:periodic_general_reflect_biorth_3}
\end{equation}
Finally, we have to evaluate $\int_{0}^L\psi_{r,m}^*(x)\psi_{l,n}(x)\mathrm{d}x$ for $m,n=1,2,3,\cdots$. Then we get
\begin{equation}
	\begin{split}
		\int_{0}^L\psi_{r,m}^*(x)\psi_{l,n}(x)\mathrm{d}x&=\frac{2}{L(\mu^2+4D\kappa_m^2)^{\frac{1}{2}}(\mu^2+4D\kappa_n^2)^{\frac{1}{2}}}\int_0^L(\mu\sin(\kappa_mx)+2D\kappa_m\cos(\kappa_mx))(\mu\sin(\kappa_nx)+2D\kappa_n\cos(\kappa_nx))\mathrm{d}x
	\end{split}
\end{equation}
It is easy to check that
\begin{equation}
	\frac{2}{L}\int_0^L\sin(\kappa_mx)\sin(\kappa_nx)=	\frac{2}{L}\int_0^L\cos(\kappa_mx)\cos(\kappa_nx)=\delta_{m,n},\quad 	\frac{2}{L}\int_0^L\sin(\kappa_mx)\cos(\kappa_nx)=	\frac{2}{L}\int_0^L\cos(\kappa_mx)\sin(\kappa_nx)=0.
\end{equation}
Therefore we can obtain
\begin{equation}
		\int_{0}^L\psi_{r,m}^*(x)\psi_{l,n}(x)\mathrm{d}x=\frac{\mu^2+4D^2\kappa_m\kappa_n}{(\mu^2+4D\kappa_m^2)^{\frac{1}{2}}(\mu^2+4D\kappa_n^2)^{\frac{1}{2}}}\delta_{m,n}=\delta_{m,n}.\label{eq:periodic_general_reflect_biorth_4}
\end{equation}

Together with Eq. \eqref{eq:periodic_general_reflect_biorth_1}, Eq. \eqref{eq:periodic_general_reflect_biorth_2}, Eq. \eqref{eq:periodic_general_reflect_biorth_3}, and Eq. \eqref{eq:periodic_general_reflect_biorth_4}, we have proved the biorthogonality relation of the right and left eigenstates presented in Eq. \eqref{eq:periodic_general_reflect_zerostates} and Eq. \eqref{eq:periodic_general_reflect_eigenstates}. Then we prove the completeness relation. We can calculate
\begin{equation}
	\begin{split}
		\psi_{r,0}(x)\psi_{l,0}^*(x^\prime)&+\sum_{n=1}^{+\infty}\psi_{r,n}(x)\psi_{l,n}^*(x^\prime)=\frac{\mu/D}{e^{\mu L/D}-1}e^{\frac{\mu}{D}x}+e^{\frac{\mu}{2D}(x-x^\prime)}\frac{2}{L} \sum_{n=1}^{+\infty}\frac{[\mu\sin(\kappa_nx)+2D\kappa_n\cos(\kappa_nx)][\mu\sin(\kappa_nx^\prime)+2D\kappa_n\cos(\kappa_nx^\prime)]}{\mu^2+4D^2\kappa_n^2}\\
		=&\frac{\mu/D}{e^{\mu L/D}-1}e^{\frac{\mu}{D}x}+e^{\frac{\mu}{2D}(x-x^\prime)}\frac{1}{L} \sum_{n=-\infty}^{+\infty}\frac{[\mu\sin(\kappa_nx)+2D\kappa_n\cos(\kappa_nx)][\mu\sin(\kappa_nx^\prime)+2D\kappa_n\cos(\kappa_nx^\prime)]}{\mu^2+4D^2\kappa_n^2}\\
		=&\frac{\mu/D}{e^{\mu L/D}-1}e^{\frac{\mu}{D}x}+e^{\frac{\mu}{2D}(x-x^\prime)}\frac{1}{4L} \sum_{n=-\infty}^{+\infty}\left( \frac{2D\kappa_n-i\mu}{2D\kappa_n+i\mu}e^{i\kappa_n(x+x^\prime)}+\frac{2D\kappa_n+i\mu}{2D\kappa_n-i\mu}e^{-i\kappa_n(x+x^\prime)}+ e^{i\kappa_n(x-x^\prime)}+e^{-i\kappa_n(x-x^\prime)}\right)\\
		=&\frac{\mu/D}{e^{\mu L/D}-1}e^{\frac{\mu}{D}x}+e^{\frac{\mu}{2D}(x-x^\prime)}\frac{1}{2L} \sum_{n=-\infty}^{+\infty}\left( \frac{2D\kappa_n-i\mu}{2D\kappa_n+i\mu}e^{i\kappa_n(x+x^\prime)}+e^{i\kappa_n(x-x^\prime)}\right)\\
		=&e^{\frac{\mu}{2D}(x-x^\prime)}\delta(x-x^\prime)+\frac{\mu/D}{e^{\mu L/D}-1}e^{\frac{\mu}{D}x}+e^{\frac{\mu}{2D}(x-x^\prime)}\frac{1}{2L} \sum_{n=-\infty}^{+\infty} \frac{2D\kappa_n-i\mu}{2D\kappa_n+i\mu}e^{i\kappa_n(x+x^\prime)}\\	=&\delta(x-x^\prime)+\frac{\mu/D}{e^{\mu L/D}-1}e^{\frac{\mu}{D}x}+e^{\frac{\mu}{2D}(x-x^\prime)}\frac{1}{2L} \sum_{n=-\infty}^{+\infty} \frac{2D\kappa_n-i\mu}{2D\kappa_n+i\mu}e^{i\kappa_n(x+x^\prime)}.
		\end{split}\label{eq:periodic_general_reflect_complete_1}
\end{equation}
\end{widetext}
In the second line, we use the fact that $[\mu\sin(\kappa_nx)+2D\kappa_n\cos(\kappa_nx)][\mu\sin(\kappa_nx^\prime)+2D\kappa_n\cos(\kappa_nx^\prime)]$ is an even function of $\kappa_n$ and vanishes at $\kappa_n=0$. In the fourth line, we combine the same terms under the symmetry $\kappa_n\to-\kappa_n$ within the summation. In the fifth line, we use the expression $\frac{1}{2L}\sum_{n=-\infty}^{+\infty}\exp(i\frac{\pi}{L} xn)=\frac{\pi}{L}\delta(\frac{\pi}{L}x)=\delta(x)$. The last term can be evaluated by  the Matsubara frequency summation:
\begin{equation}
	\begin{split}
	 \sum_{n=-\infty}^{+\infty} \frac{2D\kappa_n-i\mu}{2D\kappa_n+i\mu}&e^{i\kappa_n(x+x^\prime)}=	\sum_{n=-\infty}^{+\infty} \frac{2i\pi nD+\mu L}{2i\pi nD-\mu L}e^{2i\pi n\frac{x+x^\prime}{2L}}\\
	&= \int_{C_0}\frac{dz}{2\pi i}\frac{e^{z\frac{x+x'}{2L}}}{e^{z}-1}\frac{Dz+\mu L}{Dz-\mu L},
	\end{split}
\end{equation}
where $C_0$ is the integral contour shown in Fig. \ref{apfig:contour}. Since $x,x^\prime\in(0,L)$, we have $\frac{x+x^\prime}{2L}\in(0,1)$. Then $\frac{1}{e^z-1}e^{z\frac{x+x'}{2L}}$ converges to zero at $|z|\to\infty$. As a result, we can deform the integral contour to $C_1$ in Fig. \ref{apfig:contour} to get the residue at $z_0=\frac{\mu L}{D}$:
\begin{equation}
	\begin{split}
	\int_{C_0}\frac{dz}{2\pi i}\frac{e^{z\frac{x+x'}{2L}}}{e^{z}-1}\frac{Dz+\mu L}{Dz-\mu L}&=\int_{C_1}\frac{dz}{2\pi i}\frac{e^{z\frac{x+x'}{2L}}}{e^{z}-1}\frac{Dz+\mu L}{Dz-\mu L}\\
	&=-\frac{2L\mu/D}{e^{\mu L/D}-1}e^{\frac{\mu}{2D}(x+x^\prime)}.
	\end{split}
\end{equation}
Substituting this result into Eq. \eqref{eq:periodic_general_reflect_complete_1}, we finish the proof of completeness:
\begin{equation}
	\begin{split}
		\psi_{r,0}(x)\psi_{l,0}^*(x^\prime)+\sum_{n=1}^{+\infty}\psi_{r,n}(x)\psi_{l,n}^*(x^\prime)=\delta(x-x^\prime).
	\end{split}\label{eq:periodic_general_reflect_complete_2}
\end{equation}

Therefore we have proved the biorthogonality and completeness relations of the left and right eigenmodes in Eq. \eqref{eq:periodic_general_reflect_zerostates} and Eq. \eqref{eq:periodic_general_reflect_eigenstates}. Under this circumstance, together with Eq. \eqref{eq:periodic_general_evolution_superposition}, the time evolution of the initial state $P(x,0)=\delta(x-x_0)$ under the reflecting boundary conditions can be expressed as
\begin{equation}
	\begin{split}
		P(x,t)&=\psi_{r,0}(x)\psi^*_{l,0}(x_0)+\sum_{n=1}^{\infty}e^{-E_nt}\psi_{r,n}(x)\psi^*_{l,n}(x_0)\\
		&=\frac{\mu}{D}\frac{ e^{\frac{\mu x}{D}}}{e^{\frac{\mu L}{D}}-1}+\frac{2}{L}e^{-\frac{\mu^2}{4D}t+\frac{\mu(x-x_0)}{2D}}\sum_{n=1}^{\infty}\frac{e^{-D\kappa_n^2t}F_n(x)F_n(x_0)}{\mu^2+4\kappa_n^2D^2}.
	\end{split}
\end{equation}
which is the same as Eq. \eqref{eq:fokker_planck_reflecting_evolution} in the main text with $F_n(x)=\mu\sin(\kappa_nx)+2D\kappa_n\cos(\kappa_nx)$.
\begin{figure}[t]
	\centering
	\includegraphics*[width=5cm]{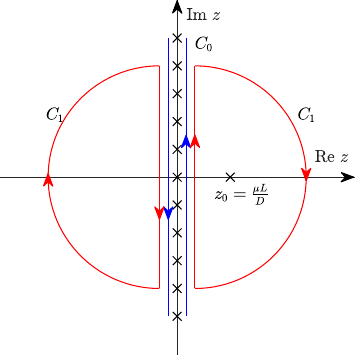}
	\caption{Poles and contours in the Matsubara frequency summation.}\label{apfig:contour}
\end{figure}

\bibliography{ref_continuum}

\begin{thebibliography}{85}%
\makeatletter
\providecommand \@ifxundefined [1]{%
 \@ifx{#1\undefined}
}%
\providecommand \@ifnum [1]{%
 \ifnum #1\expandafter \@firstoftwo
 \else \expandafter \@secondoftwo
 \fi
}%
\providecommand \@ifx [1]{%
 \ifx #1\expandafter \@firstoftwo
 \else \expandafter \@secondoftwo
 \fi
}%
\providecommand \natexlab [1]{#1}%
\providecommand \enquote  [1]{``#1''}%
\providecommand \bibnamefont  [1]{#1}%
\providecommand \bibfnamefont [1]{#1}%
\providecommand \citenamefont [1]{#1}%
\providecommand \href@noop [0]{\@secondoftwo}%
\providecommand \href [0]{\begingroup \@sanitize@url \@href}%
\providecommand \@href[1]{\@@startlink{#1}\@@href}%
\providecommand \@@href[1]{\endgroup#1\@@endlink}%
\providecommand \@sanitize@url [0]{\catcode `\\12\catcode `\$12\catcode
  `\&12\catcode `\#12\catcode `\^12\catcode `\_12\catcode `\%12\relax}%
\providecommand \@@startlink[1]{}%
\providecommand \@@endlink[0]{}%
\providecommand \url  [0]{\begingroup\@sanitize@url \@url }%
\providecommand \@url [1]{\endgroup\@href {#1}{\urlprefix }}%
\providecommand \urlprefix  [0]{URL }%
\providecommand \Eprint [0]{\href }%
\providecommand \doibase [0]{http://dx.doi.org/}%
\providecommand \selectlanguage [0]{\@gobble}%
\providecommand \bibinfo  [0]{\@secondoftwo}%
\providecommand \bibfield  [0]{\@secondoftwo}%
\providecommand \translation [1]{[#1]}%
\providecommand \BibitemOpen [0]{}%
\providecommand \bibitemStop [0]{}%
\providecommand \bibitemNoStop [0]{.\EOS\space}%
\providecommand \EOS [0]{\spacefactor3000\relax}%
\providecommand \BibitemShut  [1]{\csname bibitem#1\endcsname}%
\let\auto@bib@innerbib\@empty
\bibitem [{\citenamefont {Ashida}\ \emph {et~al.}(2020)\citenamefont {Ashida},
  \citenamefont {Gong},\ and\ \citenamefont {Ueda}}]{ashida2020non}%
  \BibitemOpen
  \bibfield  {author} {\bibinfo {author} {\bibfnamefont {Yuto}\ \bibnamefont
  {Ashida}}, \bibinfo {author} {\bibfnamefont {Zongping}\ \bibnamefont {Gong}},
  \ and\ \bibinfo {author} {\bibfnamefont {Masahito}\ \bibnamefont {Ueda}},\
  }\bibfield  {title} {\enquote {\bibinfo {title} {Non-hermitian physics},}\
  }\href@noop {} {\bibfield  {journal} {\bibinfo  {journal} {Advances in
  Physics}\ }\textbf {\bibinfo {volume} {69}},\ \bibinfo {pages} {249--435}
  (\bibinfo {year} {2020})}\BibitemShut {NoStop}%
\bibitem [{\citenamefont {Bergholtz}\ \emph {et~al.}(2021)\citenamefont
  {Bergholtz}, \citenamefont {Budich},\ and\ \citenamefont
  {Kunst}}]{bergholtz2021exceptional}%
  \BibitemOpen
  \bibfield  {author} {\bibinfo {author} {\bibfnamefont {Emil~J}\ \bibnamefont
  {Bergholtz}}, \bibinfo {author} {\bibfnamefont {Jan~Carl}\ \bibnamefont
  {Budich}}, \ and\ \bibinfo {author} {\bibfnamefont {Flore~K}\ \bibnamefont
  {Kunst}},\ }\bibfield  {title} {\enquote {\bibinfo {title} {Exceptional
  topology of non-hermitian systems},}\ }\href@noop {} {\bibfield  {journal}
  {\bibinfo  {journal} {Reviews of Modern Physics}\ }\textbf {\bibinfo {volume}
  {93}},\ \bibinfo {pages} {015005} (\bibinfo {year} {2021})}\BibitemShut
  {NoStop}%
\bibitem [{\citenamefont {Yao}\ and\ \citenamefont {Wang}(2018)}]{yao2018edge}%
  \BibitemOpen
  \bibfield  {author} {\bibinfo {author} {\bibfnamefont {Shunyu}\ \bibnamefont
  {Yao}}\ and\ \bibinfo {author} {\bibfnamefont {Zhong}\ \bibnamefont {Wang}},\
  }\bibfield  {title} {\enquote {\bibinfo {title} {Edge states and topological
  invariants of non-hermitian systems},}\ }\href {\doibase
  10.1103/PhysRevLett.121.086803} {\bibfield  {journal} {\bibinfo  {journal}
  {Phys. Rev. Lett.}\ }\textbf {\bibinfo {volume} {121}},\ \bibinfo {pages}
  {086803} (\bibinfo {year} {2018})}\BibitemShut {NoStop}%
\bibitem [{\citenamefont {Yao}\ \emph {et~al.}(2018)\citenamefont {Yao},
  \citenamefont {Song},\ and\ \citenamefont {Wang}}]{yao2018chern}%
  \BibitemOpen
  \bibfield  {author} {\bibinfo {author} {\bibfnamefont {Shunyu}\ \bibnamefont
  {Yao}}, \bibinfo {author} {\bibfnamefont {Fei}\ \bibnamefont {Song}}, \ and\
  \bibinfo {author} {\bibfnamefont {Zhong}\ \bibnamefont {Wang}},\ }\bibfield
  {title} {\enquote {\bibinfo {title} {Non-hermitian chern bands},}\ }\href
  {\doibase 10.1103/PhysRevLett.121.136802} {\bibfield  {journal} {\bibinfo
  {journal} {Phys. Rev. Lett.}\ }\textbf {\bibinfo {volume} {121}},\ \bibinfo
  {pages} {136802} (\bibinfo {year} {2018})}\BibitemShut {NoStop}%
\bibitem [{\citenamefont {Kunst}\ \emph {et~al.}(2018)\citenamefont {Kunst},
  \citenamefont {Edvardsson}, \citenamefont {Budich},\ and\ \citenamefont
  {Bergholtz}}]{kunst2018biorthogonal}%
  \BibitemOpen
  \bibfield  {author} {\bibinfo {author} {\bibfnamefont {Flore~K.}\
  \bibnamefont {Kunst}}, \bibinfo {author} {\bibfnamefont {Elisabet}\
  \bibnamefont {Edvardsson}}, \bibinfo {author} {\bibfnamefont {Jan~Carl}\
  \bibnamefont {Budich}}, \ and\ \bibinfo {author} {\bibfnamefont {Emil~J.}\
  \bibnamefont {Bergholtz}},\ }\bibfield  {title} {\enquote {\bibinfo {title}
  {Biorthogonal bulk-boundary correspondence in non-hermitian systems},}\
  }\href {\doibase 10.1103/PhysRevLett.121.026808} {\bibfield  {journal}
  {\bibinfo  {journal} {Phys. Rev. Lett.}\ }\textbf {\bibinfo {volume} {121}},\
  \bibinfo {pages} {026808} (\bibinfo {year} {2018})}\BibitemShut {NoStop}%
\bibitem [{\citenamefont {Lee}\ and\ \citenamefont
  {Thomale}(2019)}]{lee2019anatomy}%
  \BibitemOpen
  \bibfield  {author} {\bibinfo {author} {\bibfnamefont {Ching~Hua}\
  \bibnamefont {Lee}}\ and\ \bibinfo {author} {\bibfnamefont {Ronny}\
  \bibnamefont {Thomale}},\ }\bibfield  {title} {\enquote {\bibinfo {title}
  {Anatomy of skin modes and topology in non-hermitian systems},}\ }\href
  {\doibase 10.1103/PhysRevB.99.201103} {\bibfield  {journal} {\bibinfo
  {journal} {Phys. Rev. B}\ }\textbf {\bibinfo {volume} {99}},\ \bibinfo
  {pages} {201103} (\bibinfo {year} {2019})}\BibitemShut {NoStop}%
\bibitem [{\citenamefont {Martinez~Alvarez}\ \emph {et~al.}(2018)\citenamefont
  {Martinez~Alvarez}, \citenamefont {Barrios~Vargas},\ and\ \citenamefont
  {Foa~Torres}}]{martinez2018non}%
  \BibitemOpen
  \bibfield  {author} {\bibinfo {author} {\bibfnamefont {V.~M.}\ \bibnamefont
  {Martinez~Alvarez}}, \bibinfo {author} {\bibfnamefont {J.~E.}\ \bibnamefont
  {Barrios~Vargas}}, \ and\ \bibinfo {author} {\bibfnamefont {L.~E.~F.}\
  \bibnamefont {Foa~Torres}},\ }\bibfield  {title} {\enquote {\bibinfo {title}
  {Non-hermitian robust edge states in one dimension: Anomalous localization
  and eigenspace condensation at exceptional points},}\ }\href {\doibase
  10.1103/PhysRevB.97.121401} {\bibfield  {journal} {\bibinfo  {journal} {Phys.
  Rev. B}\ }\textbf {\bibinfo {volume} {97}},\ \bibinfo {pages} {121401}
  (\bibinfo {year} {2018})}\BibitemShut {NoStop}%
\bibitem [{\citenamefont {Zhang}\ \emph
  {et~al.}(2022{\natexlab{a}})\citenamefont {Zhang}, \citenamefont {Zhang},
  \citenamefont {Lu},\ and\ \citenamefont {Chen}}]{zhang2022review}%
  \BibitemOpen
  \bibfield  {author} {\bibinfo {author} {\bibfnamefont {Xiujuan}\ \bibnamefont
  {Zhang}}, \bibinfo {author} {\bibfnamefont {Tian}\ \bibnamefont {Zhang}},
  \bibinfo {author} {\bibfnamefont {Ming-Hui}\ \bibnamefont {Lu}}, \ and\
  \bibinfo {author} {\bibfnamefont {Yan-Feng}\ \bibnamefont {Chen}},\
  }\bibfield  {title} {\enquote {\bibinfo {title} {A review on non-hermitian
  skin effect},}\ }\href@noop {} {\bibfield  {journal} {\bibinfo  {journal}
  {Advances in Physics: X}\ }\textbf {\bibinfo {volume} {7}},\ \bibinfo {pages}
  {2109431} (\bibinfo {year} {2022}{\natexlab{a}})}\BibitemShut {NoStop}%
\bibitem [{\citenamefont {Okuma}\ and\ \citenamefont
  {Sato}(2023)}]{okuma2023non}%
  \BibitemOpen
  \bibfield  {author} {\bibinfo {author} {\bibfnamefont {Nobuyuki}\
  \bibnamefont {Okuma}}\ and\ \bibinfo {author} {\bibfnamefont {Masatoshi}\
  \bibnamefont {Sato}},\ }\bibfield  {title} {\enquote {\bibinfo {title}
  {Non-hermitian topological phenomena: A review},}\ }\href@noop {} {\bibfield
  {journal} {\bibinfo  {journal} {Annual Review of Condensed Matter Physics}\
  }\textbf {\bibinfo {volume} {14}},\ \bibinfo {pages} {83--107} (\bibinfo
  {year} {2023})}\BibitemShut {NoStop}%
\bibitem [{\citenamefont {Lin}\ \emph {et~al.}(2023)\citenamefont {Lin},
  \citenamefont {Tai}, \citenamefont {Li},\ and\ \citenamefont
  {Lee}}]{lin2023topological}%
  \BibitemOpen
  \bibfield  {author} {\bibinfo {author} {\bibfnamefont {Rijia}\ \bibnamefont
  {Lin}}, \bibinfo {author} {\bibfnamefont {Tommy}\ \bibnamefont {Tai}},
  \bibinfo {author} {\bibfnamefont {Linhu}\ \bibnamefont {Li}}, \ and\ \bibinfo
  {author} {\bibfnamefont {Ching~Hua}\ \bibnamefont {Lee}},\ }\bibfield
  {title} {\enquote {\bibinfo {title} {Topological non-hermitian skin
  effect},}\ }\href@noop {} {\bibfield  {journal} {\bibinfo  {journal}
  {Frontiers of Physics}\ }\textbf {\bibinfo {volume} {18}},\ \bibinfo {pages}
  {53605} (\bibinfo {year} {2023})}\BibitemShut {NoStop}%
\bibitem [{\citenamefont {Ding}\ \emph {et~al.}(2022)\citenamefont {Ding},
  \citenamefont {Fang},\ and\ \citenamefont {Ma}}]{ding2022non}%
  \BibitemOpen
  \bibfield  {author} {\bibinfo {author} {\bibfnamefont {Kun}\ \bibnamefont
  {Ding}}, \bibinfo {author} {\bibfnamefont {Chen}\ \bibnamefont {Fang}}, \
  and\ \bibinfo {author} {\bibfnamefont {Guancong}\ \bibnamefont {Ma}},\
  }\bibfield  {title} {\enquote {\bibinfo {title} {Non-hermitian topology and
  exceptional-point geometries},}\ }\href@noop {} {\bibfield  {journal}
  {\bibinfo  {journal} {Nature Reviews Physics}\ }\textbf {\bibinfo {volume}
  {4}},\ \bibinfo {pages} {745--760} (\bibinfo {year} {2022})}\BibitemShut
  {NoStop}%
\bibitem [{\citenamefont {Yokomizo}\ and\ \citenamefont
  {Murakami}(2019)}]{yokomizo2019nonbloch}%
  \BibitemOpen
  \bibfield  {author} {\bibinfo {author} {\bibfnamefont {Kazuki}\ \bibnamefont
  {Yokomizo}}\ and\ \bibinfo {author} {\bibfnamefont {Shuichi}\ \bibnamefont
  {Murakami}},\ }\bibfield  {title} {\enquote {\bibinfo {title} {Non-bloch band
  theory of non-hermitian systems},}\ }\href {\doibase
  10.1103/PhysRevLett.123.066404} {\bibfield  {journal} {\bibinfo  {journal}
  {Phys. Rev. Lett.}\ }\textbf {\bibinfo {volume} {123}},\ \bibinfo {pages}
  {066404} (\bibinfo {year} {2019})}\BibitemShut {NoStop}%
\bibitem [{\citenamefont {Zhang}\ \emph {et~al.}(2020)\citenamefont {Zhang},
  \citenamefont {Yang},\ and\ \citenamefont {Fang}}]{zhang2020correspondence}%
  \BibitemOpen
  \bibfield  {author} {\bibinfo {author} {\bibfnamefont {Kai}\ \bibnamefont
  {Zhang}}, \bibinfo {author} {\bibfnamefont {Zhesen}\ \bibnamefont {Yang}}, \
  and\ \bibinfo {author} {\bibfnamefont {Chen}\ \bibnamefont {Fang}},\
  }\bibfield  {title} {\enquote {\bibinfo {title} {Correspondence between
  winding numbers and skin modes in non-hermitian systems},}\ }\href {\doibase
  10.1103/PhysRevLett.125.126402} {\bibfield  {journal} {\bibinfo  {journal}
  {Phys. Rev. Lett.}\ }\textbf {\bibinfo {volume} {125}},\ \bibinfo {pages}
  {126402} (\bibinfo {year} {2020})}\BibitemShut {NoStop}%
\bibitem [{\citenamefont {Xiao}\ \emph {et~al.}(2020)\citenamefont {Xiao},
  \citenamefont {Deng}, \citenamefont {Wang}, \citenamefont {Zhu},
  \citenamefont {Wang}, \citenamefont {Yi},\ and\ \citenamefont
  {Xue}}]{xiao2020non}%
  \BibitemOpen
  \bibfield  {author} {\bibinfo {author} {\bibfnamefont {Lei}\ \bibnamefont
  {Xiao}}, \bibinfo {author} {\bibfnamefont {Tianshu}\ \bibnamefont {Deng}},
  \bibinfo {author} {\bibfnamefont {Kunkun}\ \bibnamefont {Wang}}, \bibinfo
  {author} {\bibfnamefont {Gaoyan}\ \bibnamefont {Zhu}}, \bibinfo {author}
  {\bibfnamefont {Zhong}\ \bibnamefont {Wang}}, \bibinfo {author}
  {\bibfnamefont {Wei}\ \bibnamefont {Yi}}, \ and\ \bibinfo {author}
  {\bibfnamefont {Peng}\ \bibnamefont {Xue}},\ }\bibfield  {title} {\enquote
  {\bibinfo {title} {Non-{Hermitian} bulk--boundary correspondence in quantum
  dynamics},}\ }\href {\doibase 10.1038/s41567-020-0836-6} {\bibfield
  {journal} {\bibinfo  {journal} {Nature Physics}\ }\textbf {\bibinfo {volume}
  {16}},\ \bibinfo {pages} {761--766} (\bibinfo {year} {2020})}\BibitemShut
  {NoStop}%
\bibitem [{\citenamefont {Ghatak}\ \emph {et~al.}(2020)\citenamefont {Ghatak},
  \citenamefont {Brandenbourger}, \citenamefont {van Wezel},\ and\
  \citenamefont {Coulais}}]{ghatak2020observation}%
  \BibitemOpen
  \bibfield  {author} {\bibinfo {author} {\bibfnamefont {Ananya}\ \bibnamefont
  {Ghatak}}, \bibinfo {author} {\bibfnamefont {Martin}\ \bibnamefont
  {Brandenbourger}}, \bibinfo {author} {\bibfnamefont {Jasper}\ \bibnamefont
  {van Wezel}}, \ and\ \bibinfo {author} {\bibfnamefont {Corentin}\
  \bibnamefont {Coulais}},\ }\bibfield  {title} {\enquote {\bibinfo {title}
  {Observation of non-{Hermitian} topology and its bulk--edge correspondence in
  an active mechanical metamaterial},}\ }\href {\doibase
  https://doi.org/10.1073/pnas.2010580117} {\bibfield  {journal} {\bibinfo
  {journal} {Proceedings of the National Academy of Sciences}\ }\textbf
  {\bibinfo {volume} {117}},\ \bibinfo {pages} {29561--29568} (\bibinfo {year}
  {2020})}\BibitemShut {NoStop}%
\bibitem [{\citenamefont {Helbig}\ \emph {et~al.}(2020)\citenamefont {Helbig},
  \citenamefont {Hofmann}, \citenamefont {Imhof}, \citenamefont {Abdelghany},
  \citenamefont {Kiessling}, \citenamefont {Molenkamp}, \citenamefont {Lee},
  \citenamefont {Szameit}, \citenamefont {Greiter},\ and\ \citenamefont
  {Thomale}}]{helbig2020generalized}%
  \BibitemOpen
  \bibfield  {author} {\bibinfo {author} {\bibfnamefont {T}~\bibnamefont
  {Helbig}}, \bibinfo {author} {\bibfnamefont {T}~\bibnamefont {Hofmann}},
  \bibinfo {author} {\bibfnamefont {S}~\bibnamefont {Imhof}}, \bibinfo {author}
  {\bibfnamefont {M}~\bibnamefont {Abdelghany}}, \bibinfo {author}
  {\bibfnamefont {T}~\bibnamefont {Kiessling}}, \bibinfo {author}
  {\bibfnamefont {LW}~\bibnamefont {Molenkamp}}, \bibinfo {author}
  {\bibfnamefont {CH}~\bibnamefont {Lee}}, \bibinfo {author} {\bibfnamefont
  {A}~\bibnamefont {Szameit}}, \bibinfo {author} {\bibfnamefont
  {M}~\bibnamefont {Greiter}}, \ and\ \bibinfo {author} {\bibfnamefont
  {R}~\bibnamefont {Thomale}},\ }\bibfield  {title} {\enquote {\bibinfo {title}
  {Generalized bulk--boundary correspondence in non-{Hermitian} topolectrical
  circuits},}\ }\href {\doibase https://doi.org/10.1038/s41567-020-0922-9}
  {\bibfield  {journal} {\bibinfo  {journal} {Nature Physics}\ }\textbf
  {\bibinfo {volume} {16}},\ \bibinfo {pages} {747--750} (\bibinfo {year}
  {2020})}\BibitemShut {NoStop}%
\bibitem [{\citenamefont {Weidemann}\ \emph {et~al.}(2020)\citenamefont
  {Weidemann}, \citenamefont {Kremer}, \citenamefont {Helbig}, \citenamefont
  {Hofmann}, \citenamefont {Stegmaier}, \citenamefont {Greiter}, \citenamefont
  {Thomale},\ and\ \citenamefont {Szameit}}]{weidemann2020topological}%
  \BibitemOpen
  \bibfield  {author} {\bibinfo {author} {\bibfnamefont {Sebastian}\
  \bibnamefont {Weidemann}}, \bibinfo {author} {\bibfnamefont {Mark}\
  \bibnamefont {Kremer}}, \bibinfo {author} {\bibfnamefont {Tobias}\
  \bibnamefont {Helbig}}, \bibinfo {author} {\bibfnamefont {Tobias}\
  \bibnamefont {Hofmann}}, \bibinfo {author} {\bibfnamefont {Alexander}\
  \bibnamefont {Stegmaier}}, \bibinfo {author} {\bibfnamefont {Martin}\
  \bibnamefont {Greiter}}, \bibinfo {author} {\bibfnamefont {Ronny}\
  \bibnamefont {Thomale}}, \ and\ \bibinfo {author} {\bibfnamefont {Alexander}\
  \bibnamefont {Szameit}},\ }\bibfield  {title} {\enquote {\bibinfo {title}
  {Topological funneling of light},}\ }\href {\doibase 10.1126/science.aaz8727}
  {\bibfield  {journal} {\bibinfo  {journal} {Science}\ }\textbf {\bibinfo
  {volume} {368}},\ \bibinfo {pages} {311--314} (\bibinfo {year}
  {2020})}\BibitemShut {NoStop}%
\bibitem [{\citenamefont {Xue}\ \emph {et~al.}(2021)\citenamefont {Xue},
  \citenamefont {Li}, \citenamefont {Hu}, \citenamefont {Song},\ and\
  \citenamefont {Wang}}]{xue2021simple}%
  \BibitemOpen
  \bibfield  {author} {\bibinfo {author} {\bibfnamefont {Wen-Tan}\ \bibnamefont
  {Xue}}, \bibinfo {author} {\bibfnamefont {Ming-Rui}\ \bibnamefont {Li}},
  \bibinfo {author} {\bibfnamefont {Yu-Min}\ \bibnamefont {Hu}}, \bibinfo
  {author} {\bibfnamefont {Fei}\ \bibnamefont {Song}}, \ and\ \bibinfo {author}
  {\bibfnamefont {Zhong}\ \bibnamefont {Wang}},\ }\bibfield  {title} {\enquote
  {\bibinfo {title} {Simple formulas of directional amplification from
  non-bloch band theory},}\ }\href {\doibase 10.1103/PhysRevB.103.L241408}
  {\bibfield  {journal} {\bibinfo  {journal} {Phys. Rev. B}\ }\textbf {\bibinfo
  {volume} {103}},\ \bibinfo {pages} {L241408} (\bibinfo {year}
  {2021})}\BibitemShut {NoStop}%
\bibitem [{\citenamefont {Wanjura}\ \emph {et~al.}(2020)\citenamefont
  {Wanjura}, \citenamefont {Brunelli},\ and\ \citenamefont
  {Nunnenkamp}}]{wanjura2020topological}%
  \BibitemOpen
  \bibfield  {author} {\bibinfo {author} {\bibfnamefont {Clara~C}\ \bibnamefont
  {Wanjura}}, \bibinfo {author} {\bibfnamefont {Matteo}\ \bibnamefont
  {Brunelli}}, \ and\ \bibinfo {author} {\bibfnamefont {Andreas}\ \bibnamefont
  {Nunnenkamp}},\ }\bibfield  {title} {\enquote {\bibinfo {title} {Topological
  framework for directional amplification in driven-dissipative cavity
  arrays},}\ }\href@noop {} {\bibfield  {journal} {\bibinfo  {journal} {Nature
  communications}\ }\textbf {\bibinfo {volume} {11}},\ \bibinfo {pages} {3149}
  (\bibinfo {year} {2020})}\BibitemShut {NoStop}%
\bibitem [{\citenamefont {Zirnstein}\ \emph {et~al.}(2021)\citenamefont
  {Zirnstein}, \citenamefont {Refael},\ and\ \citenamefont
  {Rosenow}}]{Zirnstein2021bulk}%
  \BibitemOpen
  \bibfield  {author} {\bibinfo {author} {\bibfnamefont {Heinrich-Gregor}\
  \bibnamefont {Zirnstein}}, \bibinfo {author} {\bibfnamefont {Gil}\
  \bibnamefont {Refael}}, \ and\ \bibinfo {author} {\bibfnamefont {Bernd}\
  \bibnamefont {Rosenow}},\ }\bibfield  {title} {\enquote {\bibinfo {title}
  {Bulk-boundary correspondence for non-hermitian hamiltonians via green
  functions},}\ }\href {\doibase 10.1103/PhysRevLett.126.216407} {\bibfield
  {journal} {\bibinfo  {journal} {Phys. Rev. Lett.}\ }\textbf {\bibinfo
  {volume} {126}},\ \bibinfo {pages} {216407} (\bibinfo {year}
  {2021})}\BibitemShut {NoStop}%
\bibitem [{\citenamefont {Okuma}\ and\ \citenamefont
  {Sato}(2021)}]{okuma2021non}%
  \BibitemOpen
  \bibfield  {author} {\bibinfo {author} {\bibfnamefont {Nobuyuki}\
  \bibnamefont {Okuma}}\ and\ \bibinfo {author} {\bibfnamefont {Masatoshi}\
  \bibnamefont {Sato}},\ }\bibfield  {title} {\enquote {\bibinfo {title}
  {Non-hermitian skin effects in hermitian correlated or disordered systems:
  Quantities sensitive or insensitive to boundary effects and
  pseudo-quantum-number},}\ }\href {\doibase 10.1103/PhysRevLett.126.176601}
  {\bibfield  {journal} {\bibinfo  {journal} {Phys. Rev. Lett.}\ }\textbf
  {\bibinfo {volume} {126}},\ \bibinfo {pages} {176601} (\bibinfo {year}
  {2021})}\BibitemShut {NoStop}%
\bibitem [{\citenamefont {Hu}\ and\ \citenamefont {Wang}(2023)}]{hu2023greens}%
  \BibitemOpen
  \bibfield  {author} {\bibinfo {author} {\bibfnamefont {Yu-Min}\ \bibnamefont
  {Hu}}\ and\ \bibinfo {author} {\bibfnamefont {Zhong}\ \bibnamefont {Wang}},\
  }\bibfield  {title} {\enquote {\bibinfo {title} {Green's functions of
  multiband non-hermitian systems},}\ }\href {\doibase
  10.1103/PhysRevResearch.5.043073} {\bibfield  {journal} {\bibinfo  {journal}
  {Phys. Rev. Res.}\ }\textbf {\bibinfo {volume} {5}},\ \bibinfo {pages}
  {043073} (\bibinfo {year} {2023})}\BibitemShut {NoStop}%
\bibitem [{\citenamefont {Song}\ \emph
  {et~al.}(2019{\natexlab{a}})\citenamefont {Song}, \citenamefont {Yao},\ and\
  \citenamefont {Wang}}]{song2019non}%
  \BibitemOpen
  \bibfield  {author} {\bibinfo {author} {\bibfnamefont {Fei}\ \bibnamefont
  {Song}}, \bibinfo {author} {\bibfnamefont {Shunyu}\ \bibnamefont {Yao}}, \
  and\ \bibinfo {author} {\bibfnamefont {Zhong}\ \bibnamefont {Wang}},\
  }\bibfield  {title} {\enquote {\bibinfo {title} {Non-hermitian skin effect
  and chiral damping in open quantum systems},}\ }\href {\doibase
  10.1103/PhysRevLett.123.170401} {\bibfield  {journal} {\bibinfo  {journal}
  {Phys. Rev. Lett.}\ }\textbf {\bibinfo {volume} {123}},\ \bibinfo {pages}
  {170401} (\bibinfo {year} {2019}{\natexlab{a}})}\BibitemShut {NoStop}%
\bibitem [{\citenamefont {Liu}\ \emph {et~al.}(2020)\citenamefont {Liu},
  \citenamefont {Zhang}, \citenamefont {Yang},\ and\ \citenamefont
  {Chen}}]{liu2020helical}%
  \BibitemOpen
  \bibfield  {author} {\bibinfo {author} {\bibfnamefont {Chun-Hui}\
  \bibnamefont {Liu}}, \bibinfo {author} {\bibfnamefont {Kai}\ \bibnamefont
  {Zhang}}, \bibinfo {author} {\bibfnamefont {Zhesen}\ \bibnamefont {Yang}}, \
  and\ \bibinfo {author} {\bibfnamefont {Shu}\ \bibnamefont {Chen}},\
  }\bibfield  {title} {\enquote {\bibinfo {title} {Helical damping and
  dynamical critical skin effect in open quantum systems},}\ }\href {\doibase
  10.1103/PhysRevResearch.2.043167} {\bibfield  {journal} {\bibinfo  {journal}
  {Phys. Rev. Res.}\ }\textbf {\bibinfo {volume} {2}},\ \bibinfo {pages}
  {043167} (\bibinfo {year} {2020})}\BibitemShut {NoStop}%
\bibitem [{\citenamefont {Longhi}(2019{\natexlab{a}})}]{longhi2019probing}%
  \BibitemOpen
  \bibfield  {author} {\bibinfo {author} {\bibfnamefont {Stefano}\ \bibnamefont
  {Longhi}},\ }\bibfield  {title} {\enquote {\bibinfo {title} {Probing
  non-hermitian skin effect and non-bloch phase transitions},}\ }\href
  {\doibase 10.1103/PhysRevResearch.1.023013} {\bibfield  {journal} {\bibinfo
  {journal} {Phys. Rev. Res.}\ }\textbf {\bibinfo {volume} {1}},\ \bibinfo
  {pages} {023013} (\bibinfo {year} {2019}{\natexlab{a}})}\BibitemShut
  {NoStop}%
\bibitem [{\citenamefont {Longhi}(2019{\natexlab{b}})}]{longhi2019nonbloch}%
  \BibitemOpen
  \bibfield  {author} {\bibinfo {author} {\bibfnamefont {Stefano}\ \bibnamefont
  {Longhi}},\ }\bibfield  {title} {\enquote {\bibinfo {title} {Non-bloch
  {\textdollar}$\lbrace${\textbackslash}cal
  p$\rbrace$$\lbrace${\textbackslash}cal t$\rbrace${\textdollar}{PT} symmetry
  breaking in non-hermitian photonic quantum walks},}\ }\href {\doibase
  10.1364/ol.44.005804} {\bibfield  {journal} {\bibinfo  {journal} {Optics
  Letters}\ }\textbf {\bibinfo {volume} {44}},\ \bibinfo {pages} {5804}
  (\bibinfo {year} {2019}{\natexlab{b}})}\BibitemShut {NoStop}%
\bibitem [{\citenamefont {Xiao}\ \emph {et~al.}(2021)\citenamefont {Xiao},
  \citenamefont {Deng}, \citenamefont {Wang}, \citenamefont {Wang},
  \citenamefont {Yi},\ and\ \citenamefont {Xue}}]{xiao2021observation}%
  \BibitemOpen
  \bibfield  {author} {\bibinfo {author} {\bibfnamefont {Lei}\ \bibnamefont
  {Xiao}}, \bibinfo {author} {\bibfnamefont {Tianshu}\ \bibnamefont {Deng}},
  \bibinfo {author} {\bibfnamefont {Kunkun}\ \bibnamefont {Wang}}, \bibinfo
  {author} {\bibfnamefont {Zhong}\ \bibnamefont {Wang}}, \bibinfo {author}
  {\bibfnamefont {Wei}\ \bibnamefont {Yi}}, \ and\ \bibinfo {author}
  {\bibfnamefont {Peng}\ \bibnamefont {Xue}},\ }\bibfield  {title} {\enquote
  {\bibinfo {title} {Observation of non-bloch parity-time symmetry and
  exceptional points},}\ }\href {\doibase 10.1103/PhysRevLett.126.230402}
  {\bibfield  {journal} {\bibinfo  {journal} {Phys. Rev. Lett.}\ }\textbf
  {\bibinfo {volume} {126}},\ \bibinfo {pages} {230402} (\bibinfo {year}
  {2021})}\BibitemShut {NoStop}%
\bibitem [{\citenamefont {Xue}\ \emph {et~al.}(2022)\citenamefont {Xue},
  \citenamefont {Hu}, \citenamefont {Song},\ and\ \citenamefont
  {Wang}}]{xue2022non}%
  \BibitemOpen
  \bibfield  {author} {\bibinfo {author} {\bibfnamefont {Wen-Tan}\ \bibnamefont
  {Xue}}, \bibinfo {author} {\bibfnamefont {Yu-Min}\ \bibnamefont {Hu}},
  \bibinfo {author} {\bibfnamefont {Fei}\ \bibnamefont {Song}}, \ and\ \bibinfo
  {author} {\bibfnamefont {Zhong}\ \bibnamefont {Wang}},\ }\bibfield  {title}
  {\enquote {\bibinfo {title} {Non-hermitian edge burst},}\ }\href {\doibase
  10.1103/PhysRevLett.128.120401} {\bibfield  {journal} {\bibinfo  {journal}
  {Phys. Rev. Lett.}\ }\textbf {\bibinfo {volume} {128}},\ \bibinfo {pages}
  {120401} (\bibinfo {year} {2022})}\BibitemShut {NoStop}%
\bibitem [{\citenamefont {Xiao}\ \emph {et~al.}(2023)\citenamefont {Xiao},
  \citenamefont {Xue}, \citenamefont {Song}, \citenamefont {Hu}, \citenamefont
  {Yi}, \citenamefont {Wang},\ and\ \citenamefont {Xue}}]{xiao2023observation}%
  \BibitemOpen
  \bibfield  {author} {\bibinfo {author} {\bibfnamefont {Lei}\ \bibnamefont
  {Xiao}}, \bibinfo {author} {\bibfnamefont {Wen-Tan}\ \bibnamefont {Xue}},
  \bibinfo {author} {\bibfnamefont {Fei}\ \bibnamefont {Song}}, \bibinfo
  {author} {\bibfnamefont {Yu-Min}\ \bibnamefont {Hu}}, \bibinfo {author}
  {\bibfnamefont {Wei}\ \bibnamefont {Yi}}, \bibinfo {author} {\bibfnamefont
  {Zhong}\ \bibnamefont {Wang}}, \ and\ \bibinfo {author} {\bibfnamefont
  {Peng}\ \bibnamefont {Xue}},\ }\href@noop {} {\enquote {\bibinfo {title}
  {Observation of non-hermitian edge burst in quantum dynamics},}\ } (\bibinfo
  {year} {2023}),\ \Eprint {http://arxiv.org/abs/2303.12831} {arXiv:2303.12831
  [cond-mat.mes-hall]} \BibitemShut {NoStop}%
\bibitem [{\citenamefont {Liu}\ \emph {et~al.}(2019)\citenamefont {Liu},
  \citenamefont {Jiang},\ and\ \citenamefont {Chen}}]{liu2019topological}%
  \BibitemOpen
  \bibfield  {author} {\bibinfo {author} {\bibfnamefont {Chun-Hui}\
  \bibnamefont {Liu}}, \bibinfo {author} {\bibfnamefont {Hui}\ \bibnamefont
  {Jiang}}, \ and\ \bibinfo {author} {\bibfnamefont {Shu}\ \bibnamefont
  {Chen}},\ }\bibfield  {title} {\enquote {\bibinfo {title} {Topological
  classification of non-hermitian systems with reflection symmetry},}\ }\href
  {\doibase 10.1103/PhysRevB.99.125103} {\bibfield  {journal} {\bibinfo
  {journal} {Phys. Rev. B}\ }\textbf {\bibinfo {volume} {99}},\ \bibinfo
  {pages} {125103} (\bibinfo {year} {2019})}\BibitemShut {NoStop}%
\bibitem [{\citenamefont {Okuma}\ \emph {et~al.}(2020)\citenamefont {Okuma},
  \citenamefont {Kawabata}, \citenamefont {Shiozaki},\ and\ \citenamefont
  {Sato}}]{okuma2020topological}%
  \BibitemOpen
  \bibfield  {author} {\bibinfo {author} {\bibfnamefont {Nobuyuki}\
  \bibnamefont {Okuma}}, \bibinfo {author} {\bibfnamefont {Kohei}\ \bibnamefont
  {Kawabata}}, \bibinfo {author} {\bibfnamefont {Ken}\ \bibnamefont
  {Shiozaki}}, \ and\ \bibinfo {author} {\bibfnamefont {Masatoshi}\
  \bibnamefont {Sato}},\ }\bibfield  {title} {\enquote {\bibinfo {title}
  {Topological origin of non-hermitian skin effects},}\ }\href {\doibase
  10.1103/PhysRevLett.124.086801} {\bibfield  {journal} {\bibinfo  {journal}
  {Phys. Rev. Lett.}\ }\textbf {\bibinfo {volume} {124}},\ \bibinfo {pages}
  {086801} (\bibinfo {year} {2020})}\BibitemShut {NoStop}%
\bibitem [{\citenamefont {Yi}\ and\ \citenamefont
  {Yang}(2020)}]{yi2020non-hermitian}%
  \BibitemOpen
  \bibfield  {author} {\bibinfo {author} {\bibfnamefont {Yifei}\ \bibnamefont
  {Yi}}\ and\ \bibinfo {author} {\bibfnamefont {Zhesen}\ \bibnamefont {Yang}},\
  }\bibfield  {title} {\enquote {\bibinfo {title} {Non-hermitian skin modes
  induced by on-site dissipations and chiral tunneling effect},}\ }\href
  {\doibase 10.1103/PhysRevLett.125.186802} {\bibfield  {journal} {\bibinfo
  {journal} {Phys. Rev. Lett.}\ }\textbf {\bibinfo {volume} {125}},\ \bibinfo
  {pages} {186802} (\bibinfo {year} {2020})}\BibitemShut {NoStop}%
\bibitem [{\citenamefont {Kawabata}\ \emph
  {et~al.}(2020{\natexlab{a}})\citenamefont {Kawabata}, \citenamefont {Okuma},\
  and\ \citenamefont {Sato}}]{kawabata2020nonbloch}%
  \BibitemOpen
  \bibfield  {author} {\bibinfo {author} {\bibfnamefont {Kohei}\ \bibnamefont
  {Kawabata}}, \bibinfo {author} {\bibfnamefont {Nobuyuki}\ \bibnamefont
  {Okuma}}, \ and\ \bibinfo {author} {\bibfnamefont {Masatoshi}\ \bibnamefont
  {Sato}},\ }\bibfield  {title} {\enquote {\bibinfo {title} {Non-bloch band
  theory of non-hermitian hamiltonians in the symplectic class},}\ }\href
  {\doibase 10.1103/PhysRevB.101.195147} {\bibfield  {journal} {\bibinfo
  {journal} {Phys. Rev. B}\ }\textbf {\bibinfo {volume} {101}},\ \bibinfo
  {pages} {195147} (\bibinfo {year} {2020}{\natexlab{a}})}\BibitemShut
  {NoStop}%
\bibitem [{\citenamefont {Yang}(2020)}]{yang2020nonperturbative}%
  \BibitemOpen
  \bibfield  {author} {\bibinfo {author} {\bibfnamefont {Zhesen}\ \bibnamefont
  {Yang}},\ }\href@noop {} {\enquote {\bibinfo {title} {Non-perturbative
  breakdown of bloch's theorem and hermitian skin effects},}\ } (\bibinfo
  {year} {2020}),\ \Eprint {http://arxiv.org/abs/2012.03333} {arXiv:2012.03333
  [cond-mat.mes-hall]} \BibitemShut {NoStop}%
\bibitem [{\citenamefont {Yokomizo}\ and\ \citenamefont
  {Murakami}(2021)}]{yokomizo2021non}%
  \BibitemOpen
  \bibfield  {author} {\bibinfo {author} {\bibfnamefont {Kazuki}\ \bibnamefont
  {Yokomizo}}\ and\ \bibinfo {author} {\bibfnamefont {Shuichi}\ \bibnamefont
  {Murakami}},\ }\bibfield  {title} {\enquote {\bibinfo {title} {Non-bloch band
  theory in bosonic bogoliubov--de gennes systems},}\ }\href {\doibase
  10.1103/PhysRevB.103.165123} {\bibfield  {journal} {\bibinfo  {journal}
  {Phys. Rev. B}\ }\textbf {\bibinfo {volume} {103}},\ \bibinfo {pages}
  {165123} (\bibinfo {year} {2021})}\BibitemShut {NoStop}%
\bibitem [{\citenamefont {Kawabata}\ \emph
  {et~al.}(2020{\natexlab{b}})\citenamefont {Kawabata}, \citenamefont {Sato},\
  and\ \citenamefont {Shiozaki}}]{kawabata2020higher-order}%
  \BibitemOpen
  \bibfield  {author} {\bibinfo {author} {\bibfnamefont {Kohei}\ \bibnamefont
  {Kawabata}}, \bibinfo {author} {\bibfnamefont {Masatoshi}\ \bibnamefont
  {Sato}}, \ and\ \bibinfo {author} {\bibfnamefont {Ken}\ \bibnamefont
  {Shiozaki}},\ }\bibfield  {title} {\enquote {\bibinfo {title} {Higher-order
  non-hermitian skin effect},}\ }\href {\doibase 10.1103/PhysRevB.102.205118}
  {\bibfield  {journal} {\bibinfo  {journal} {Phys. Rev. B}\ }\textbf {\bibinfo
  {volume} {102}},\ \bibinfo {pages} {205118} (\bibinfo {year}
  {2020}{\natexlab{b}})}\BibitemShut {NoStop}%
\bibitem [{\citenamefont {Zhang}\ \emph
  {et~al.}(2022{\natexlab{b}})\citenamefont {Zhang}, \citenamefont {Yang},\
  and\ \citenamefont {Fang}}]{zhang2022universal}%
  \BibitemOpen
  \bibfield  {author} {\bibinfo {author} {\bibfnamefont {Kai}\ \bibnamefont
  {Zhang}}, \bibinfo {author} {\bibfnamefont {Zhesen}\ \bibnamefont {Yang}}, \
  and\ \bibinfo {author} {\bibfnamefont {Chen}\ \bibnamefont {Fang}},\
  }\bibfield  {title} {\enquote {\bibinfo {title} {Universal non-hermitian skin
  effect in two and higher dimensions},}\ }\href@noop {} {\bibfield  {journal}
  {\bibinfo  {journal} {Nature communications}\ }\textbf {\bibinfo {volume}
  {13}},\ \bibinfo {pages} {2496} (\bibinfo {year}
  {2022}{\natexlab{b}})}\BibitemShut {NoStop}%
\bibitem [{\citenamefont {Wang}\ \emph {et~al.}(2024)\citenamefont {Wang},
  \citenamefont {Song},\ and\ \citenamefont {Wang}}]{wang2022amoeba}%
  \BibitemOpen
  \bibfield  {author} {\bibinfo {author} {\bibfnamefont {Hong-Yi}\ \bibnamefont
  {Wang}}, \bibinfo {author} {\bibfnamefont {Fei}\ \bibnamefont {Song}}, \ and\
  \bibinfo {author} {\bibfnamefont {Zhong}\ \bibnamefont {Wang}},\ }\bibfield
  {title} {\enquote {\bibinfo {title} {Amoeba formulation of non-bloch band
  theory in arbitrary dimensions},}\ }\href {\doibase
  10.1103/PhysRevX.14.021011} {\bibfield  {journal} {\bibinfo  {journal} {Phys.
  Rev. X}\ }\textbf {\bibinfo {volume} {14}},\ \bibinfo {pages} {021011}
  (\bibinfo {year} {2024})}\BibitemShut {NoStop}%
\bibitem [{\citenamefont {Ding}\ \emph {et~al.}(2015)\citenamefont {Ding},
  \citenamefont {Zhang},\ and\ \citenamefont {Chan}}]{Ding2015coalescence}%
  \BibitemOpen
  \bibfield  {author} {\bibinfo {author} {\bibfnamefont {Kun}\ \bibnamefont
  {Ding}}, \bibinfo {author} {\bibfnamefont {Z.~Q.}\ \bibnamefont {Zhang}}, \
  and\ \bibinfo {author} {\bibfnamefont {C.~T.}\ \bibnamefont {Chan}},\
  }\bibfield  {title} {\enquote {\bibinfo {title} {Coalescence of exceptional
  points and phase diagrams for one-dimensional {PT}-symmetric photonic
  crystals},}\ }\href {\doibase 10.1103/PhysRevB.92.235310} {\bibfield
  {journal} {\bibinfo  {journal} {Phys. Rev. B}\ }\textbf {\bibinfo {volume}
  {92}},\ \bibinfo {pages} {235310} (\bibinfo {year} {2015})}\BibitemShut
  {NoStop}%
\bibitem [{\citenamefont {Hahn}\ \emph {et~al.}(2016)\citenamefont {Hahn},
  \citenamefont {Choi}, \citenamefont {Yoon}, \citenamefont {Song},
  \citenamefont {Oh},\ and\ \citenamefont {Berini}}]{hahn2016observation}%
  \BibitemOpen
  \bibfield  {author} {\bibinfo {author} {\bibfnamefont {Choloong}\
  \bibnamefont {Hahn}}, \bibinfo {author} {\bibfnamefont {Youngsun}\
  \bibnamefont {Choi}}, \bibinfo {author} {\bibfnamefont {Jae~Woong}\
  \bibnamefont {Yoon}}, \bibinfo {author} {\bibfnamefont {Seok~Ho}\
  \bibnamefont {Song}}, \bibinfo {author} {\bibfnamefont {Cha~Hwan}\
  \bibnamefont {Oh}}, \ and\ \bibinfo {author} {\bibfnamefont {Pierre}\
  \bibnamefont {Berini}},\ }\bibfield  {title} {\enquote {\bibinfo {title}
  {Observation of exceptional points in reconfigurable non-{Hermitian}
  vector-field holographic lattices},}\ }\href {\doibase 10.1038/ncomms12201}
  {\bibfield  {journal} {\bibinfo  {journal} {Nature communications}\ }\textbf
  {\bibinfo {volume} {7}},\ \bibinfo {pages} {1--6} (\bibinfo {year}
  {2016})}\BibitemShut {NoStop}%
\bibitem [{\citenamefont {Cerjan}\ \emph {et~al.}(2016)\citenamefont {Cerjan},
  \citenamefont {Raman},\ and\ \citenamefont {Fan}}]{Cerjan2016exceptional}%
  \BibitemOpen
  \bibfield  {author} {\bibinfo {author} {\bibfnamefont {Alexander}\
  \bibnamefont {Cerjan}}, \bibinfo {author} {\bibfnamefont {Aaswath}\
  \bibnamefont {Raman}}, \ and\ \bibinfo {author} {\bibfnamefont {Shanhui}\
  \bibnamefont {Fan}},\ }\bibfield  {title} {\enquote {\bibinfo {title}
  {Exceptional contours and band structure design in parity-time symmetric
  photonic crystals},}\ }\href {\doibase 10.1103/PhysRevLett.116.203902}
  {\bibfield  {journal} {\bibinfo  {journal} {Phys. Rev. Lett.}\ }\textbf
  {\bibinfo {volume} {116}},\ \bibinfo {pages} {203902} (\bibinfo {year}
  {2016})}\BibitemShut {NoStop}%
\bibitem [{\citenamefont {Feng}\ \emph {et~al.}(2017)\citenamefont {Feng},
  \citenamefont {El-Ganainy},\ and\ \citenamefont {Ge}}]{feng2017non}%
  \BibitemOpen
  \bibfield  {author} {\bibinfo {author} {\bibfnamefont {Liang}\ \bibnamefont
  {Feng}}, \bibinfo {author} {\bibfnamefont {Ramy}\ \bibnamefont {El-Ganainy}},
  \ and\ \bibinfo {author} {\bibfnamefont {Li}~\bibnamefont {Ge}},\ }\bibfield
  {title} {\enquote {\bibinfo {title} {Non-{Hermitian} photonics based on
  parity--time symmetry},}\ }\href {\doibase 10.1038/s41566-017-0031-1}
  {\bibfield  {journal} {\bibinfo  {journal} {Nature Photonics}\ }\textbf
  {\bibinfo {volume} {11}},\ \bibinfo {pages} {752--762} (\bibinfo {year}
  {2017})}\BibitemShut {NoStop}%
\bibitem [{\citenamefont {El-Ganainy}\ \emph {et~al.}(2018)\citenamefont
  {El-Ganainy}, \citenamefont {Makris}, \citenamefont {Khajavikhan},
  \citenamefont {Musslimani}, \citenamefont {Rotter},\ and\ \citenamefont
  {Christodoulides}}]{el2018non}%
  \BibitemOpen
  \bibfield  {author} {\bibinfo {author} {\bibfnamefont {Ramy}\ \bibnamefont
  {El-Ganainy}}, \bibinfo {author} {\bibfnamefont {Konstantinos~G}\
  \bibnamefont {Makris}}, \bibinfo {author} {\bibfnamefont {Mercedeh}\
  \bibnamefont {Khajavikhan}}, \bibinfo {author} {\bibfnamefont {Ziad~H}\
  \bibnamefont {Musslimani}}, \bibinfo {author} {\bibfnamefont {Stefan}\
  \bibnamefont {Rotter}}, \ and\ \bibinfo {author} {\bibfnamefont
  {Demetrios~N}\ \bibnamefont {Christodoulides}},\ }\bibfield  {title}
  {\enquote {\bibinfo {title} {Non-{Hermitian} physics and {PT} symmetry},}\
  }\href {\doibase https://doi.org/10.1038/nphys4323} {\bibfield  {journal}
  {\bibinfo  {journal} {Nature Physics}\ }\textbf {\bibinfo {volume} {14}},\
  \bibinfo {pages} {11--19} (\bibinfo {year} {2018})}\BibitemShut {NoStop}%
\bibitem [{\citenamefont {Zhou}\ \emph {et~al.}(2018)\citenamefont {Zhou},
  \citenamefont {Peng}, \citenamefont {Yoon}, \citenamefont {Hsu},
  \citenamefont {Nelson}, \citenamefont {Fu}, \citenamefont {Joannopoulos},
  \citenamefont {Solja{\v{c}}i{\'c}},\ and\ \citenamefont
  {Zhen}}]{zhou2018observation}%
  \BibitemOpen
  \bibfield  {author} {\bibinfo {author} {\bibfnamefont {Hengyun}\ \bibnamefont
  {Zhou}}, \bibinfo {author} {\bibfnamefont {Chao}\ \bibnamefont {Peng}},
  \bibinfo {author} {\bibfnamefont {Yoseob}\ \bibnamefont {Yoon}}, \bibinfo
  {author} {\bibfnamefont {Chia~Wei}\ \bibnamefont {Hsu}}, \bibinfo {author}
  {\bibfnamefont {Keith~A}\ \bibnamefont {Nelson}}, \bibinfo {author}
  {\bibfnamefont {Liang}\ \bibnamefont {Fu}}, \bibinfo {author} {\bibfnamefont
  {John~D}\ \bibnamefont {Joannopoulos}}, \bibinfo {author} {\bibfnamefont
  {Marin}\ \bibnamefont {Solja{\v{c}}i{\'c}}}, \ and\ \bibinfo {author}
  {\bibfnamefont {Bo}~\bibnamefont {Zhen}},\ }\bibfield  {title} {\enquote
  {\bibinfo {title} {Observation of bulk {Fermi} arc and polarization half
  charge from paired exceptional points},}\ }\href {\doibase
  10.1126/science.aap9859} {\bibfield  {journal} {\bibinfo  {journal}
  {Science}\ }\textbf {\bibinfo {volume} {359}},\ \bibinfo {pages} {1009--1012}
  (\bibinfo {year} {2018})}\BibitemShut {NoStop}%
\bibitem [{\citenamefont {Bandres}\ \emph {et~al.}(2018)\citenamefont
  {Bandres}, \citenamefont {Wittek}, \citenamefont {Harari}, \citenamefont
  {Parto}, \citenamefont {Ren}, \citenamefont {Segev}, \citenamefont
  {Christodoulides},\ and\ \citenamefont
  {Khajavikhan}}]{bandres2018topological}%
  \BibitemOpen
  \bibfield  {author} {\bibinfo {author} {\bibfnamefont {Miguel~A}\
  \bibnamefont {Bandres}}, \bibinfo {author} {\bibfnamefont {Steffen}\
  \bibnamefont {Wittek}}, \bibinfo {author} {\bibfnamefont {Gal}\ \bibnamefont
  {Harari}}, \bibinfo {author} {\bibfnamefont {Midya}\ \bibnamefont {Parto}},
  \bibinfo {author} {\bibfnamefont {Jinhan}\ \bibnamefont {Ren}}, \bibinfo
  {author} {\bibfnamefont {Mordechai}\ \bibnamefont {Segev}}, \bibinfo {author}
  {\bibfnamefont {Demetrios~N}\ \bibnamefont {Christodoulides}}, \ and\
  \bibinfo {author} {\bibfnamefont {Mercedeh}\ \bibnamefont {Khajavikhan}},\
  }\bibfield  {title} {\enquote {\bibinfo {title} {Topological insulator laser:
  Experiments},}\ }\href {\doibase 10.1126/science.aar4005} {\bibfield
  {journal} {\bibinfo  {journal} {Science}\ }\textbf {\bibinfo {volume} {359}}
  (\bibinfo {year} {2018}),\ 10.1126/science.aar4005}\BibitemShut {NoStop}%
\bibitem [{\citenamefont {Harari}\ \emph {et~al.}(2018)\citenamefont {Harari},
  \citenamefont {Bandres}, \citenamefont {Lumer}, \citenamefont {Rechtsman},
  \citenamefont {Chong}, \citenamefont {Khajavikhan}, \citenamefont
  {Christodoulides},\ and\ \citenamefont {Segev}}]{harari2018topological}%
  \BibitemOpen
  \bibfield  {author} {\bibinfo {author} {\bibfnamefont {Gal}\ \bibnamefont
  {Harari}}, \bibinfo {author} {\bibfnamefont {Miguel~A.}\ \bibnamefont
  {Bandres}}, \bibinfo {author} {\bibfnamefont {Yaakov}\ \bibnamefont {Lumer}},
  \bibinfo {author} {\bibfnamefont {Mikael~C.}\ \bibnamefont {Rechtsman}},
  \bibinfo {author} {\bibfnamefont {Y.~D.}\ \bibnamefont {Chong}}, \bibinfo
  {author} {\bibfnamefont {Mercedeh}\ \bibnamefont {Khajavikhan}}, \bibinfo
  {author} {\bibfnamefont {Demetrios~N.}\ \bibnamefont {Christodoulides}}, \
  and\ \bibinfo {author} {\bibfnamefont {Mordechai}\ \bibnamefont {Segev}},\
  }\bibfield  {title} {\enquote {\bibinfo {title} {Topological insulator laser:
  {Theory}},}\ }\href {\doibase 10.1126/science.aar4003} {\bibfield  {journal}
  {\bibinfo  {journal} {Science}\ }\textbf {\bibinfo {volume} {359}} (\bibinfo
  {year} {2018}),\ 10.1126/science.aar4003}\BibitemShut {NoStop}%
\bibitem [{\citenamefont {Silveirinha}(2019)}]{Silveirinha2019topological}%
  \BibitemOpen
  \bibfield  {author} {\bibinfo {author} {\bibfnamefont {M\'ario~G.}\
  \bibnamefont {Silveirinha}},\ }\bibfield  {title} {\enquote {\bibinfo {title}
  {Topological theory of non-{Hermitian} photonic systems},}\ }\href {\doibase
  10.1103/PhysRevB.99.125155} {\bibfield  {journal} {\bibinfo  {journal} {Phys.
  Rev. B}\ }\textbf {\bibinfo {volume} {99}},\ \bibinfo {pages} {125155}
  (\bibinfo {year} {2019})}\BibitemShut {NoStop}%
\bibitem [{\citenamefont {Lu}\ \emph {et~al.}(2014)\citenamefont {Lu},
  \citenamefont {Joannopoulos},\ and\ \citenamefont
  {Solja{\v{c}}i{\'{c}}}}]{Lu2014topological}%
  \BibitemOpen
  \bibfield  {author} {\bibinfo {author} {\bibfnamefont {Ling}\ \bibnamefont
  {Lu}}, \bibinfo {author} {\bibfnamefont {John~D.}\ \bibnamefont
  {Joannopoulos}}, \ and\ \bibinfo {author} {\bibfnamefont {Marin}\
  \bibnamefont {Solja{\v{c}}i{\'{c}}}},\ }\bibfield  {title} {\enquote
  {\bibinfo {title} {Topological photonics},}\ }\href {\doibase
  10.1038/nphoton.2014.248} {\bibfield  {journal} {\bibinfo  {journal} {Nature
  Photonics}\ }\textbf {\bibinfo {volume} {8}},\ \bibinfo {pages} {821--829}
  (\bibinfo {year} {2014})}\BibitemShut {NoStop}%
\bibitem [{\citenamefont {Ozawa}\ \emph {et~al.}(2019)\citenamefont {Ozawa},
  \citenamefont {Price}, \citenamefont {Amo}, \citenamefont {Goldman},
  \citenamefont {Hafezi}, \citenamefont {Lu}, \citenamefont {Rechtsman},
  \citenamefont {Schuster}, \citenamefont {Simon}, \citenamefont {Zilberberg},\
  and\ \citenamefont {Carusotto}}]{Ozawa2019topological}%
  \BibitemOpen
  \bibfield  {author} {\bibinfo {author} {\bibfnamefont {Tomoki}\ \bibnamefont
  {Ozawa}}, \bibinfo {author} {\bibfnamefont {Hannah~M.}\ \bibnamefont
  {Price}}, \bibinfo {author} {\bibfnamefont {Alberto}\ \bibnamefont {Amo}},
  \bibinfo {author} {\bibfnamefont {Nathan}\ \bibnamefont {Goldman}}, \bibinfo
  {author} {\bibfnamefont {Mohammad}\ \bibnamefont {Hafezi}}, \bibinfo {author}
  {\bibfnamefont {Ling}\ \bibnamefont {Lu}}, \bibinfo {author} {\bibfnamefont
  {Mikael~C.}\ \bibnamefont {Rechtsman}}, \bibinfo {author} {\bibfnamefont
  {David}\ \bibnamefont {Schuster}}, \bibinfo {author} {\bibfnamefont
  {Jonathan}\ \bibnamefont {Simon}}, \bibinfo {author} {\bibfnamefont {Oded}\
  \bibnamefont {Zilberberg}}, \ and\ \bibinfo {author} {\bibfnamefont {Iacopo}\
  \bibnamefont {Carusotto}},\ }\bibfield  {title} {\enquote {\bibinfo {title}
  {Topological photonics},}\ }\href {\doibase 10.1103/RevModPhys.91.015006}
  {\bibfield  {journal} {\bibinfo  {journal} {Rev. Mod. Phys.}\ }\textbf
  {\bibinfo {volume} {91}},\ \bibinfo {pages} {015006} (\bibinfo {year}
  {2019})}\BibitemShut {NoStop}%
\bibitem [{\citenamefont {Shmuel}\ and\ \citenamefont
  {Moiseyev}(2020)}]{Shmuel2020linking}%
  \BibitemOpen
  \bibfield  {author} {\bibinfo {author} {\bibfnamefont {Gal}\ \bibnamefont
  {Shmuel}}\ and\ \bibinfo {author} {\bibfnamefont {Nimrod}\ \bibnamefont
  {Moiseyev}},\ }\bibfield  {title} {\enquote {\bibinfo {title} {Linking scalar
  elastodynamics and non-{Hermitian} quantum mechanics},}\ }\href {\doibase
  10.1103/PhysRevApplied.13.024074} {\bibfield  {journal} {\bibinfo  {journal}
  {Phys. Rev. Applied}\ }\textbf {\bibinfo {volume} {13}},\ \bibinfo {pages}
  {024074} (\bibinfo {year} {2020})}\BibitemShut {NoStop}%
\bibitem [{\citenamefont {Scheibner}\ \emph
  {et~al.}(2020{\natexlab{a}})\citenamefont {Scheibner}, \citenamefont
  {Souslov}, \citenamefont {Banerjee}, \citenamefont {Surowka}, \citenamefont
  {Irvine},\ and\ \citenamefont {Vitelli}}]{scheibner2020odd}%
  \BibitemOpen
  \bibfield  {author} {\bibinfo {author} {\bibfnamefont {Colin}\ \bibnamefont
  {Scheibner}}, \bibinfo {author} {\bibfnamefont {Anton}\ \bibnamefont
  {Souslov}}, \bibinfo {author} {\bibfnamefont {Debarghya}\ \bibnamefont
  {Banerjee}}, \bibinfo {author} {\bibfnamefont {Piotr}\ \bibnamefont
  {Surowka}}, \bibinfo {author} {\bibfnamefont {William~TM}\ \bibnamefont
  {Irvine}}, \ and\ \bibinfo {author} {\bibfnamefont {Vincenzo}\ \bibnamefont
  {Vitelli}},\ }\bibfield  {title} {\enquote {\bibinfo {title} {Odd
  elasticity},}\ }\href {\doibase 10.1038/s41567-020-0795-y} {\bibfield
  {journal} {\bibinfo  {journal} {Nature Physics}\ }\textbf {\bibinfo {volume}
  {16}},\ \bibinfo {pages} {475--480} (\bibinfo {year}
  {2020}{\natexlab{a}})}\BibitemShut {NoStop}%
\bibitem [{\citenamefont {Scheibner}\ \emph
  {et~al.}(2020{\natexlab{b}})\citenamefont {Scheibner}, \citenamefont
  {Irvine},\ and\ \citenamefont {Vitelli}}]{scheibner2020non}%
  \BibitemOpen
  \bibfield  {author} {\bibinfo {author} {\bibfnamefont {Colin}\ \bibnamefont
  {Scheibner}}, \bibinfo {author} {\bibfnamefont {William T.~M.}\ \bibnamefont
  {Irvine}}, \ and\ \bibinfo {author} {\bibfnamefont {Vincenzo}\ \bibnamefont
  {Vitelli}},\ }\bibfield  {title} {\enquote {\bibinfo {title} {Non-{Hermitian}
  band topology and skin modes in active elastic media},}\ }\href {\doibase
  10.1103/PhysRevLett.125.118001} {\bibfield  {journal} {\bibinfo  {journal}
  {Phys. Rev. Lett.}\ }\textbf {\bibinfo {volume} {125}},\ \bibinfo {pages}
  {118001} (\bibinfo {year} {2020}{\natexlab{b}})}\BibitemShut {NoStop}%
\bibitem [{\citenamefont {Chen}\ \emph {et~al.}(2021)\citenamefont {Chen},
  \citenamefont {Li}, \citenamefont {Scheibner}, \citenamefont {Vitelli},\ and\
  \citenamefont {Huang}}]{chen2021realization}%
  \BibitemOpen
  \bibfield  {author} {\bibinfo {author} {\bibfnamefont {Yangyang}\
  \bibnamefont {Chen}}, \bibinfo {author} {\bibfnamefont {Xiaopeng}\
  \bibnamefont {Li}}, \bibinfo {author} {\bibfnamefont {Colin}\ \bibnamefont
  {Scheibner}}, \bibinfo {author} {\bibfnamefont {Vincenzo}\ \bibnamefont
  {Vitelli}}, \ and\ \bibinfo {author} {\bibfnamefont {Guoliang}\ \bibnamefont
  {Huang}},\ }\bibfield  {title} {\enquote {\bibinfo {title} {Realization of
  active metamaterials with odd micropolar elasticity},}\ }\href {\doibase
  10.1038/s41467-021-26034-z} {\bibfield  {journal} {\bibinfo  {journal}
  {Nature communications}\ }\textbf {\bibinfo {volume} {12}},\ \bibinfo {pages}
  {5935} (\bibinfo {year} {2021})}\BibitemShut {NoStop}%
\bibitem [{\citenamefont {Braverman}\ \emph {et~al.}(2021)\citenamefont
  {Braverman}, \citenamefont {Scheibner}, \citenamefont {VanSaders},\ and\
  \citenamefont {Vitelli}}]{Braverman2021topological}%
  \BibitemOpen
  \bibfield  {author} {\bibinfo {author} {\bibfnamefont {Lara}\ \bibnamefont
  {Braverman}}, \bibinfo {author} {\bibfnamefont {Colin}\ \bibnamefont
  {Scheibner}}, \bibinfo {author} {\bibfnamefont {Bryan}\ \bibnamefont
  {VanSaders}}, \ and\ \bibinfo {author} {\bibfnamefont {Vincenzo}\
  \bibnamefont {Vitelli}},\ }\bibfield  {title} {\enquote {\bibinfo {title}
  {Topological defects in solids with odd elasticity},}\ }\href {\doibase
  10.1103/PhysRevLett.127.268001} {\bibfield  {journal} {\bibinfo  {journal}
  {Phys. Rev. Lett.}\ }\textbf {\bibinfo {volume} {127}},\ \bibinfo {pages}
  {268001} (\bibinfo {year} {2021})}\BibitemShut {NoStop}%
\bibitem [{\citenamefont {Fruchart}\ \emph {et~al.}(2023)\citenamefont
  {Fruchart}, \citenamefont {Scheibner},\ and\ \citenamefont
  {Vitelli}}]{fruchart2023odd}%
  \BibitemOpen
  \bibfield  {author} {\bibinfo {author} {\bibfnamefont {Michel}\ \bibnamefont
  {Fruchart}}, \bibinfo {author} {\bibfnamefont {Colin}\ \bibnamefont
  {Scheibner}}, \ and\ \bibinfo {author} {\bibfnamefont {Vincenzo}\
  \bibnamefont {Vitelli}},\ }\bibfield  {title} {\enquote {\bibinfo {title}
  {Odd viscosity and odd elasticity},}\ }\href@noop {} {\bibfield  {journal}
  {\bibinfo  {journal} {Annual Review of Condensed Matter Physics}\ }\textbf
  {\bibinfo {volume} {14}},\ \bibinfo {pages} {471--510} (\bibinfo {year}
  {2023})}\BibitemShut {NoStop}%
\bibitem [{\citenamefont {Zhong}\ \emph {et~al.}(2021)\citenamefont {Zhong},
  \citenamefont {Wang}, \citenamefont {Park}, \citenamefont {Asadchy},
  \citenamefont {Wojcik}, \citenamefont {Dutt},\ and\ \citenamefont
  {Fan}}]{zhong2021nontrivial}%
  \BibitemOpen
  \bibfield  {author} {\bibinfo {author} {\bibfnamefont {Janet}\ \bibnamefont
  {Zhong}}, \bibinfo {author} {\bibfnamefont {Kai}\ \bibnamefont {Wang}},
  \bibinfo {author} {\bibfnamefont {Yubin}\ \bibnamefont {Park}}, \bibinfo
  {author} {\bibfnamefont {Viktar}\ \bibnamefont {Asadchy}}, \bibinfo {author}
  {\bibfnamefont {Charles~C.}\ \bibnamefont {Wojcik}}, \bibinfo {author}
  {\bibfnamefont {Avik}\ \bibnamefont {Dutt}}, \ and\ \bibinfo {author}
  {\bibfnamefont {Shanhui}\ \bibnamefont {Fan}},\ }\bibfield  {title} {\enquote
  {\bibinfo {title} {Nontrivial point-gap topology and non-hermitian skin
  effect in photonic crystals},}\ }\href {\doibase 10.1103/PhysRevB.104.125416}
  {\bibfield  {journal} {\bibinfo  {journal} {Phys. Rev. B}\ }\textbf {\bibinfo
  {volume} {104}},\ \bibinfo {pages} {125416} (\bibinfo {year}
  {2021})}\BibitemShut {NoStop}%
\bibitem [{\citenamefont {Ochiai}(2022)}]{Ochiai2022non}%
  \BibitemOpen
  \bibfield  {author} {\bibinfo {author} {\bibfnamefont {Tetsuyuki}\
  \bibnamefont {Ochiai}},\ }\bibfield  {title} {\enquote {\bibinfo {title}
  {Non-hermitian skin effect and lasing of absorbing open-boundary modes in
  photonic crystals},}\ }\href {\doibase 10.1103/PhysRevB.106.195412}
  {\bibfield  {journal} {\bibinfo  {journal} {Phys. Rev. B}\ }\textbf {\bibinfo
  {volume} {106}},\ \bibinfo {pages} {195412} (\bibinfo {year}
  {2022})}\BibitemShut {NoStop}%
\bibitem [{\citenamefont {Yan}\ \emph {et~al.}(2021)\citenamefont {Yan},
  \citenamefont {Chen},\ and\ \citenamefont {Yang}}]{yan2021nonhermitian}%
  \BibitemOpen
  \bibfield  {author} {\bibinfo {author} {\bibfnamefont {Qinghui}\ \bibnamefont
  {Yan}}, \bibinfo {author} {\bibfnamefont {Hongsheng}\ \bibnamefont {Chen}}, \
  and\ \bibinfo {author} {\bibfnamefont {Yihao}\ \bibnamefont {Yang}},\
  }\href@noop {} {\enquote {\bibinfo {title} {Non-hermitian skin effect and
  delocalized edge states in photonic crystals with anomalous parity-time
  symmetry},}\ } (\bibinfo {year} {2021}),\ \Eprint
  {http://arxiv.org/abs/2111.08213} {arXiv:2111.08213 [physics.optics]}
  \BibitemShut {NoStop}%
\bibitem [{\citenamefont {Yokomizo}\ \emph {et~al.}(2022)\citenamefont
  {Yokomizo}, \citenamefont {Yoda},\ and\ \citenamefont
  {Murakami}}]{Yokomizo2022non-hermitian_wave}%
  \BibitemOpen
  \bibfield  {author} {\bibinfo {author} {\bibfnamefont {Kazuki}\ \bibnamefont
  {Yokomizo}}, \bibinfo {author} {\bibfnamefont {Taiki}\ \bibnamefont {Yoda}},
  \ and\ \bibinfo {author} {\bibfnamefont {Shuichi}\ \bibnamefont {Murakami}},\
  }\bibfield  {title} {\enquote {\bibinfo {title} {Non-hermitian waves in a
  continuous periodic model and application to photonic crystals},}\ }\href
  {\doibase 10.1103/PhysRevResearch.4.023089} {\bibfield  {journal} {\bibinfo
  {journal} {Phys. Rev. Res.}\ }\textbf {\bibinfo {volume} {4}},\ \bibinfo
  {pages} {023089} (\bibinfo {year} {2022})}\BibitemShut {NoStop}%
\bibitem [{\citenamefont {Fang}\ \emph {et~al.}(2022)\citenamefont {Fang},
  \citenamefont {Hu}, \citenamefont {Zhou},\ and\ \citenamefont
  {Ding}}]{FangHu2022Geometry}%
  \BibitemOpen
  \bibfield  {author} {\bibinfo {author} {\bibfnamefont {Zhening}\ \bibnamefont
  {Fang}}, \bibinfo {author} {\bibfnamefont {Mengying}\ \bibnamefont {Hu}},
  \bibinfo {author} {\bibfnamefont {Lei}\ \bibnamefont {Zhou}}, \ and\ \bibinfo
  {author} {\bibfnamefont {Kun}\ \bibnamefont {Ding}},\ }\bibfield  {title}
  {\enquote {\bibinfo {title} {Geometry-dependent skin effects in reciprocal
  photonic crystals},}\ }\href {\doibase doi:10.1515/nanoph-2022-0211}
  {\bibfield  {journal} {\bibinfo  {journal} {Nanophotonics}\ }\textbf
  {\bibinfo {volume} {11}},\ \bibinfo {pages} {3447--3456} (\bibinfo {year}
  {2022})}\BibitemShut {NoStop}%
\bibitem [{\citenamefont {Liu}\ \emph {et~al.}(2023)\citenamefont {Liu},
  \citenamefont {Mandal}, \citenamefont {Zhou}, \citenamefont {Xi},
  \citenamefont {Banerjee}, \citenamefont {Hu}, \citenamefont {Wei},
  \citenamefont {Wang}, \citenamefont {Wang}, \citenamefont {Gao},
  \citenamefont {Chen}, \citenamefont {Yang}, \citenamefont {Chong},\ and\
  \citenamefont {Zhang}}]{liu2023localization}%
  \BibitemOpen
  \bibfield  {author} {\bibinfo {author} {\bibfnamefont {Gui-Geng}\
  \bibnamefont {Liu}}, \bibinfo {author} {\bibfnamefont {Subhaskar}\
  \bibnamefont {Mandal}}, \bibinfo {author} {\bibfnamefont {Peiheng}\
  \bibnamefont {Zhou}}, \bibinfo {author} {\bibfnamefont {Xiang}\ \bibnamefont
  {Xi}}, \bibinfo {author} {\bibfnamefont {Rimi}\ \bibnamefont {Banerjee}},
  \bibinfo {author} {\bibfnamefont {Yuan-Hang}\ \bibnamefont {Hu}}, \bibinfo
  {author} {\bibfnamefont {Minggui}\ \bibnamefont {Wei}}, \bibinfo {author}
  {\bibfnamefont {Maoren}\ \bibnamefont {Wang}}, \bibinfo {author}
  {\bibfnamefont {Qiang}\ \bibnamefont {Wang}}, \bibinfo {author}
  {\bibfnamefont {Zhen}\ \bibnamefont {Gao}}, \bibinfo {author} {\bibfnamefont
  {Hongsheng}\ \bibnamefont {Chen}}, \bibinfo {author} {\bibfnamefont {Yihao}\
  \bibnamefont {Yang}}, \bibinfo {author} {\bibfnamefont {Yidong}\ \bibnamefont
  {Chong}}, \ and\ \bibinfo {author} {\bibfnamefont {Baile}\ \bibnamefont
  {Zhang}},\ }\href@noop {} {\enquote {\bibinfo {title} {Localization of chiral
  edge states by the non-hermitian skin effect},}\ } (\bibinfo {year} {2023}),\
  \Eprint {http://arxiv.org/abs/2305.13139} {arXiv:2305.13139
  [cond-mat.mes-hall]} \BibitemShut {NoStop}%
\bibitem [{\citenamefont {Zhu}\ and\ \citenamefont
  {Gong}(2023)}]{zhu2023photonic}%
  \BibitemOpen
  \bibfield  {author} {\bibinfo {author} {\bibfnamefont {Weiwei}\ \bibnamefont
  {Zhu}}\ and\ \bibinfo {author} {\bibfnamefont {Jiangbin}\ \bibnamefont
  {Gong}},\ }\bibfield  {title} {\enquote {\bibinfo {title} {Photonic corner
  skin modes in non-hermitian photonic crystals},}\ }\href {\doibase
  10.1103/PhysRevB.108.035406} {\bibfield  {journal} {\bibinfo  {journal}
  {Phys. Rev. B}\ }\textbf {\bibinfo {volume} {108}},\ \bibinfo {pages}
  {035406} (\bibinfo {year} {2023})}\BibitemShut {NoStop}%
\bibitem [{\citenamefont {Kokhanchik}\ \emph {et~al.}(2023)\citenamefont
  {Kokhanchik}, \citenamefont {Solnyshkov},\ and\ \citenamefont
  {Malpuech}}]{kokhanchik2023nonhermitian}%
  \BibitemOpen
  \bibfield  {author} {\bibinfo {author} {\bibfnamefont {Pavel}\ \bibnamefont
  {Kokhanchik}}, \bibinfo {author} {\bibfnamefont {Dmitry}\ \bibnamefont
  {Solnyshkov}}, \ and\ \bibinfo {author} {\bibfnamefont {Guillaume}\
  \bibnamefont {Malpuech}},\ }\bibfield  {title} {\enquote {\bibinfo {title}
  {Non-hermitian skin effect induced by rashba-dresselhaus spin-orbit
  coupling},}\ }\href {\doibase 10.1103/PhysRevB.108.L041403} {\bibfield
  {journal} {\bibinfo  {journal} {Phys. Rev. B}\ }\textbf {\bibinfo {volume}
  {108}},\ \bibinfo {pages} {L041403} (\bibinfo {year} {2023})}\BibitemShut
  {NoStop}%
\bibitem [{\citenamefont {Guo}\ \emph {et~al.}(2022)\citenamefont {Guo},
  \citenamefont {Dong}, \citenamefont {Zhang}, \citenamefont {Hu},\ and\
  \citenamefont {Yang}}]{guo2022theoretical}%
  \BibitemOpen
  \bibfield  {author} {\bibinfo {author} {\bibfnamefont {Sibo}\ \bibnamefont
  {Guo}}, \bibinfo {author} {\bibfnamefont {Chenxiao}\ \bibnamefont {Dong}},
  \bibinfo {author} {\bibfnamefont {Fuchun}\ \bibnamefont {Zhang}}, \bibinfo
  {author} {\bibfnamefont {Jiangping}\ \bibnamefont {Hu}}, \ and\ \bibinfo
  {author} {\bibfnamefont {Zhesen}\ \bibnamefont {Yang}},\ }\bibfield  {title}
  {\enquote {\bibinfo {title} {Theoretical prediction of a non-hermitian skin
  effect in ultracold-atom systems},}\ }\href {\doibase
  10.1103/PhysRevA.106.L061302} {\bibfield  {journal} {\bibinfo  {journal}
  {Phys. Rev. A}\ }\textbf {\bibinfo {volume} {106}},\ \bibinfo {pages}
  {L061302} (\bibinfo {year} {2022})}\BibitemShut {NoStop}%
\bibitem [{\citenamefont {Li}\ \emph {et~al.}(2020)\citenamefont {Li},
  \citenamefont {Lee},\ and\ \citenamefont {Gong}}]{li2020topological}%
  \BibitemOpen
  \bibfield  {author} {\bibinfo {author} {\bibfnamefont {Linhu}\ \bibnamefont
  {Li}}, \bibinfo {author} {\bibfnamefont {Ching~Hua}\ \bibnamefont {Lee}}, \
  and\ \bibinfo {author} {\bibfnamefont {Jiangbin}\ \bibnamefont {Gong}},\
  }\bibfield  {title} {\enquote {\bibinfo {title} {Topological switch for
  non-hermitian skin effect in cold-atom systems with loss},}\ }\href {\doibase
  10.1103/PhysRevLett.124.250402} {\bibfield  {journal} {\bibinfo  {journal}
  {Phys. Rev. Lett.}\ }\textbf {\bibinfo {volume} {124}},\ \bibinfo {pages}
  {250402} (\bibinfo {year} {2020})}\BibitemShut {NoStop}%
\bibitem [{\citenamefont {Liang}\ \emph {et~al.}(2022)\citenamefont {Liang},
  \citenamefont {Xie}, \citenamefont {Dong}, \citenamefont {Li}, \citenamefont
  {Li}, \citenamefont {Gadway}, \citenamefont {Yi},\ and\ \citenamefont
  {Yan}}]{liang2022dynamic}%
  \BibitemOpen
  \bibfield  {author} {\bibinfo {author} {\bibfnamefont {Qian}\ \bibnamefont
  {Liang}}, \bibinfo {author} {\bibfnamefont {Dizhou}\ \bibnamefont {Xie}},
  \bibinfo {author} {\bibfnamefont {Zhaoli}\ \bibnamefont {Dong}}, \bibinfo
  {author} {\bibfnamefont {Haowei}\ \bibnamefont {Li}}, \bibinfo {author}
  {\bibfnamefont {Hang}\ \bibnamefont {Li}}, \bibinfo {author} {\bibfnamefont
  {Bryce}\ \bibnamefont {Gadway}}, \bibinfo {author} {\bibfnamefont {Wei}\
  \bibnamefont {Yi}}, \ and\ \bibinfo {author} {\bibfnamefont {Bo}~\bibnamefont
  {Yan}},\ }\bibfield  {title} {\enquote {\bibinfo {title} {Dynamic signatures
  of non-hermitian skin effect and topology in ultracold atoms},}\ }\href
  {\doibase 10.1103/PhysRevLett.129.070401} {\bibfield  {journal} {\bibinfo
  {journal} {Phys. Rev. Lett.}\ }\textbf {\bibinfo {volume} {129}},\ \bibinfo
  {pages} {070401} (\bibinfo {year} {2022})}\BibitemShut {NoStop}%
\bibitem [{\citenamefont {Longhi}(2021)}]{longhi2021non}%
  \BibitemOpen
  \bibfield  {author} {\bibinfo {author} {\bibfnamefont {Stefano}\ \bibnamefont
  {Longhi}},\ }\bibfield  {title} {\enquote {\bibinfo {title} {Non-hermitian
  skin effect beyond the tight-binding models},}\ }\href {\doibase
  10.1103/PhysRevB.104.125109} {\bibfield  {journal} {\bibinfo  {journal}
  {Phys. Rev. B}\ }\textbf {\bibinfo {volume} {104}},\ \bibinfo {pages}
  {125109} (\bibinfo {year} {2021})}\BibitemShut {NoStop}%
\bibitem [{\citenamefont {Yuce}\ and\ \citenamefont
  {Ramezani}(2022)}]{yuce2022non}%
  \BibitemOpen
  \bibfield  {author} {\bibinfo {author} {\bibfnamefont {Cem}\ \bibnamefont
  {Yuce}}\ and\ \bibinfo {author} {\bibfnamefont {Hamidreza}\ \bibnamefont
  {Ramezani}},\ }\bibfield  {title} {\enquote {\bibinfo {title} {Non-hermitian
  skin effect in two dimensional continuous systems},}\ }\href@noop {}
  {\bibfield  {journal} {\bibinfo  {journal} {Physica Scripta}\ }\textbf
  {\bibinfo {volume} {98}},\ \bibinfo {pages} {015005} (\bibinfo {year}
  {2022})}\BibitemShut {NoStop}%
\bibitem [{\citenamefont {Hatano}\ and\ \citenamefont
  {Nelson}(1996)}]{hatano1996localization}%
  \BibitemOpen
  \bibfield  {author} {\bibinfo {author} {\bibfnamefont {Naomichi}\
  \bibnamefont {Hatano}}\ and\ \bibinfo {author} {\bibfnamefont {David~R.}\
  \bibnamefont {Nelson}},\ }\bibfield  {title} {\enquote {\bibinfo {title}
  {Localization transitions in non-{Hermitian} quantum mechanics},}\ }\href
  {\doibase 10.1103/PhysRevLett.77.570} {\bibfield  {journal} {\bibinfo
  {journal} {Phys. Rev. Lett.}\ }\textbf {\bibinfo {volume} {77}},\ \bibinfo
  {pages} {570--573} (\bibinfo {year} {1996})}\BibitemShut {NoStop}%
\bibitem [{\citenamefont {Hatano}\ and\ \citenamefont
  {Nelson}(1997)}]{hatano1997vortex}%
  \BibitemOpen
  \bibfield  {author} {\bibinfo {author} {\bibfnamefont {Naomichi}\
  \bibnamefont {Hatano}}\ and\ \bibinfo {author} {\bibfnamefont {David~R.}\
  \bibnamefont {Nelson}},\ }\bibfield  {title} {\enquote {\bibinfo {title}
  {Vortex pinning and non-{Hermitian} quantum mechanics},}\ }\href {\doibase
  10.1103/PhysRevB.56.8651} {\bibfield  {journal} {\bibinfo  {journal} {Phys.
  Rev. B}\ }\textbf {\bibinfo {volume} {56}},\ \bibinfo {pages} {8651--8673}
  (\bibinfo {year} {1997})}\BibitemShut {NoStop}%
\bibitem [{\citenamefont {Risken}(1996)}]{risken1996fokker}%
  \BibitemOpen
  \bibfield  {author} {\bibinfo {author} {\bibfnamefont {Hannes}\ \bibnamefont
  {Risken}},\ }\href@noop {} {\emph {\bibinfo {title} {The Fokker-Planck
  Equation: Methods of Solution and Applications}}}\ (\bibinfo  {publisher}
  {Springer},\ \bibinfo {year} {1996})\BibitemShut {NoStop}%
\bibitem [{\citenamefont {Bonneau}\ \emph {et~al.}(2001)\citenamefont
  {Bonneau}, \citenamefont {Faraut},\ and\ \citenamefont
  {Valent}}]{bonneau2001self}%
  \BibitemOpen
  \bibfield  {author} {\bibinfo {author} {\bibfnamefont {Guy}\ \bibnamefont
  {Bonneau}}, \bibinfo {author} {\bibfnamefont {Jacques}\ \bibnamefont
  {Faraut}}, \ and\ \bibinfo {author} {\bibfnamefont {Galliano}\ \bibnamefont
  {Valent}},\ }\bibfield  {title} {\enquote {\bibinfo {title} {Self-adjoint
  extensions of operators and the teaching of quantum mechanics},}\ }\href@noop
  {} {\bibfield  {journal} {\bibinfo  {journal} {American Journal of physics}\
  }\textbf {\bibinfo {volume} {69}},\ \bibinfo {pages} {322--331} (\bibinfo
  {year} {2001})}\BibitemShut {NoStop}%
\bibitem [{\citenamefont {Blank}\ \emph {et~al.}(2008)\citenamefont {Blank},
  \citenamefont {Exner},\ and\ \citenamefont {Havlicek}}]{blank2008hilbert}%
  \BibitemOpen
  \bibfield  {author} {\bibinfo {author} {\bibfnamefont {Jiri}\ \bibnamefont
  {Blank}}, \bibinfo {author} {\bibfnamefont {Pavel}\ \bibnamefont {Exner}}, \
  and\ \bibinfo {author} {\bibfnamefont {Miloslav}\ \bibnamefont {Havlicek}},\
  }\href@noop {} {\emph {\bibinfo {title} {Hilbert space operators in quantum
  physics}}}\ (\bibinfo  {publisher} {Springer Science \& Business Media},\
  \bibinfo {year} {2008})\BibitemShut {NoStop}%
\bibitem [{\citenamefont {Bauchau}\ and\ \citenamefont
  {Craig}(2009)}]{Euler-Bernoulli}%
  \BibitemOpen
  \bibfield  {author} {\bibinfo {author} {\bibfnamefont {Oliver~A}\
  \bibnamefont {Bauchau}}\ and\ \bibinfo {author} {\bibfnamefont {James~I}\
  \bibnamefont {Craig}},\ }\bibfield  {title} {\enquote {\bibinfo {title}
  {Euler-bernoulli beam theory},}\ }in\ \href@noop {} {\emph {\bibinfo
  {booktitle} {Structural analysis}}}\ (\bibinfo  {publisher} {Springer},\
  \bibinfo {year} {2009})\ pp.\ \bibinfo {pages} {173--221}\BibitemShut
  {NoStop}%
\bibitem [{\citenamefont {Yang}\ \emph {et~al.}(2020)\citenamefont {Yang},
  \citenamefont {Zhang}, \citenamefont {Fang},\ and\ \citenamefont
  {Hu}}]{yang2020non}%
  \BibitemOpen
  \bibfield  {author} {\bibinfo {author} {\bibfnamefont {Zhesen}\ \bibnamefont
  {Yang}}, \bibinfo {author} {\bibfnamefont {Kai}\ \bibnamefont {Zhang}},
  \bibinfo {author} {\bibfnamefont {Chen}\ \bibnamefont {Fang}}, \ and\
  \bibinfo {author} {\bibfnamefont {Jiangping}\ \bibnamefont {Hu}},\ }\bibfield
   {title} {\enquote {\bibinfo {title} {Non-{Hermitian} bulk-boundary
  correspondence and auxiliary generalized {Brillouin} zone theory},}\ }\href
  {\doibase 10.1103/PhysRevLett.125.226402} {\bibfield  {journal} {\bibinfo
  {journal} {Phys. Rev. Lett.}\ }\textbf {\bibinfo {volume} {125}},\ \bibinfo
  {pages} {226402} (\bibinfo {year} {2020})}\BibitemShut {NoStop}%
\bibitem [{\citenamefont {Bender}\ and\ \citenamefont
  {Boettcher}(1998)}]{bender1998real}%
  \BibitemOpen
  \bibfield  {author} {\bibinfo {author} {\bibfnamefont {Carl~M.}\ \bibnamefont
  {Bender}}\ and\ \bibinfo {author} {\bibfnamefont {Stefan}\ \bibnamefont
  {Boettcher}},\ }\bibfield  {title} {\enquote {\bibinfo {title} {Real spectra
  in non-hermitian hamiltonians having {PT} symmetry},}\ }\href {\doibase
  10.1103/PhysRevLett.80.5243} {\bibfield  {journal} {\bibinfo  {journal}
  {Phys. Rev. Lett.}\ }\textbf {\bibinfo {volume} {80}},\ \bibinfo {pages}
  {5243--5246} (\bibinfo {year} {1998})}\BibitemShut {NoStop}%
\bibitem [{\citenamefont {Bender}\ \emph {et~al.}(2002)\citenamefont {Bender},
  \citenamefont {Berry},\ and\ \citenamefont
  {Mandilara}}]{bender2002generalized}%
  \BibitemOpen
  \bibfield  {author} {\bibinfo {author} {\bibfnamefont {Carl~M}\ \bibnamefont
  {Bender}}, \bibinfo {author} {\bibfnamefont {MV}~\bibnamefont {Berry}}, \
  and\ \bibinfo {author} {\bibfnamefont {Aikaterini}\ \bibnamefont
  {Mandilara}},\ }\bibfield  {title} {\enquote {\bibinfo {title} {Generalized
  pt symmetry and real spectra},}\ }\href@noop {} {\bibfield  {journal}
  {\bibinfo  {journal} {Journal of Physics A: Mathematical and General}\
  }\textbf {\bibinfo {volume} {35}},\ \bibinfo {pages} {L467} (\bibinfo {year}
  {2002})}\BibitemShut {NoStop}%
\bibitem [{\citenamefont
  {{Mostafazadeh}}(2002{\natexlab{a}})}]{Mostafazadeh2002i}%
  \BibitemOpen
  \bibfield  {author} {\bibinfo {author} {\bibfnamefont {A.}~\bibnamefont
  {{Mostafazadeh}}},\ }\bibfield  {title} {\enquote {\bibinfo {title}
  {{Pseudo-Hermiticity versus PT symmetry: The necessary condition for the
  reality of the spectrum of a non-Hermitian Hamiltonian}},}\ }\href {\doibase
  10.1063/1.1418246} {\bibfield  {journal} {\bibinfo  {journal} {Journal of
  Mathematical Physics}\ }\textbf {\bibinfo {volume} {43}},\ \bibinfo {pages}
  {205--214} (\bibinfo {year} {2002}{\natexlab{a}})}\BibitemShut {NoStop}%
\bibitem [{\citenamefont
  {{Mostafazadeh}}(2002{\natexlab{b}})}]{Mostafazadeh2002ii}%
  \BibitemOpen
  \bibfield  {author} {\bibinfo {author} {\bibfnamefont {A.}~\bibnamefont
  {{Mostafazadeh}}},\ }\bibfield  {title} {\enquote {\bibinfo {title}
  {{Pseudo-Hermiticity versus PT-symmetry. II. A complete characterization of
  non-Hermitian Hamiltonians with a real spectrum}},}\ }\href {\doibase
  10.1063/1.1461427} {\bibfield  {journal} {\bibinfo  {journal} {Journal of
  Mathematical Physics}\ }\textbf {\bibinfo {volume} {43}},\ \bibinfo {pages}
  {2814--2816} (\bibinfo {year} {2002}{\natexlab{b}})}\BibitemShut {NoStop}%
\bibitem [{\citenamefont {Bender}(2007)}]{bender2007making}%
  \BibitemOpen
  \bibfield  {author} {\bibinfo {author} {\bibfnamefont {Carl~M}\ \bibnamefont
  {Bender}},\ }\bibfield  {title} {\enquote {\bibinfo {title} {Making sense of
  non-hermitian hamiltonians},}\ }\href@noop {} {\bibfield  {journal} {\bibinfo
   {journal} {Reports on Progress in Physics}\ }\textbf {\bibinfo {volume}
  {70}},\ \bibinfo {pages} {947} (\bibinfo {year} {2007})}\BibitemShut
  {NoStop}%
\bibitem [{\citenamefont {Song}\ \emph
  {et~al.}(2019{\natexlab{b}})\citenamefont {Song}, \citenamefont {Yao},\ and\
  \citenamefont {Wang}}]{song2019real}%
  \BibitemOpen
  \bibfield  {author} {\bibinfo {author} {\bibfnamefont {Fei}\ \bibnamefont
  {Song}}, \bibinfo {author} {\bibfnamefont {Shunyu}\ \bibnamefont {Yao}}, \
  and\ \bibinfo {author} {\bibfnamefont {Zhong}\ \bibnamefont {Wang}},\
  }\bibfield  {title} {\enquote {\bibinfo {title} {Non-hermitian topological
  invariants in real space},}\ }\href {\doibase 10.1103/PhysRevLett.123.246801}
  {\bibfield  {journal} {\bibinfo  {journal} {Phys. Rev. Lett.}\ }\textbf
  {\bibinfo {volume} {123}},\ \bibinfo {pages} {246801} (\bibinfo {year}
  {2019}{\natexlab{b}})}\BibitemShut {NoStop}%
\bibitem [{\citenamefont {Hu}\ \emph {et~al.}(2024)\citenamefont {Hu},
  \citenamefont {Wang}, \citenamefont {Wang},\ and\ \citenamefont
  {Song}}]{hu2024geometric}%
  \BibitemOpen
  \bibfield  {author} {\bibinfo {author} {\bibfnamefont {Yu-Min}\ \bibnamefont
  {Hu}}, \bibinfo {author} {\bibfnamefont {Hong-Yi}\ \bibnamefont {Wang}},
  \bibinfo {author} {\bibfnamefont {Zhong}\ \bibnamefont {Wang}}, \ and\
  \bibinfo {author} {\bibfnamefont {Fei}\ \bibnamefont {Song}},\ }\bibfield
  {title} {\enquote {\bibinfo {title} {Geometric origin of non-bloch
  $\mathcal{P}\mathcal{T}$ symmetry breaking},}\ }\href {\doibase
  10.1103/PhysRevLett.132.050402} {\bibfield  {journal} {\bibinfo  {journal}
  {Phys. Rev. Lett.}\ }\textbf {\bibinfo {volume} {132}},\ \bibinfo {pages}
  {050402} (\bibinfo {year} {2024})}\BibitemShut {NoStop}%
\bibitem [{\citenamefont {Kunst}\ and\ \citenamefont
  {Dwivedi}(2019)}]{kunst2019non}%
  \BibitemOpen
  \bibfield  {author} {\bibinfo {author} {\bibfnamefont {Flore~K.}\
  \bibnamefont {Kunst}}\ and\ \bibinfo {author} {\bibfnamefont {Vatsal}\
  \bibnamefont {Dwivedi}},\ }\bibfield  {title} {\enquote {\bibinfo {title}
  {Non-hermitian systems and topology: A transfer-matrix perspective},}\ }\href
  {\doibase 10.1103/PhysRevB.99.245116} {\bibfield  {journal} {\bibinfo
  {journal} {Phys. Rev. B}\ }\textbf {\bibinfo {volume} {99}},\ \bibinfo
  {pages} {245116} (\bibinfo {year} {2019})}\BibitemShut {NoStop}%
\bibitem [{Note1()}]{Note1}%
  \BibitemOpen
  \bibinfo {note} {It is arbitrary to choose the initial reference point $x_0$
  in the definition of the transfer matrix. Here we fix $x_0=0$ for
  simplicity.}\BibitemShut {Stop}%
\bibitem [{Note2()}]{Note2}%
  \BibitemOpen
  \bibinfo {note} {Eq. \protect \textup {\hbox {\mathsurround \z@ \protect
  \normalfont (\ignorespaces \ref {eq:periodic_hatano_nelson_det}\unskip
  \@@italiccorr )}} stems from the fact that $\protect \qopname \relax
  m{det}[\protect \mathcal {P}_xe^{i\DOTSI \intop \ilimits@ _0^a
  A(x')dx'}]=\protect \qopname \relax m{det}[\protect \qopname \relax
  m{lim}\limits _{N\to +\infty }\DOTSB \prod@ \slimits@ _{n=1}^Ne^{iA(\protect
  \frac {na}{N})\protect \frac {a}{N}}]=\protect \qopname \relax m{lim}\limits
  _{N\to +\infty }\DOTSB \prod@ \slimits@ _{n=1}^N\protect \qopname \relax
  m{det}[e^{iA(\protect \frac {na}{N})\protect \frac {a}{N}}]=\protect \qopname
  \relax m{lim}\limits _{N\to +\infty }e^{i\DOTSB \sum@ \slimits@
  _{n=1}^N\protect \operatorname {Tr}[A(\protect \frac {na}{N})]\protect \frac
  {a}{N}}=e^{i\DOTSI \intop \ilimits@ _0^a\protect \operatorname
  {Tr}[A(x^\prime )]\protect \mathrm {d}x^\prime }$.}\BibitemShut {Stop}%
\end{thebibliography}%
\end{document}